\documentclass{aa}  

\usepackage{graphicx}
\usepackage{txfonts}
\usepackage{longtable}
\begin{document}

   \title{Gas and dust from extremely metal-poor AGB stars}

%  \subtitle{I. Gas }

   \author{P. Ventura\inst{1}, F. Dell'Agli\inst{1}, D. Romano$^{2}$, S. Tosi$^{3}$, M. Limongi$^{1,4,5}$, A. Chieffi$^{6}$, M. Castellani$^{1}$, \\
   M. Tailo$^{7}$, M. Lugaro$^{8,9,10}$, E. Marini$^{11,12,13}$, A. Yag\"ue Lopez$^{8}$
   }

   \institute{INAF, Observatory of Rome, Via Frascati 33, 00077 Monte Porzio Catone (RM), Italy \and
   INAF, Astrophysics and Space Science Observatory, Via Piero Gobetti 93/3, 40129 Bologna, Italy  \and
   Dipartimento di Matematica e Fisica, Universit\`a degli Studi Roma Tre, 
   Via della Vasca Navale 84, 00100, Roma, Italy \and
   Kavli Institute for the Physics and Mathematics of the Universe, Todai Institutes for Advanced Study, the University of Tokyo, Kashiwa, Japan 277-8583 (Kavli IPMU, WPI) \and
   INFN. Sezione di Perugia, via A. Pascoli s/n, 06125 Perugia, Italy. \and
   INAF - Istituto di Astrofisica e Planetologia Spaziali, Via Fosso del Cavaliere 100, I-00133, Roma, Italy  \and
   Dipartimento di Fisica e Astronomia 'Galileo Galilei', Universit\`a di Padova, Vicolo dell'Osservatorio 3, I-35122 Padova, Italy  \and
   Konkoly Observatory, Research Centre for Astronomy and Earth Sciences, 
   Konkoly Thege Mikl\'{o}s \'{u}t 15-17, H-1121 Budapest, Hungary \and
   ELTE E\"{o}tv\"{o}s Lor\'and University, Institute of Physics, Budapest 1117, P\'azm\'any P\'eter s\'et\'any 1/A, Hungary 
   \and
   School of Physics and Astronomy, Monash University, VIC 3800, Australia 
   \and
   Nordita, KTH Royal Institute of Technology and Stockholm University Hannes Alfvens v\"ag 12, 114 21 Stockholm, Sweden
   \and
   Instituto de Astrofisica de Canarias (IAC), E-38200 La Laguna, Tenerife, Spain
   \and
   Departamento de Astrofisica, Universidad de La Laguna (ULL), E-38206 La Laguna, Tenerife, Spain
   }

   \date{Received September 15, 1996; accepted March 16, 1997}

% \abstract{}{}{}{}{} 
% 5 {} token are mandatory

  \abstract
   {The study of stars evolving through the asymptotic giant branch (AGB) proves crucial in several astrophysical contexts, given the important feedback provided by these objects to the host system, in terms of the gas, poured into the interstellar medium after being exposed to   contamination from nucleosynthesis processes, and the dust formed in their wind. Most of the studies conducted so far have been focused on AGB stars with solar and sub-solar chemical composition, whereas the extremely metal-poor domain has been poorly explored.}
   {We study the evolution of extremely metal-poor AGB stars, with metallicities down to $[$Fe$/$H$]=-5$, to understand the main evolutionary
   properties, the efficiency of the processes able to alter their surface 
   chemical composition and to determine the gas and dust yields.}
   {We calculate two sets of evolutionary sequences of stars in the 
   $1-7.5~\rm M_{\odot}$ mass range, evolved from the pre-main sequence to
   the end of the AGB phase. To explore the extremely metal-poor chemistries
   we adopted the metallicities $Z=3\times 10^{-5}$ and $Z=3\times 10^{-7}$
   which correspond, respectively to $[$Fe$/$H$]=-3$ and $[$Fe$/$H$]=-5$.
   The results from stellar evolution modelling are used to calculate
   the yields of the individual chemical species. We also modelled dust formation in the wind, to determine the dust produced by these objects.
   }
   {The evolution of AGB stars in the extremely metal-poor domain explored
   here proves tremendously sensitive to the initial mass of the star. 
   $\rm M \leq 2~\rm M_{\odot}$ stars experience several third dredge-up events, which favour the gradual surface enrichment of $^{12}$C and the formation of significant quantities of carbonaceous dust, of the order of $\sim 0.01~\rm M_{\odot}$. The $^{13}$C and nitrogen yields
   are found to be significantly smaller than in previous explorations of low-mass, metal-poor
   AGB stars, owing to 
   the weaker proton ingestion episodes experienced during the initial AGB phases.
   $\rm M \geq 5~\rm M_{\odot}$ stars experience hot bottom burning and their surface chemistry reflects the equilibria of a very advanced proton-capture 
   nucleosynthesis; little dust production takes place in their wind.
   Intermediate mass stars experience both third dredge-up and hot bottom
   burning: they prove efficient producers of nitrogen, which is formed
   by proton captures on $^{12}$C nuclei of primary origin dredged-up
   from the internal regions.}
   {}


   \titlerunning{Extremely metal-poor AGB stars}
   \authorrunning{Ventura et al.}
   \maketitle
%
%-------------------------------------------------------------------

\section{Introduction}
The last decades have witnessed a growing interest towards the evolution
of the stars evolving through the AGB. This evolutionary
phase, which follows core helium burning and precedes the white dwarf 
cooling, is crossed by all the stars with initial mass in the $0.8-8~\rm M_{\odot}$ range, 
the exact values of the threshold mass limits being sensitive to the metallicity and 
to the still debated extension of the convective core during the main sequence phase.

The reason why AGB stars have attracted the interest of the astrophysical
community is the important role played in different contexts, such as 
the chemical evolution of the Milky Way \citep{romano10}, the chemical patterns
observed in star-forming galaxies in the Local Universe \citep{vincenzo16}, the 
formation of multiple populations in globular clusters \citep{ventura01, dercole08},
the contribution to the overall dust budget in the Local Universe and at
high redshift \citep{valiante09}. Important areas where AGB stars have 
proven useful include inferring the masses of galaxies at high redshifts, owing 
to their large infrared luminosities \citep{maraston06}, and the interpretation of
the chemical composition of planetary nebulae \citep{ventura15}.

Various research groups have decided to develop libraries of AGB models,
which provide the evolution of the main physical and chemical properties of
the stars as they evolve through the AGB and global quantities, such as the
gas yields. Among the most widely used, we believe important to mention
the FRUITY database \citep{cristallo11, cristallo15} and the MONASH models
\citep{karakas10, karakas14a, karakas14b, karakas16}.

Once the potential role of AGB stars as important dust manufactures was 
proposed, some research groups included the description of dust formation in
the winds of AGB stars. This opened the possibility to simulate the evolution of the spectral
energy distribution at infrared wavelengths and to provide dust yields from
stars of different mass and metallicity \citep{ventura12, ventura14b, ventura18, 
flavia19, nanni13, nanni14}. These models have been recently used to characterize the evolved stellar populations of the Magellanic Clouds \citep{flavia14a, flavia14b, flavia15a, flavia15b, nanni16, nanni18, nanni19a, nanni19b} and of Local Group galaxies \citep{flavia16, flavia18, flavia19}.

A feature common to all the studies conducted so far is the fact that they are limited to stars of metallicity above $10^{-4}$, with the only exceptions of 
an early exploration by Campbell \& Lattanzio (2008, CL08), who studied $\rm M <3~\rm M_{\odot}$ stars
in the very metal-poor domain, down to almost zero metallicity, and the recent
work by Gil-Pons et al. (2021, GP21), who presented detailed models of $\rm M >3~\rm M_{\odot}$ stars
of metallicity $Z=10^{-5}$. The reasons for the scarcity of AGB models in
the extremely metal-poor domain are the difficulty in obtaining observations
of very low metallicity evolved stars and the intrinsic difficulties of these
computations, considering that the number of thermal pulses (TP) experienced is much
higher than the metal-rich counterparts \citep{lau08} and the occurrence of hard-to-model phenomena, such as the dual-shell flashes \citep{campbell08}.

Nevertheless the availability of extremely metal-poor AGB models turns out
extremely important in the context of galactic archaeology, because the comparison
between theoretical gas yields and the observations of carbon-enhanced metal-poor (CEMP) stars in the Halo and dwarf galaxies might shed new light on the chemistry of the early Universe.
Furthermore, the gas yields from stars in the very metal-poor domain would be a 
valuable input for the chemical evolution models, which can currently rely only on
AGB models with [Fe$/$H$]>-2.15$. This is a major limitation, especially for studies dealing with the most metal-poor galaxies known, namely, the ultrafaint dwarf galaxies (UFDs) that are now found numerous in and around the haloes of larger galaxies such as the Milky Way and Andromeda. The bulk of the stars in the UFDs have metallicities between $[$Fe$/$H$]=-3.2$ and $[$Fe$/$H$]=-2.0$ \citep{simon19}. This piece of observational evidence, combined with theoretical expectation that low- and intermediate-mass star formation is favoured in UFDs at the expenses of massive stars \citep{yan20} makes it clear why detailed modeling of extremely metal-poor AGB stars is desperately needed.

The main motivation of the present work is to fill this gap, presenting detailed
results from AGB modelling, extended to extremely metal-poor chemistries. 
We calculated a set of models of metallicity $Z=3\times 10^{-5}$, which corresponds to [Fe$/$H]$=-3$. This is the lowest chemical composition currently available in the data-base of massive stars presented by Limongi \& Chieffi (2018, hereinfater LC18), which allows, for the first time, to present gas yields from chemically homogeneous model stars, spanning the $1-120~\rm M_{\odot}$ mass range. To explore lower metallicities and to understand the trend of the stellar properties and of the gas and dust yields with metallicity, we also present a set of $Z=3\times 10^{-7}$ AGB model stars.

The paper is organized as follows: the numerical ingredients used to
calculate the stellar evolutionary sequences and to model dust formation, and the physical assumption adopted in the computations are given in 
section \ref{modinput}; the evolutionary phases before the AGB and the AGB
phase itself are discussed in section \ref{preagb} and \ref{agb}, respectively; the gas yields are commented in section \ref{yields};
in section \ref{comp} the results obtained in the present work are
compared with previous investigations available in the literature; the relevance
of the present study for the interpretation of the chemical composition
of CEMP stars is discussed in section \ref{cemp}.
the dust budget from the AGB stars are discussed in section \ref{dust};
finally, the conclusions are given in section \ref{end}.

\section{Physical and numerical input}
\label{modinput}
\subsection{Stellar evolution modelling}
\label{agbinput}
The stellar models presented and discussed in this work were calculated by means of the ATON code for stellar evolution \citep{ventura98}. An exhaustive description of the numerical details of the code and the most recent updates can be found in \citet{ventura13}. We remind here the physical and chemical input most relevant for the results obtained in the present work.

\subsubsection{The chemical composition}
We calculated two sets of models, with metallicity $Z=3\times 10^{-5}$ and
$Z=3\times 10^{-7}$. The initial helium mass fraction was assumed $Y=0.25$ in both cases. We used the solar mixture by \citet{a09}, with the exception of the
distribution of $^{12}$C and the $\alpha-$elements, which were taken from \citet{cayrel04}. With these assumptions, the two metallicities
given above correspond to $[$Fe$/$H$]=-3, -5$.

The evolutionary sequences are run from the pre-MS to the late phases, preceding
the begin of the post-AGB phase. Low-mass models ($\rm M  < 2~\rm M_{\odot}$) experiencing the helium flash were evolved from the horizontal branch, starting from the core mass and surface chemical composition calculated until the tip of the red giant branch (RGB). 

\subsubsection{Equation of State}
Tables of the equation of state are generated in the (gas) pressure-temperature plane, according to the OPAL EOS (2005), overwritten in the pressure ionization regime by the EOS by \citet{saumon95}, and extended to the high-density, high-temperature domain according to the treatment by \citet{stolz00}.
\subsubsection{Convection}
The temperature gradient within regions unstable to convection was calculated via the Full Spectrum of Turbulence (FST) model \citep{cm91}. Overshoot of convective eddies within radiatively stable regions is modelled by assuming that the velocity of convective elements decays exponentially beyond the neutrality point, fixed via the Schwartzschild criterion. The overshoot scale (hereinafter OS), i.e. the e-folding distance of the velocity decays during the core (hydrogen and helium) burning phases and during the AGB phase, is taken as $0.02H_{\rm P}$ and $0.002H_{\rm P}$, respectively; $H_{\rm P}$ is the pressure scale height at the 
formal convective border. The latter values reflect the calibrations discussed, respectively, in \citet{ventura98} and \citet{ventura14b}.

\subsubsection{Mass-loss}
The mass-loss rate during the phases when the star is oxygen-rich was determined via the mass-loss period relationship by \citet{vw93}. For carbon stars we used the \citet{vw93} recipe when C$-$O$<8.4$\footnote{We indicate with C$-$O the carbon excess with respect to oxygen, defined as C$-$O$=12+\log[($n$($C$)-$n$($O$))/$n$($H$)]$, where n(X) is the abundance by number of species X.}, whereas for higher carbon excesses we adopted the mass-loss description from the Berlin group \citep{wachter02, wachter08}. 

\subsubsection{Opacities}
The radiative opacities are calculated according to the OPAL release, in the version documented by \citet{opal}. The molecular opacities in the low-temperature regime ($T < 10^4$ K) are calculated with the AESOPUS tool \citep{marigo09}. The opacities are constructed self-consistently, by following the changes in the chemical composition of the envelope, particularly of the individual abundances of carbon, nitrogen, and oxygen.
 
\subsubsection{Nuclear network}
The nuclear network includes 30 elements (up to $^{31}$P) and 64 reactions. The 
rates of the $3\alpha$ reactions were taken from \citet{fynbo05}, wherease for
the proton capture process by $^{14}$N we used \citet{formicola04}. The other
reactions were taken from the REACLIB dataset.

\subsection{Dust production}
\label{dustmod}
We modelled the formation and growth of dust particles in the wind of AGB stars according to  the schematization proposed by the Heidelberg group \citep{fg06}, similarly to previous works by our team \citep{ventura12, ventura14b, ventura15, ventura16} and used in a series of papers by  \citet{nanni13, nanni14, nanni16, nanni18, nanni19a, nanni20}. The interested reader can find all the relevant equations in \citet{ventura12}. Here we provide a brief description of the methodology used.

Dust particles are assumed to form and grow in the wind, which expands isotropically from the photosphere of the star. In carbon-rich environments we consider the formation of solid carbon and silicon carbide (SiC), whereas in the winds of oxygen-rich stars we assume that the dust species formed are silicates and alumina dust (Al$_2$O$_3$); solid iron is formed in either cases. Each solid compound is characterized by a key species, defined among the least abundant among the chemical elements entering the formation reaction from gaseous molecules to solid particles: for the species considered here the key species are silicon (silicates and SiC), carbon (solid carbon), aluminium (alumina dust) and iron (solid iron). Each dust species starts to form in the region where the growth rate exceeds the vaporisation rate. The former is connected with the thermal velocity of the key species and the sticking coefficient, which represents the tendency of the gaseous molecules to stick to the
already formed solid grains. The vaporisation rate depends on the thermodynamic properties of the dust species, mainly on the formation enthalpies of the solid compounds and of the gaseous molecules involved in the formation reaction \citep{fg06}.

The dynamics of the wind is described by the momentum equation, where the acceleration is determined by the competition between gravity and radiation pressure
acting on the newly formed dust grains. The coupling between grain growth and wind dynamics is given by the extinction coefficients, describing absorption and scattering of the radiation by dust particles. For the species considered here the extinction coefficients were found by using the optical constants from \citet{zubko} (amorphous carbon), \citet{peg} (SiC), \citet{begemann94} (alumina dust), \citet{ossenkopf} (silicates) and \citet{ordal} (solid iron).

The modelling of dust formation, as described above, allows the determination of the size reached by the grains of the various species,
an estimate of the surface fraction of gaseous silicon, carbon, aluminium and
iron condensed into dust (see eq. 20-23 and 34-35 in \citet{fg06}) and the dust production rate for each dust species. The latter depends on the gas mass-loss rate, the surface mass fractions of the afore mentioned chemical elements, and the fraction of the latter species condensed into dust \citep[see Section 5.2 in][]{fg06}.  

\begin{table*}
\caption{The main properties of the model stars discussed in this work,
for the (initial) mass range $1-7.5~\rm M_{\odot}$ and metallicities
$Z=3\times 10^{-5}$ (upper part of the table) and $Z=3\times 10^{-7}$
(lower part). The quantities reported in the various columns are
the initial mass of the star (col. 1), the duration (Myr units) of the core
hydrogen (2) and helium (3) burning phases, of the early AGB (4, kyr unit) and
TP-AGB phases (5), and of the inter-pulse period (6), the largest luminosity (6, solar luminosity units) and temperature at the base of the convective envelope (7, kelvin) attained during the AGB evolution, the largest value attained by the
TDU parameter $\lambda$ (9), the core mass (in solar masses) at the first TP 
(10) and at end of the AGB phase (11), the final surface C$/$O (12) and the
fraction of the AGB life-time during which the star has a C-star chemical composition (13).
}             
\label{tabgen}      
\centering
\begin{tabular}{c c c c c c c c c c c c c}    
\hline      
M$/$M$_{\odot}$ & $\tau_{\rm H}$ & $\tau_{\rm He}$ & $\tau_{\rm E-AGB}$ & $\tau_{\rm TP-AGB}$ & $\tau_{\rm int}$ & L$_{\rm max}$ & 
T$_{\rm b,max}$ & $\lambda_{\rm max}$ & M$_{\rm c,i}$ & M$_{\rm c,f}$ & 
(C$/$O)$_{\rm f}$ & $\%$(C-star) \\
\hline 
\multicolumn{13}{c}{$Z=3\times 10^{-5}$} \\
\hline
1.0 & 5260 & 72 & 17000 & 1320 & 155 & $6.07\times 10^3$ & 1.94 & 0.18 & 0.53 & 0.57 & 14.37 & 0.53 \\
1.25 & 2460 & 77 & 8840 & 1300 & 110 & $7.99\times 10^3$ & 3.26 & 0.44 & 0.55 & 0.59 & 8.00 & 0.68 \\
1.5  & 1340 & 92 & 9670 & 1740 & 100 & $9.74\times 10^3$ & 5.30 & 0.54 & 0.54 & 0.61 & 8.35 & 0.77 \\
2.0  & 557 & 120 & 5820 & 1090 & 61 & $1.48\times 10^4$ & 22.6  & 0.70 & 0.62 & 0.69 & 2.87 & 0.63 \\
2.5 & 321 & 77 & 3510 & 722 & 26 & $3.25\times 10^4$    & 80.5  & 0.58 & 0.72 & 0.82 & 0.38 & $-$ \\
3.0 & 219 & 43 & 2270 & 407 & 13 & $4.44\times 10^4$    & 89    & 0.54 & 0.79 & 0.86 & 0.47 & $-$ \\
3.5 & 161 & 27 & 1460 & 372 & 9  & $5.80\times 10^4$    & 95.4  & 0.50 & 0.82 & 0.89 & 0.94 & $-$ \\
4.0 & 124 & 19 & 1010 & 309 & 6  & $6.94\times 10^4$    & 101   & 0.46 & 0.85 & 0.92 & 0.93 & $-$ \\
4.5 & 98 & 14 & 724 & 266 & 4    & $8.51\times 10^4$    & 110   & 0.40 & 0.89 & 0.96 & 0.93 & $-$ \\
5.0 & 80 & 11 & 533 & 453 & 2    & $1.10\times 10^5$    & 150   &  -   & 0.93 & 1.10 & 9.66 & 0.73 \\
6.0 & 57 & 6.8 & 307 & 249 & 0.9 & $1.40\times 10^5$    & 150   &  -   & 1.03 & 1.14 & 10.06 & 0.91 \\
7.0 & 43 & 4.8 & 184 & 109 & 0.3 & $1.86\times 10^5$    & 171   &  -   & 1.22 & 1.25 & 15.16 & 0.85 \\
7.5 & 38 & 4.1 & 155 & 24 & 0.2  & $2.16\times 10^5$    & 158   &  -   & 1.31 & 1.32 & 17.39 & 0.83 \\
\hline 
\multicolumn{13}{c}{$Z=3\times 10^{-7}$} \\
\hline
1.0 & 5090 & 64 & 9440 & 1770 & 200   & $6.11\times 10^3$ & 1.78  & 0.17 & 0.52 & 0.58 & 23.36 & 0.69 \\
1.25 & 2450 & 81 & 12600 & 1770 & 120 & $6.94\times 10^3$ & 2.66  & 0.40 & 0.52 & 0.58 & 17.97 & 0.51 \\
1.5 & 1380 & 207 & 1350 & 3320 & 110  & $9.55\times 10^3$ & 5.38  & 0.42 & 0.46 & 0.62 & 4.54 & 0.69 \\
2.0 & 561 & 64 & 4290 & 778 & 50      & $1.65\times 10^4$ & 29.4  & 0.71 & 0.70 & 0.72 & 6.34 & 0.85 \\
2.5 & 286 & 44 & 2410 & 532 & 15      & $3.59\times 10^4$ & 83.5  & 0.57 & 0.76 & 0.84 & 0.78 & $-$ \\
3.0 & 178 & 33 & 1850 & 377 & 9       & $4.95\times 10^4$ & 90.9  & 0.53 & 0.81 & 0.87 & 0.58 & $-$ \\
3.5 & 125 & 27 & 1410 & 710 & 6 &     $7.67\times 10^4$   & 116   & 0.39 & 0.83 & 0.99 & 1.09 & $-$ \\
4.0 & 97 & 21 & 1110 & 1120 & 4 &     $9.28\times 10^4$   & 185   & 0.08 & 0.85 & 1.14 & $ - $ & $-$ \\
5.0 & 67 & 11 & 685 & 847 & 2.2 &     $1.44\times 10^5$   & 194   &  -   & 0.90 & 1.16 & $ - $ & $-$ \\
6.0 & 49 & 6.7 & 385 & 449 & 0.6 &    $1.62\times 10^5$   & 175   &  -   & 0.99 & 1.19 & 14.67 & 0.88 \\
7.0 & 38 & 4.6 & 197 & 130 & 0.3 &    $2.07\times 10^5$   & 166   &  -   & 1.17 & 1.21 & 10.33 & 1.00 \\
7.5 & 34 & 3.9 & 174 & 44 & 0.2  &    $2.40\times 10^5$   & 157   &  -   & 1.14 & 1.26 & 11.94 & 1.00 
\\
\hline
\end{tabular}
\end{table*}

\begin{table*}
\caption{Modification of the surface mass fractions of the individual species as a consequence of the first and second dredge-up episodes, for the model stars of metallicity Z$=3\times 10^{-5}$ (upper part of the table) and Z$=3\times 10^{-7}$ (lower side of the table). The mass fractions of
the different species (with the exception of helium) are given in
the $X(Y)$ format, which stands for $X^Y$. The line below the metallicity label reports the initial mass fractions.}             
\label{tabdup}      
\centering          
%\begin{tabular}{c c c c c c c c c c c c c c c c c}     % 10 columns 
%\begin{tabular}{p{0.5cm} p{0.5cm} p{1.0cm} p{1.0cm} p{1.0cm} p{0.6cm} p{0.6cm} p{0.6cm} p{0.6cm} %p{0.6cm} p{0.6cm} p{0.6cm} p{0.6cm} p{0.6cm} p{0.6cm} p{0.6cm} p{0.6cm}} 
\begin{tabular}{c@{\hspace{0.02cm}}c@{\hspace{0.18cm}}c@{\hspace{0.18cm}}c@{\hspace{0.18cm}}c@{\hspace{0.18cm}}c@{\hspace{0.18cm}}c@{\hspace{0.18cm}}c@{\hspace{0.18cm}}c@{\hspace{0.18cm}}c@{\hspace{0.18cm}}c@{\hspace{0.18cm}}c@{\hspace{0.18cm}}c@{\hspace{0.18cm}}c@{\hspace{0.18cm}}c@{\hspace{0.18cm}}c}
\hline
M & $^4$He& $^{12}$C & $^{13}$C & $^{14}$N & $^{15}$N &
$^{16}$O & $^{17}$O & $^{18}$O & $^{20}$Ne & $^{22}$Ne & $^{23}$Na &
$^{24}$Mg & $^{25}$Mg & $^{26}$Mg & $^{27}$Al  \\ 
\hline
\multicolumn{16}{c}{Z$=3\times 10^{-5}$} \\
\hline        
init &.25 & 4.1(-6) & 4.5(-8) & 6.8(-7) & 3.1(-9) & 1.8(-5) & 6.8(-9) & 3.7(-8) & 1.3(-6) & 9.7(-8) & 3.1(-8) & 9.2(-7) & 1.2(-7) & 1.3(-7) & 5.4(-8)  \\
\hline 
\multicolumn{16}{c}{FIRST DREDGE-UP} \\
\hline
 1.0  &.26 & 3.4(-6) & 1.2(-7) & 1.4(-6) & 1.5(-9) & 1.8(-5) & 7.2(-9) & 3.4(-8) & 1.3(-6) & 9.7(-8) & 3.1(-8) & 9.2(-7) & 1.2(-7) & 1.3(-7) & 5.4(-8)  \\
 1.25 &.27 & 2.8(-6) & 1.1(-7) & 2.1(-6) & 1.5(-9) & 1.8(-5) & 3.4(-8) & 2.9(-8) & 1.3(-6) & 9.7(-8) & 3.1(-8) & 9.2(-7) & 1.2(-7) & 1.3(-7) & 5.4(-8)  \\
 1.5  &.28 & 2.4(-6) & 1.0(-7) & 2.6(-6) & 1.5(-9) & 1.8(-5) & 2.4(-7) & 2.5(-8) & 1.3(-6) & 9.7(-8) & 3.1(-8) & 9.2(-7) & 1.2(-7) & 1.3(-7) & 5.4(-8)  \\
 2.0  &.26 & 2.6(-6) & 1.2(-7) & 2.3(-6) & 1.5(-9) & 1.8(-5) & 2.2(-8) & 2.7(-8) & 1.3(-6) & 9.2(-8) & 3.4(-8) & 9.2(-7) & 1.2(-7) & 1.3(-7) & 5.4(-8)  \\
 2.5  &.25 & 4.1(-6) & 4.5(-8) & 6.8(-7) & 3.1(-9) & 1.8(-5) & 6.8(-9) & 3.7(-8) & 1.3(-6) & 9.7(-8) & 3.1(-8) & 9.2(-7) & 1.2(-7) & 1.3(-7) & 5.4(-8)   \\
 3.0  &.25 & 4.1(-6) & 4.5(-8) & 6.8(-7) & 3.1(-9) & 1.8(-5) & 6.8(-9) & 3.7(-8) & 1.3(-6) & 9.7(-8) & 3.1(-8) & 9.2(-7) & 1.2(-7) & 1.3(-7) & 5.4(-8)  \\
 3.5  &.25 & 4.1(-6) & 4.5(-8) & 6.8(-7) & 3.1(-9) & 1.8(-5) & 6.8(-9) & 3.7(-8) & 1.3(-6) & 9.7(-8) & 3.1(-8) & 9.2(-7) & 1.2(-7) & 1.3(-7) & 5.4(-8)  \\
 4.0  &.25 & 4.1(-6) & 4.5(-8) & 6.8(-7) & 3.1(-9) & 1.8(-5) & 6.8(-9) & 3.7(-8) & 1.3(-6) & 9.7(-8) & 3.1(-8) & 9.2(-7) & 1.2(-7) & 1.3(-7) & 5.4(-8)  \\
 4.5  &.25 & 4.1(-6) & 4.5(-8) & 6.8(-7) & 3.1(-9) & 1.8(-5) & 6.8(-9) & 3.7(-8) & 1.3(-6) & 9.7(-8) & 3.1(-8) & 9.2(-7) & 1.2(-7) & 1.3(-7) & 5.4(-8)  \\
 5.0  &.25 & 4.1(-6) & 4.5(-8) & 6.8(-7) & 3.1(-9) & 1.8(-5) & 6.8(-9) & 3.7(-8) & 1.3(-6) & 9.7(-8) & 3.1(-8) & 9.2(-7) & 1.2(-7) & 1.3(-7) & 5.4(-8)  \\
 6.0  &.25 & 4.1(-6) & 4.5(-8) & 6.8(-7) & 3.1(-9) & 1.8(-5) & 6.8(-9) & 3.7(-8) & 1.3(-6) & 9.7(-8) & 3.1(-8) & 9.2(-7) & 1.2(-7) & 1.3(-7) & 5.4(-8)  \\
 7.0  &.25 & 4.1(-6) & 4.5(-8) & 6.8(-7) & 3.1(-9) & 1.8(-5) & 6.8(-9) & 3.7(-8) & 1.3(-6) & 9.7(-8) & 3.1(-8) & 9.2(-7) & 1.2(-7) & 1.3(-7) & 5.4(-8)  \\
 7.5  &.25 & 4.1(-6) & 4.5(-8) & 6.8(-7) & 3.1(-9) & 1.8(-5) & 6.8(-9) & 3.7(-8) & 1.3(-6) & 9.7(-8) & 3.1(-8) & 9.2(-7) & 1.2(-7) & 1.3(-7) & 5.4(-8)  \\
\hline
\multicolumn{16}{c}{SECOND DREDGE-UP} \\
\hline
 1.0  & .26  & 3.4(-6) & 1.2(-7) & 1.4(-6) & 1.5(-9)  & 1.8(-5) & 7.2(-9) & 3.4(-8) & 1.3(-6) & 9.8(-8) & 3.1(-8) & 9.2(-7) & 1.2(-7) & 1.3(-7) & 5.4(-8)  \\
 1.25 & .27  &  2.0(-6) & 1.0(-7) & 3.8(-6) & 1.2(-9)  & 1.7(-5) & 3.0(-7) & 2.1(-8) & 1.3(-6) & 8.8(-8) & 4.1(-8) & 9.2(-7) & 1.3(-7) & 1.3(-7) & 5.4(-8)    \\
 1.5  & .28  &  2.4(-6) & 1.0(-7) & 2.6(-6) & 1.5(-9)  & 1.8(-5) & 2.4(-7) & 2.5(-8) & 1.3(-6) & 9.7(-8) & 3.1(-8) & 9.2(-7) & 1.2(-7) & 1.3(-7) & 5.4(-8) \\
 2.0  & .26  & 2.2(-6) & 1.2(-7) & 3.7(-6) & 1.2(-9)  & 1.7(-5) & 7.6(-8) & 2.4(-8) & 1.3(-6) & 8.4(-8) & 4.9(-8) & 9.2(-7) & 1.1(-7) & 1.3(-7) & 5.5(-8) \\
 2.5  & .26  & 2.0(-6) & 1.0(-7) & 4.2(-6) & 1.2(-9)  & 1.7(-5) & 1.9(-7) & 2.2(-8) & 1.3(-6) & 7.5(-8) & 6.8(-8) & 9.2(-7) & 1.1(-7) & 1.3(-7) & 5.8(-8) \\
 3.0  & .28  & 1.8(-6) & 9.7(-8) & 5.0(-6) & 1.1(-9)  & 1.6(-5) & 3.3(-7) & 1.9(-8) & 1.3(-6) & 6.8(-8) & 8.8(-8) & 9.2(-7) & 1.1(-7) & 1.3(-7) & 6.3(-8) \\
 3.5  & .30  & 1.6(-6) & 9.2(-8) & 6.1(-6) & 1.0(-9)  & 1.5(-5) & 3.3(-7) & 1.7(-8) & 1.3(-6) & 6.2(-8) & 1.1(-7) & 9.3(-7) & 1.0(-7) & 1.2(-7) & 7.1(-8) \\
 4.0  & .32  & 1.5(-6) & 9.0(-8) & 6.9(-6) & 9.3(-10) & 1.4(-5) & 3.0(-7) & 1.6(-8) & 1.3(-6) & 5.8(-8) & 1.3(-7) & 9.5(-7) & 9.9(-8) & 1.2(-7) & 7.8(-8) \\
 4.5  & .33  & 1.4(-6) & 9.9(-8) & 7.5(-6) & 8.2(-10) & 1.4(-5) & 2.7(-7) & 1.5(-8) & 1.2(-6) & 5.5(-8) & 1.4(-7) & 9.6(-7) & 9.5(-8) & 1.2(-7) & 8.3(-8) \\
 5.0  & .35  & 1.4(-6) & 9.9(-8) & 8.0(-6) & 7.9(-10) & 1.3(-5) & 2.4(-7) & 1.5(-8) & 1.2(-6) & 5.3(-8) & 1.5(-7) & 9.8(-7) & 9.3(-8) & 1.1(-7) & 9.0(-8) \\
 6.0  & .36  & 8.5(-6) & 9.2(-8) & 8.7(-6) & 7.0(-10) & 1.3(-5) & 2.1(-7) & 4.6(-8) & 1.2(-6) & 5.7(-8) & 1.6(-7) & 1.0(-6) & 8.8(-8) & 1.1(-7) & 1.0(-7) \\
 7.0  & .36  & 2.8(-5) & 1.6(-7) & 8.2(-6) & 7.5(-10) & 1.3(-5) & 1.5(-7) & 7.8(-8) & 1.2(-6) & 1.1(-7) & 1.5(-7) & 1.0(-6) & 8.4(-8) & 1.1(-7) & 1.2(-7) \\
 7.5  & .36  & 9.9(-5) & 5.4(-6) & 1.6(-5) & 6.9(-10) & 1.4(-5) & 1.6(-7) & 6.4(-8) & 1.2(-6) & 2.3(-7) & 1.6(-7) & 1.0(-6) & 8.5(-8) & 1.1(-7) & 1.2(-7) \\
 \hline
 \hline
 \multicolumn{16}{c}{Z$=3\times 10^{-7}$} \\
 \hline        
 init &.25 & 4.1(-8) & 4.5(-10) & 6.8(-9) & 3.1(-11) & 1.8(-7) & 6.8(-11) & 3.7(-10) & 1.3(-8) & 9.7(-10) & 3.1(-10) & 9.2(-9) & 1.2(-9) & 1.3(-9) & 5.4(-10)  \\
\hline 
\multicolumn{16}{c}{FIRST DREDGE-UP} \\
\hline
 1.0  & .26 & 3.8(-8) & 1.3(-9) & 9.6(-9) & 2.5(-11) & 1.8(-7) & 6.9(-11) & 3.6(-10) & 1.3(-8) & 9.8(-10) & 3.1(-10) & 9.2(-9) & 1.2(-9) & 1.3(-9) & 5.4(-10)  \\
 1.25 & .26 & 2.9(-8) & 1.1(-9) & 2.0(-8) & 1.9(-11) & 1.8(-7) & 3.0(-10) & 2.9(-10) & 1.3(-8) & 9.3(-10) & 3.5(-10) & 9.2(-9) & 1.2(-9) & 1.3(-9) & 5.4(-10) \\
 1.5  & .25 & 3.7(-8) & 1.5(-9) & 9.7(-9) & 2.3(-11) & 1.8(-7) & 6.9(-11) & 3.6(-10) & 1.3(-8) & 9.7(-10) & 3.1(-10) & 9.2(-9) & 1.2(-9) & 1.3(-9) & 5.4(-10) \\
 2.0  & .25 & 4.1(-8) & 4.5(-10)& 6.8(-9) & 3.1(-11) & 1.8(-7) & 6.8(-11) & 3.7(-10) & 1.3(-8) & 9.7(-10) & 3.1(-10) & 9.2(-9) & 1.2(-9) & 1.3(-9) & 5.4(-10)    \\
 2.5  & .25 & 4.1(-8) & 4.5(-10)& 6.8(-9) & 3.1(-11) & 1.8(-7) & 6.8(-11) & 3.7(-10) & 1.3(-8) & 9.7(-10) & 3.1(-10) & 9.2(-9) & 1.2(-9) & 1.3(-9) & 5.4(-10)\\
 3.0  & .25 & 4.1(-8) & 4.5(-10)& 6.8(-9) & 3.1(-11) & 1.8(-7) & 6.8(-11) & 3.7(-10) & 1.3(-8) & 9.7(-10) & 3.1(-10) & 9.2(-9) & 1.2(-9) & 1.3(-9) & 5.4(-10) \\
 3.5  & .25 & 4.1(-8) & 4.5(-10)& 6.8(-9) & 3.1(-11) & 1.8(-7) & 6.8(-11) & 3.7(-10) & 1.3(-8) & 9.7(-10) & 3.1(-10) & 9.2(-9) & 1.2(-9) & 1.3(-9) & 5.4(-10) \\
 4.0  & .25 & 4.1(-8) & 4.5(-10)& 6.8(-9) & 3.1(-11) & 1.8(-7) & 6.8(-11) & 3.7(-10) & 1.3(-8) & 9.7(-10) & 3.1(-10) & 9.2(-9) & 1.2(-9) & 1.3(-9) & 5.4(-10) \\
 5.0  & .25 & 4.1(-8) & 4.5(-10)& 6.8(-9) & 3.1(-11) & 1.8(-7) & 6.8(-11) & 3.7(-10) & 1.3(-8) & 9.7(-10) & 3.1(-10) & 9.2(-9) & 1.2(-9) & 1.3(-9) & 5.4(-10) \\
 6.0  & .25 & 4.1(-8) & 4.5(-10)& 6.8(-9) & 3.1(-11) & 1.8(-7) & 6.8(-11) & 3.7(-10) & 1.3(-8) & 9.7(-10) & 3.1(-10) & 9.2(-9) & 1.2(-9) & 1.3(-9) & 5.4(-10) \\
 7.0  & .25 & 4.1(-8) & 4.5(-10)& 6.8(-9) & 3.1(-11) & 1.8(-7) & 6.8(-11) & 3.7(-10) & 1.3(-8) & 9.7(-10) & 3.1(-10) & 9.2(-9) & 1.2(-9) & 1.3(-9) & 5.4(-10) \\
 7.5  & .25 & 4.1(-8) & 4.5(-10)& 6.8(-9) & 3.1(-11) & 1.8(-7) & 6.8(-11) & 3.7(-10) & 1.3(-8) & 9.7(-10) & 3.1(-10) & 9.2(-9) & 1.2(-9) & 1.3(-9) & 5.4(-10) \\
 \hline
\multicolumn{16}{c}{SECOND DREDGE-UP} \\
\hline
 1.0  & .26 & 3.2(-8) & 1.1(-9) & 3.2(-8) & 2.2(-11) & 1.7(-7) & 1.2(-9)  & 8.3(-10) & 1.3(-8) & 8.3(-10) & 4.6(-10) & 9.2(-9) & 1.1(-9) & 1.3(-9) & 5.4(-10)  \\
 1.25 & .26 & 2.5(-8) & 1.0(-9) & 3.3(-8) & 1.7(-11) & 1.7(-7) & 1.2(-9)  & 2.6(-10) & 1.3(-8) & 8.6(-10) & 4.3(-10) & 9.2(-9) & 1.1(-9) & 1.3(-9) & 5.4(-10) \\
 1.5  & .26 & 2.7(-8) & 1.5(-9) & 3.3(-8) & 1.0(-11) & 1.7(-7) & 1.0(-9)  & 2.6(-10) & 1.3(-8) & 8.3(-10) & 4.6(-10) & 9.2(-9) & 1.2(-9) & 1.3(-9) & 5.4(-10)   \\
 2.0  & .25 & 2.0(-8) & 9.2(-10)& 6.1(-8) & 1.3(-11) & 1.5(-7) & 1.2(-9)  & 2.0(-10) & 1.2(-8) & 6.8(-10) & 1.2(-9)  & 9.3(-9) & 9.8(-10)& 1.3(-9) & 7.2(-10)\\
 2.5  & .29 & 1.6(-8) & 8.6(-10)& 7.3(-8) & 1.1(-11) & 1.4(-7) & 2.0(-9)  & 1.7(-10) & 1.2(-8) & 5.9(-10) & 1.3(-9)  & 9.7(-9) & 9.4(-10)& 1.2(-9) & 8.0(-10) \\
 3.0  & .31 & 1.4(-8) & 7.6(-10)& 8.5(-8) & 9.5(-12) & 1.3(-7) & 2.5(-9)  & 1.5(-10) & 1.2(-8) & 5.2(-10) & 1.4(-9)  & 9.6(-9) & 8.9(-10)& 1.2(-9) & 8.9(-10) \\
 3.5  & .32 & 1.2(-8) & 8.9(-10)& 9.7(-8) & 7.7(-12) & 1.2(-7) & 2.6(-9)  & 1.3(-10) & 1.1(-8) & 4.6(-10) & 1.4(-9)  & 9.2(-9) & 8.4(-10)& 1.2(-9) & 8.9(-10) \\
 4.0  & .33 & 9.6(-9) & 7.7(-10)& 1.1(-7) & 6.8(-12) & 1.1(-7) & 2.5(-9)  & 1.1(-10) & 1.1(-8) & 4.1(-10) & 1.4(-9)  & 9.0(-9) & 7.8(-10)& 1.2(-9) & 8.9(-10) \\
 5.0  & .35 & 9.1(-9) & 6.2(-10)& 1.2(-7) & 5.8(-12) & 9.4(-8) & 2.3(-9)  & 7.9(-11) & 1.0(-8) & 3.1(-10) & 1.6(-9)  & 8.7(-9) & 6.8(-10)& 1.2(-9) & 9.5(-10) \\
 6.0  & .36 & 1.6(-6) & 5.8(-10)& 1.3(-7) & 5.3(-12) & 8.4(-8) & 2.2(-9)  & 1.2(-10) & 1.0(-8) & 2.5(-10) & 1.8(-9)  & 8.4(-9) & 6.1(-10)& 1.1(-9) & 1.0(-9)  \\
 7.0  & .36 & 9.4(-6) & 4.2(-10)& 1.5(-7) & 5.6(-12) & 7.4(-8) & 1.3(-9)  & 4.9(-10) & 9.6(-9) & 2.6(-10) & 2.0(-9)  & 8.5(-9) & 3.9(-10)& 1.2(-9) & 1.2(-9)  
\\
 7.5  & .36 & 5.9(-5) & 3.5(-9)& 1.5(-7) & 5.8(-12) & 4.0(-7) & 1.1(-9)  & 9.4(-10) & 9.5(-9) & 1.3(-9) & 2.1(-9)  & 8.5(-9) & 4.0(-10)& 1.1(-9) & 1.3(-9)  \\
\hline
\end{tabular}
\end{table*}

\section{The evolutionary phases before the AGB}
\label{preagb}
The main properties of the stellar models presented in this work are listed in Table \ref{tabgen}, and concern both the pre-AGB and the AGB phases. The different columns
of Table \ref{tabgen} report the duration of the most important evolutionary phases and some quantities characterizing the AGB evolution, such as the maximum luminosity reached, the variation of the core mass and the fraction of the AGB life-time during which the star evolves as C-star.

Before the beginning of the series of thermal pulses all the model stars evolve across the two major core nuclear burning phases, whose time scale, as is evident by looking at the numbers in the second and third column of Table \ref{tabgen}, is sensitive to the stellar mass. 

The duration of the core hydrogen burning phase of the Z$=3\times 10^{-5}$ stars increases from  $\tau_{\rm H} \sim 38$ Myr, for M$=7.5~\rm M_{\odot}$, to $\tau_{\rm H} \sim 5.3$ Gyr, for M$=1 \rm M_{\odot}$. All the stars with initial mass M$ \geq 1.25~\rm M_{\odot}$ develop a convective core, which disappears towards the final stages of core H-burning. Z$=3\times 10^{-7}$ stars evolve slightly faster, with $\tau_{\rm H}$ spanning the 34.5 Myr - 5.1 Gyr range for the same mass range. 

The conditions under which helium is ignited in the core depends on the stellar mass, with M$\leq 1.5~\rm M_{\odot}$ stars experiencing the helium flash, whereas their M$\geq 2~\rm M_{\odot}$ counterparts undergoing quiescent helium burning. The duration of this evolutionary phase, $\tau_{\rm He}$, exhibits a non-linear behaviour with the mass of the star. In the Z$=3\times 10^{-5}$ (Z$=3\times 10^{-7}$) case $\tau_{\rm He}$ first increases, from 72 (64) Myr, for solar mass stars, to 120 (200) Myr, for $2~\rm M_{\odot}$ ($1.5~\rm M_{\odot}$); in the higher mass domain $\tau_{\rm He}$ decreases with the mass of the star, down to 4.1 (3.9) Myr, for $\rm M=7.5~\rm M_{\odot}$.

The evolution of the surface chemistry of the stars can be potentially altered by the two dredge-up episodes, occurring during the RGB ascending (first dredge-up, hereinafter FDU) and after the end of the core helium-burning phase (second dredge-up, SDU), when the surface convection penetrates inwards, reaching internal regions of the star. The modification of the surface chemistry consequently to FDU and SDU are reported in Table \ref{tabdup} for all the stars considered.

The effects of FDU are found in $\rm M \leq 2~\rm M_{\odot}$ stars\footnote{The threshold mass of $2~\rm M_{\odot}$ holds for the Z$=3\times 10^{-5}$ metallicity. In the Z$=3\times 10^{-7}$ case the effects of the first dredge-up are found in $\rm M \leq 1.5~\rm M_{\odot}$ stars.}, with the drop ($\sim 20-30\%$)  in the surface mass fraction of $^{12}$C and the parallel increase in the abundances of $^{13}$C (by a factor $\sim 2$) and $^{14}$N (within a factor $\sim 3$); contamination of the surface regions with CN processed matter is confirmed by the depletion (by a factor $\sim 2$) in the surface $^{15}$N. Traces of mild, full CNO processing are also found, with the enhancement of the surface $^{17}$O (at most by a factor $\sim 5$) and the depletion of the surface $^{18}$O ($\sim 30-50\%$).

Unlike the first dredge-up episode, SDU is efficient in modifying the surface
chemical composition of all the stars, particularly those of higher mass \citep{boothroyd99}. The most noticeable effect of this event is the rise in the surface helium, which increases by $\delta Y \sim 0.02$ in low-mass stars and reaches values as large as $Y=0.36$ in the $\rm M  \geq 6~\rm M_{\odot}$ domain. Regarding the species participating to the CNO nucleosynthesis, it is remarkable the increase in the surface $^{14}$N (up to a factor $\sim 20$ in the $7.5~\rm M_{\odot}$ model star) and the depletion of the surface $^{16}$O, which is left practically unchanged by the FDU. Further species affected by the SDU are sodium, whose mass fraction increases by a factor $\sim 5$ in the highest mass models, and $^{27}$Al, which becomes twice more abundant in the surface regions. The increase in the mass fraction of the latter species is mainly due to matter contaminated by proton captures by the heavy magnesium isotopes $^{25}$Mg and $^{26}$Mg, whose surface abundance is decreased by $\sim 10\%$ as a consequence of SDU in the stars of initial mass above $\sim 4~\rm M_{\odot}$. On the other hand $^{24}$Mg is practically unaffected by the SDU.

The outcome of helium burning in the central regions of the stars is the formation of a CO core, which is supported by the pressure of degenerate electrons. In M$ \geq 7~\rm M_{\odot}$\footnote{The $7~M_{\odot}$ limit holds
for $Z=3\times 10^{-5}$. In the $Z=3\times 10^{-7}$ case off-centre carbon
ignition takes place for $M\geq 6~M_{\odot}$.} stars off-centre ignition of carbon in conditions of partial degeneracy takes place: the main episode of carbon burning is followed by the formation of a convective flame which moves inwards until reaching the centre of the star, which causes the formation of a ONe core \citep{garcia94,siess06,doherty10}. The stars which develop a ONe core experience a deep SDU event, which is commonly referred to as dredge-out. In this case the ignition of a helium burning shell, which follows the exhaustion of helium in the central zone, is accompanied by the formation of a convective region, which moves outwards, until merging with the descending convective envelope \citep{iben97, siess07}. Unlike the classic SDU, in this case the surface chemistry is heavily affected by the results of helium-burning nucleosynthesis, which reflects into a significant increase in the surface abundance of $^{12}$C.

After experiencing the SDU (or dregde-out, in the stars experiencing carbon
ignition),
the stars will move to the next step of their evolutionary history, 
the AGB phase, characterized by a series of thermal pulses, with the shell ignition of helium in conditions of thermal instability \citep{sh65}, and the gradual loss of the external envelope. One of the most relevant quantities for the AGB evolution is the core mass of the star \citep{paczynski}, defined as
the mass of the H- and He-free region, whose initial value is reported in the 8th column of Table \ref{tabgen} for all the model stars
presented in this work. We note the little variation of the (initial) core mass among the stars experiencing the helium flash and the generally increasing trend with the mass of the star, up to value of the order
of $\sim 1.3~\rm M_{\odot}$ for the largest masses investigated.

\begin{figure*}
\begin{minipage}{0.48\textwidth}
\resizebox{1.\hsize}{!}{\includegraphics{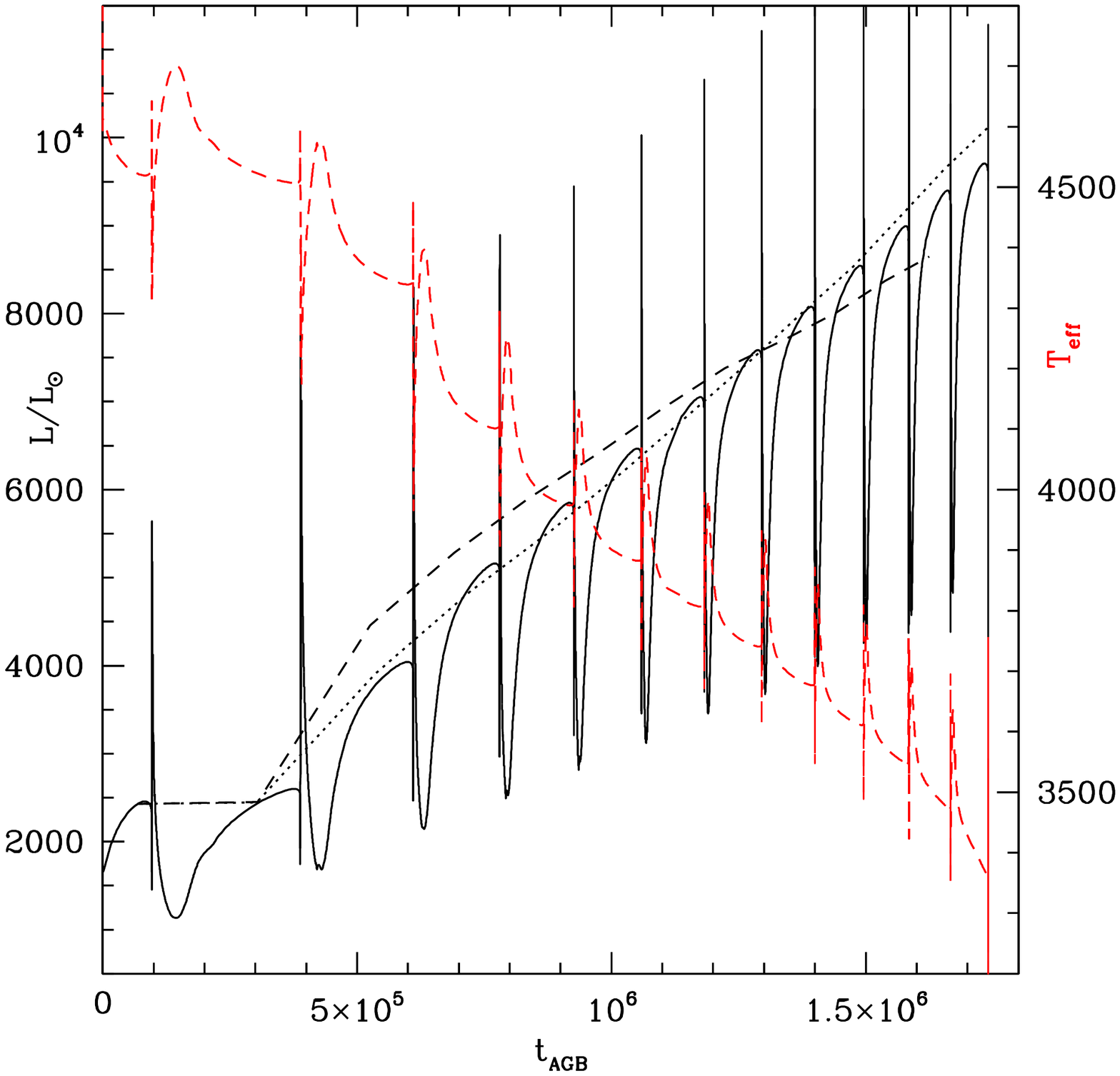}}
\end{minipage}
\begin{minipage}{0.48\textwidth}
\resizebox{1.\hsize}{!}{\includegraphics{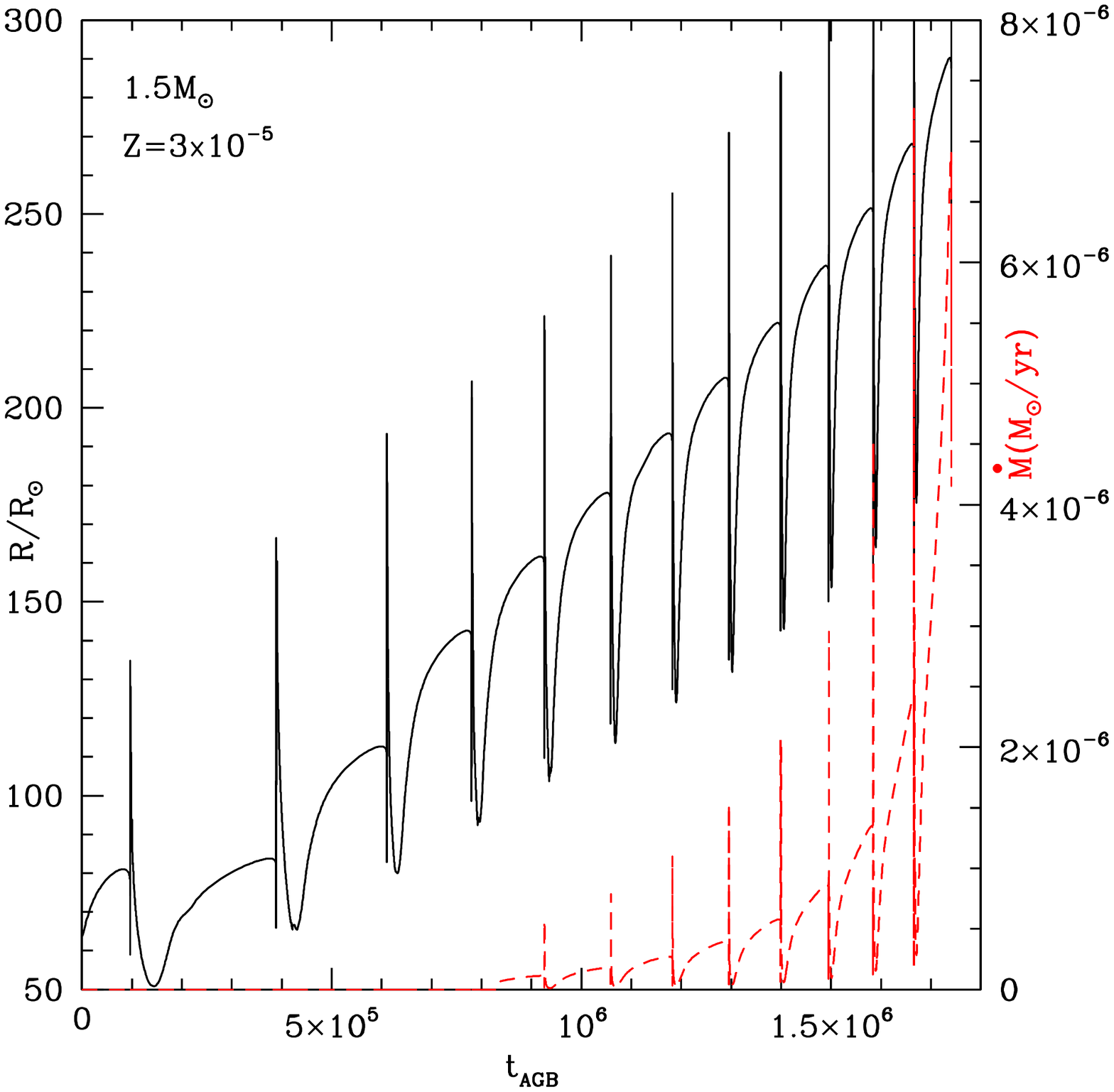}}
\end{minipage}
\vskip-90pt
\begin{minipage}{0.48\textwidth}
\resizebox{1.\hsize}{!}{\includegraphics{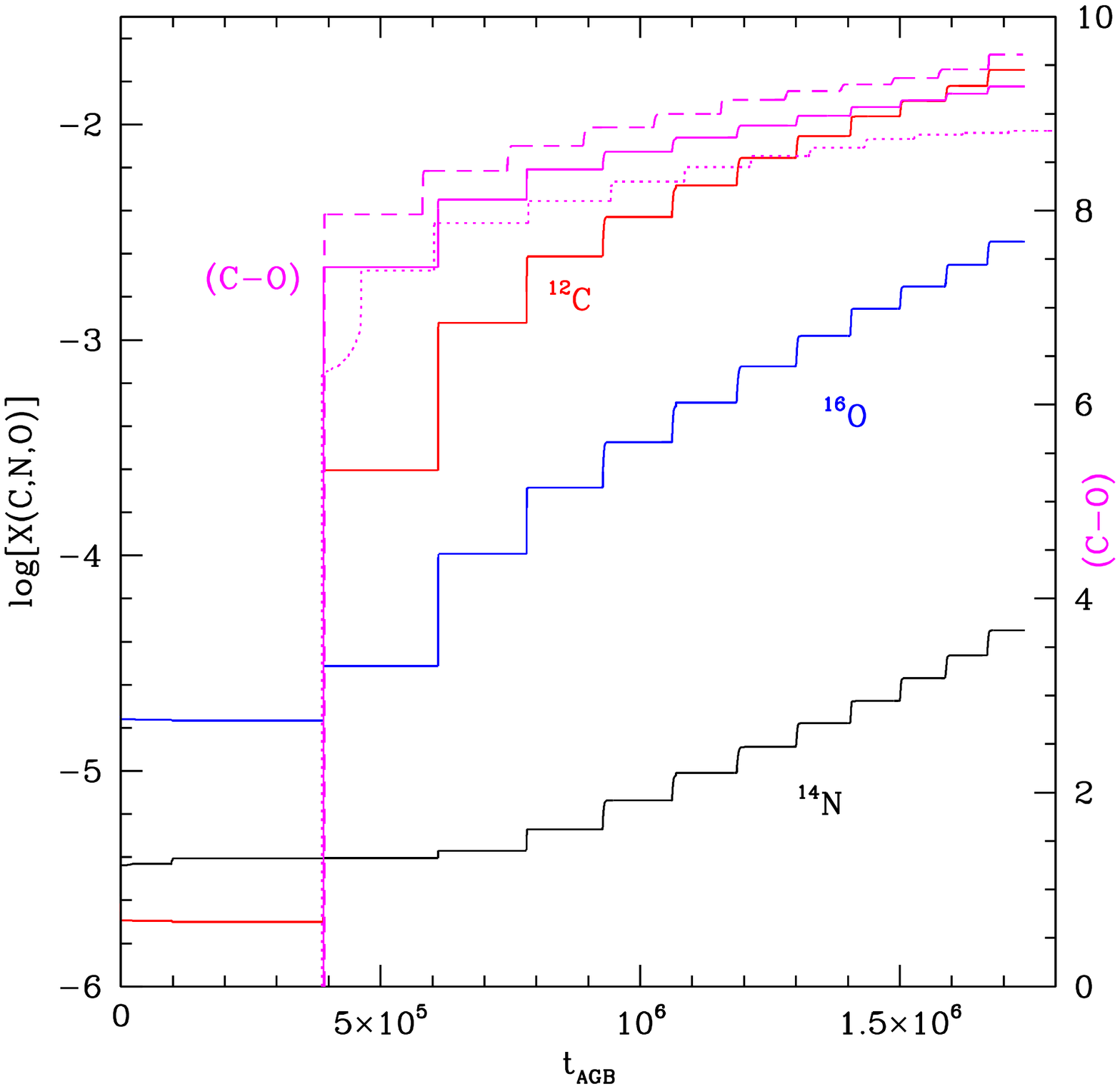}}
\end{minipage}
\begin{minipage}{0.48\textwidth}
\resizebox{1.\hsize}{!}{\includegraphics{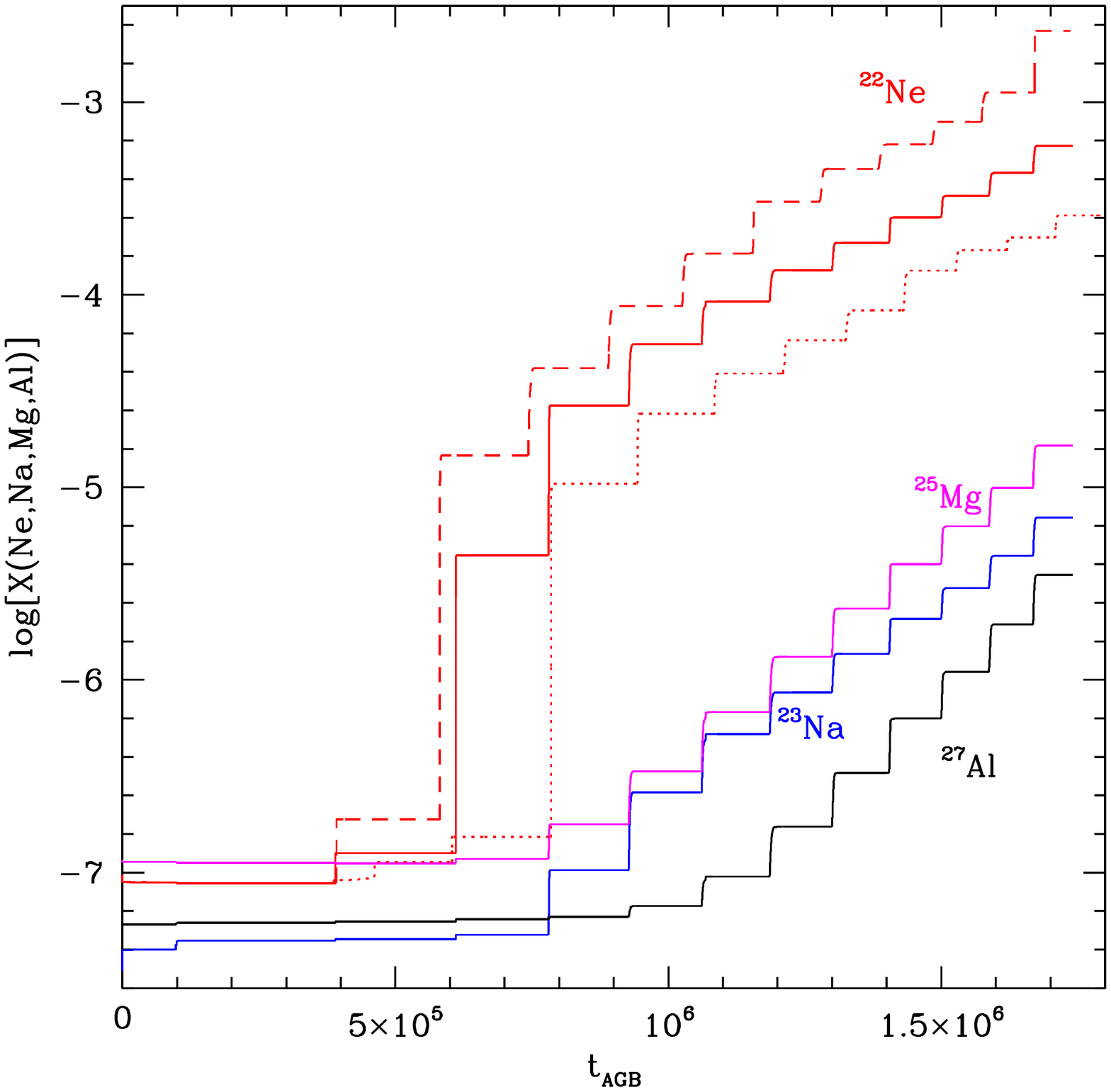}}
\end{minipage}
\vskip-50pt
\caption{The evolution of the $1.5~\rm M_{\odot}$ model star of metallicity 
$Z=3\times 10^{-5}$ as a function of the time, counted from the beginning of the
TP-AGB phase. The top-left panel reports the variation of the luminosity
(black, solid track, scale on the left) and of the effective temperature (red, 
dashed, scale on the right); the evolution of the stellar radius (scale on the left) 
and of the mass-loss rate (scale on the right) are shown in the top, right panel, 
as black, solid and red, dashed lines, respectively; the variation of the CNO surface 
abundances are shown in the bottom, left panels, where the different lines refer 
to the mass fractions of $^{12}$C (red), $^{14}$N (black), $^{16}$O (blue) and 
(C-O) (magenta, dashed, scale on the right; see text for the definition); 
the bottom, right panel reports the variation of the surface mass fractions of 
$^{22}$Ne (red line), $^{23}$Na (blue), $^{25}$Mg (magenta) and $^{27}$Al (black). 
The black, dotted (dashed) line in the top, left panel
connects the luminosity values taken in the middle of each inter-pulse
of a model star of the same initial mass and metallicity, calculated by assuming 
that during the AGB phase the OS is a factor 2 lower (higher) than in the
standard case; the same holds for the magenta dotted and dashed lines in the
bottom, left panel, indicating the evolution of the surface (C$-$O), and the red
lines in the bottom, right panel, which refer to the surface $^{22}$Ne.
} 
\label{f15m}
\end{figure*}

\section{The evolution across the asymptotic giant branch}
\label{agb}
\begin{figure*}
\begin{minipage}{0.48\textwidth}
\resizebox{1.\hsize}{!}{\includegraphics{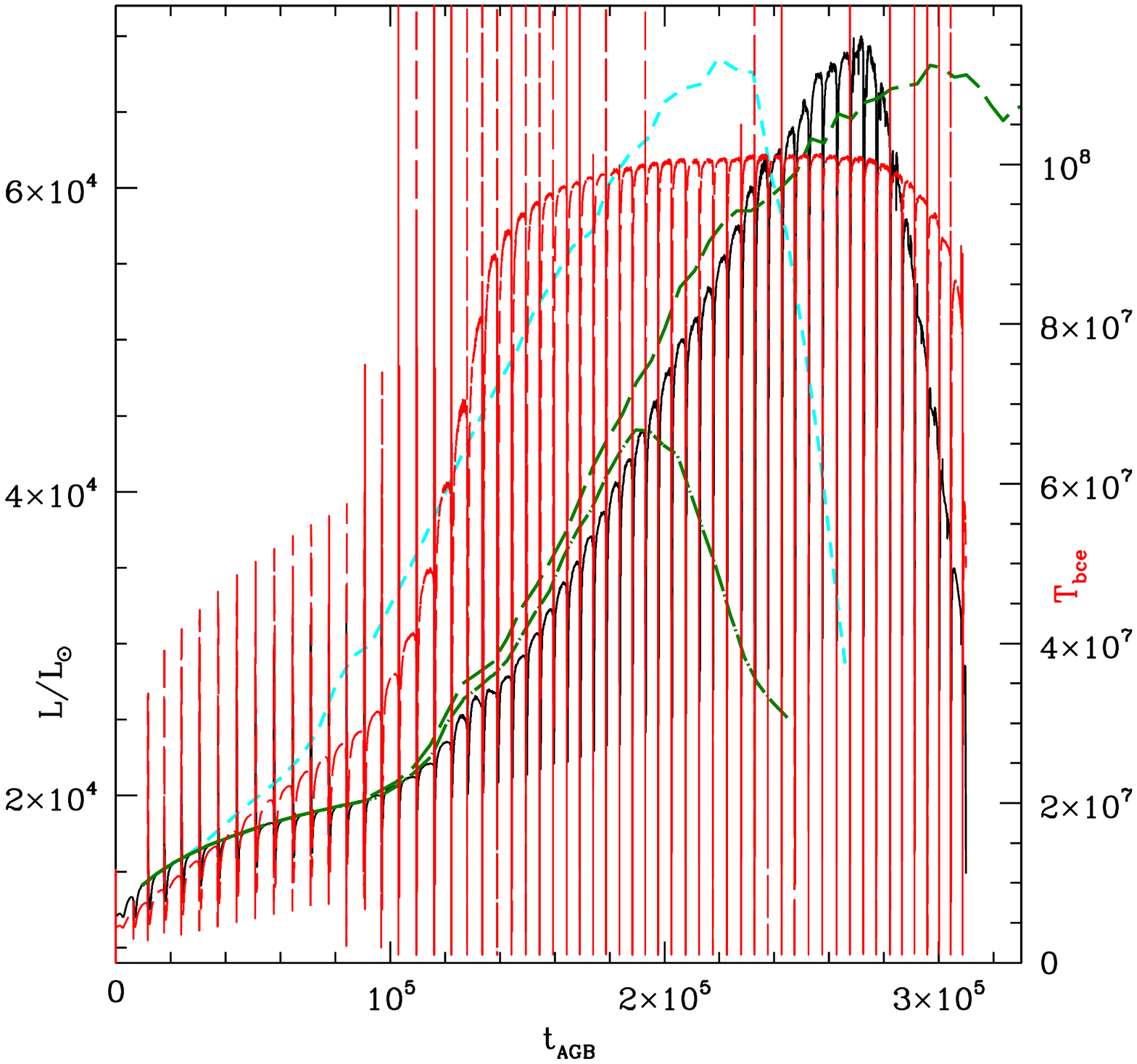}}
\end{minipage}
\begin{minipage}{0.48\textwidth}
\resizebox{1.\hsize}{!}{\includegraphics{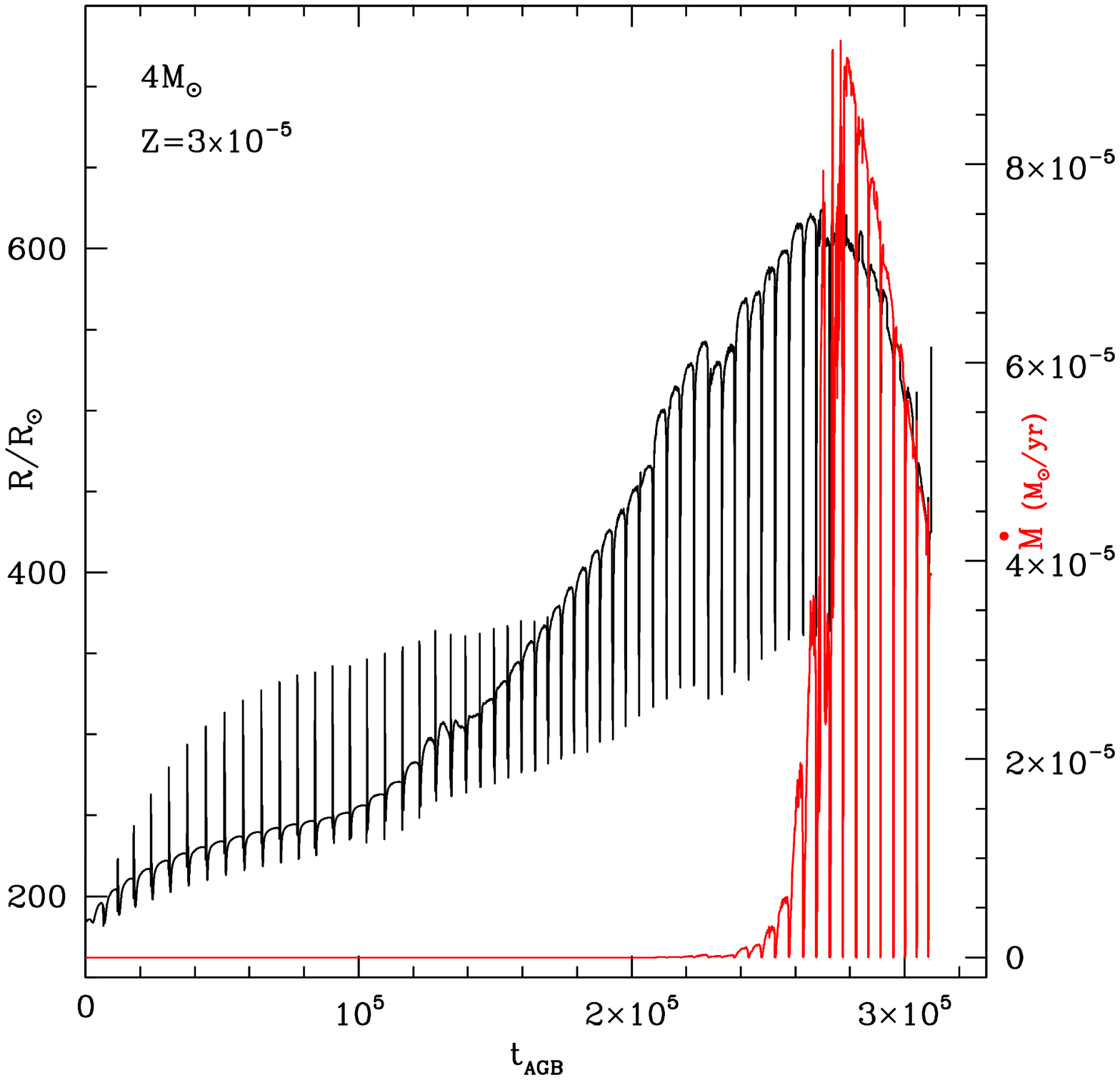}}
\end{minipage}
\vskip-70pt
\begin{minipage}{0.48\textwidth}
\resizebox{1.\hsize}{!}{\includegraphics{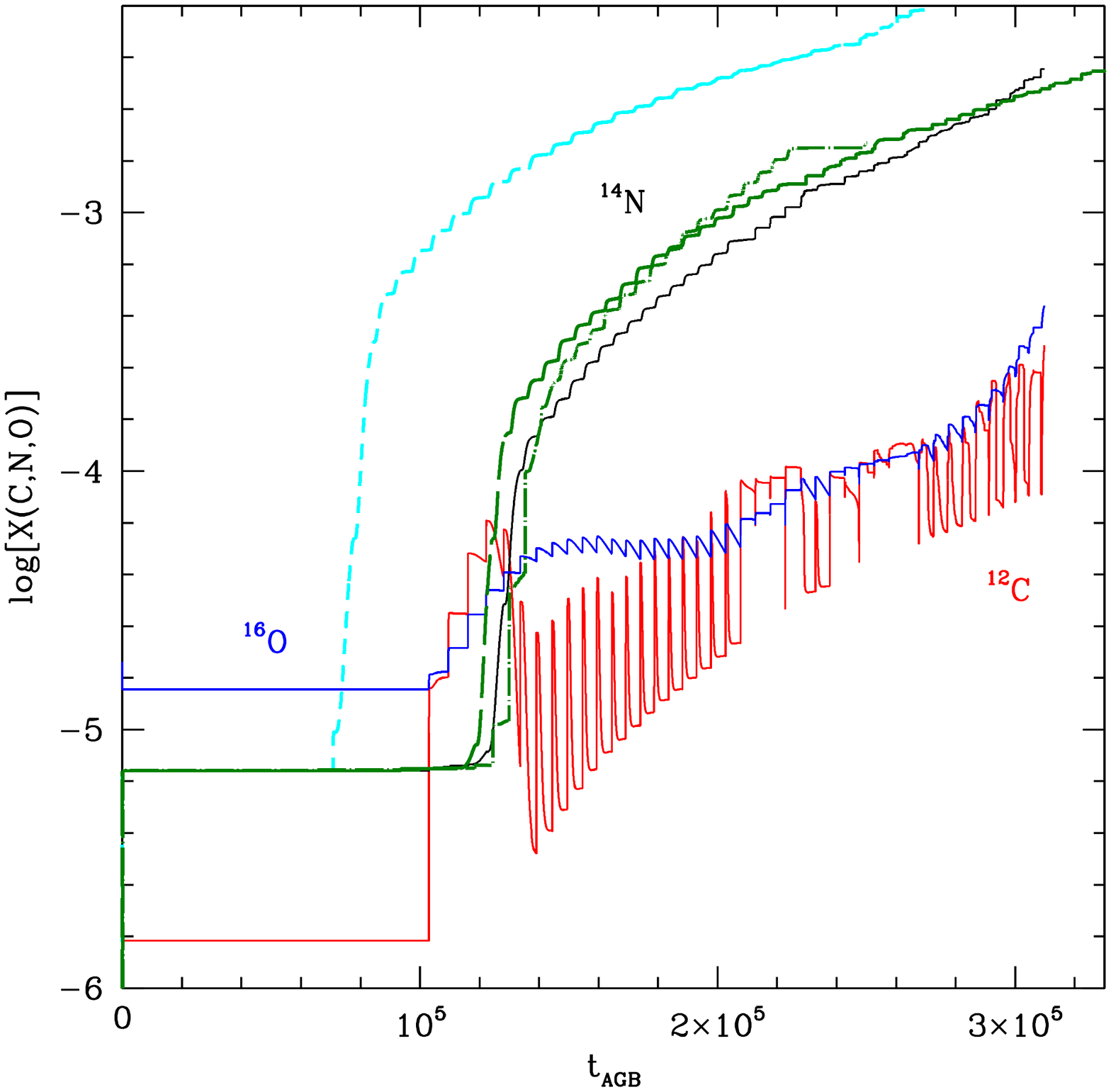}}
\end{minipage}
\begin{minipage}{0.48\textwidth}
\resizebox{1.\hsize}{!}{\includegraphics{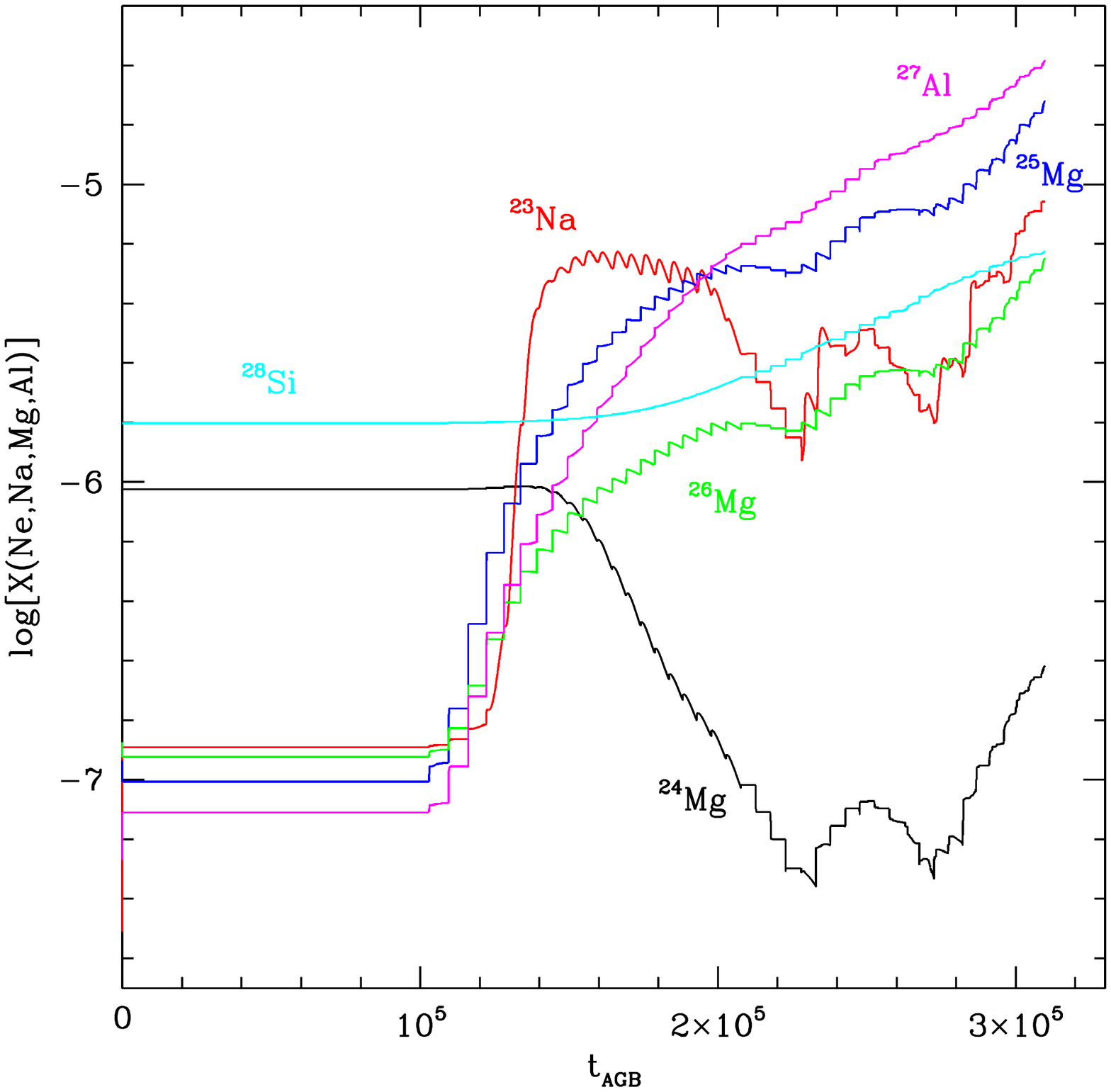}}
\end{minipage}
\vskip-50pt
\caption{The evolution of the $4~\rm M_{\odot}$ model star of metallicity 
$Z=3\times 10^{-5}$ as a function of the time, counted from the beginning of the TP-AGB phase. The top, left panels reports the variation of the luminosity (black, solid track, scale on the left) and of the temperature at the base of the convective envelope (red lines, scale on the right), whereas the top, right panel shows the evolution of the stellar radius (scale on the left) and of the mass-loss rate (red, scale on the right); the variation of the CNO surface abundances are shown in the bottom, left panels, where the different lines refer to the mass fractions of $^{12}$C (red), $^{14}$N (black), $^{16}$O (blue); the bottom, right panel reports the variation of the surface mass fractions of various elements involved in the Ne-Na-Mg-Al-Si nucleosynthesis. The green long-dashed and
dotted-dashed lines in the left panels refer to evolution of the luminosity (top, left panel) and surface $^{14}$N (bottom, left) of a model star calculated by assuming a \citet{blocker95} mass-loss rate, with the free parameter entering the formula taken, respectively, as $\eta_R=0.005$ and $\eta_R=0.02$. The cyan dashed lines in the same panels refer to the luminosity and the surface $^{14}$N of a model star calculated by doubling the OS.} 
\label{f40m}
\end{figure*}

The variation of the physical properties of the stars and the changes in the surface chemical composition taking place during the AGB phase are extremely sensitive to the initial mass of the star \citep[see e.g.][]{karakas14b}. We introduce a gross classification among: a) low-mass stars (initial mass below $2~\rm M_{\odot}$), whose surface abundance is influenced by third dredge-up (TDU) only; b) $2-5~\rm M_{\odot}$ stars, which experience both TDU and hot bottom burning (HBB); c) massive AGBs ($\rm M \geq 5~\rm M_{\odot}$), in which the surface chemical composition reflects the effects of HBB. In the following we will discuss the evolution of the three groups separately. Our analysis will be driven by the results shown
in Fig.~\ref{f15m}, \ref{f40m} and \ref{f60m}, where we report, for
three model stars, taken as representative of each of the groups considered, 
the variation of the most relevant chemical and physical properties.\footnote{The evolutionary sequences for all
the stars investigated in the present work are available at  www.oa-roma.inaf.it/arca/} 

\subsection{Low mass AGB stars}
\label{lowmass}
The evolution of low-mass stars in the metal-free or in the extremely metal-poor domain might
be affected by violent episodes of proton ingestion, during which protons from the external regions
of the star are transported into hot regions, owing to the formation and expansion of
convective zones, triggered by the ignition of helium burning. 

The first of these events occurs during the core He flash \citep{fujimoto00}, when the convective zone formed at the ignition of the flash
expands outwards until reaching H-rich material; as a consequence, hydrogen is transported to regions of high temperature, which triggers a H-flash and the split of the convective shell into two regions.
During the following phases the surface convective regions reaches layers previously exposed to
helium burning, so that the envelope is enriched in carbon and the star becomes a carbon star. As
discussed in section \ref{agbinput} we did not model the helium flash of low-mass stars, rather
we resumed the computations from the HB; however, this is likely not a major issue for the results
presented in the following, as \citet{fujimoto90} and \citet{fujimoto00} showed that the development
of the double flash at the ignition of helium burning in the core occurs only in ultra-metal poor
([Fe$/$H]$\leq -8$), $M<1~M_{\odot}$ stars (see also Fig.~4 in CL08), thus in a range of mass and metallicity not explored in the present work.

Similar episodes, related to violent ignition of hydrogen burning, are found in stellar models of higher mass and metallicity, during the AGB phase. In this case the pulse driven convective zone extends past the He/H discontinuity and transports H-rich material into hot regions of the star, where the violent
ignition of hydrogen favours the occurrence of a H-flash, concomitant with the He-flash. Also in this
case the subsequent inwards penetration of the envelope reaches layers touched by $3\alpha$ 
nucleosynthesis, with the consequent carbon enrichment of the surface. This physical mechanism 
is discussed in detail by CL08, who showed that such a deep mixing is expected to take
place in $M\leq 3~M_{\odot}$ stars of metallicity ([Fe$/$H]$< -3$). During these deep TDU events very high neutron densities are attained, with the efficient activation of s-process nucleosynthesis \citep{cristallo09, fujimoto00, iwamoto04, suda04, choplin21}.

Both the core H-flash taking place at the ignition of the helium flash and the violent ignition of
hydrogen occurring during thermal pulses have the same physical origin, related to the significant
expansion of the convective shell formed as a consequence of thermally unstable ignition of helium,
which reaches regions of the star with some hydrogen, which is transported to hot regions. This is
the reason why CL08 proposed the ‘Dual Core Flash’ terminology for both mechanisms.

Fig.~\ref{f15m} shows the evolution of a model star of initial mass $1.5~\rm M_{\odot}$ and metallicity $Z=3\times 10^{-5}$, in terms of the variation of luminosity, effective temperature, radius, mass-loss rate, surface abundance of the CNO elements and other species involved in the Ne-Na and Mg-Al nucleosynthesis. As discussed previously, the evolution of low-mass AGB stars in the metal-poor domain
explored in the present work is characterized by the ingestion of protons from the envelope in the convective shell formed as a consequence of the first TP, followed by a deep TDU episode, which makes the star to reach the C-star stage. 

The luminosity, shown in the top, left panel of Fig.~\ref{f15m}, increases from $\sim 2000~L_{\odot}$ to $\sim 4500~L_{\odot}$, owing the growth in the core mass, which is the driver of this evolutionary phase. 
The results reported in the bottom-left panel of Fig.~\ref{f15m} show that $^{16}$O is also dredged-up to the surface after each TDU episode, though at lower extent than $^{12}$C, and some $^{14}$N, synthesized via proton capture in the upper regions of the pulse driven convective shell \citep{straniero95}.

Further effects of TDU are the gradual increase in the surface mass fraction 
of other species involved in the $\alpha-$capture nucleosynthesis activated at
each TP, primarily $^{22}$Ne, $^{25}$Mg and $^{26}$Mg, and of those synthesized by proton captures, mainly $^{23}$Na and $^{27}$Al (see bottom, right panel of
Fig.~\ref{f15m}).

The progressive increase in the overall surface metallicity, caused by the 
repeated TDU events, favours the general expansion of the external regions
of the star, with the stellar radius increasing, until reaching
size of the order of $300~R_{\odot}$, as shown in the top, right panel of
Fig.~\ref{f15m}; this is accompanied by a gradual cooling of the surface
layers, with the effective temperature decreasing from $\sim 4700$ K, at the
first TP, to $\sim 3400$ K, during the final AGB phases; this can be seen in
the top, left panel of Fig.~\ref{f15m}.

The expansion of the star and the increase in the luminosity determine a
significant increase in the mass-loss rate, which is almost null during the
first TP and increases until reaching almost $10^{-5}~\rm M_{\odot}/$yr at the
very end of the AGB evolution (see top, right panel of Fig.~\ref{f15m}).
Unlike higher metallicity, low-mass stars, which experience significant mass
loss since the first TP \citep{karakas10, ventura16}, in the low-metallicity models studied here the mass-loss is initially negligible, and becomes relevant only during the last 2-3 inter-pulse periods. This is due to the small surface opacity of these stars, which makes the stars to evolve at smaller radii than the higher metallicity counterparts, thus exposed to smaller mass-loss rates. Only after a significant increase in the overall surface metallicity (primarily $^{12}$C) takes place, mass-loss assumes significant values.

An important consequence of this behaviour is that the average chemistry of the gas ejected by 
low-mass stars into the interstellar medium reflects the final surface 
chemical composition, because almost the totality of the envelope is lost during the 
last two inter-pulse phases. The final surface C$/$O spans the 3-14 (6-24) range in 
low-mass AGBs of metallicity $Z=3\times 10^{-5}$ ($Z=3\times 10^{-7}$) and is generally 
higher the lower the mass of the star. This apparently anomalous behaviour is motivated
be the fact that stars of higher mass experience deeper TDU episodes, with the result
that a significant increase in the surface $^{16}$O takes place. As far as the surface 
$^{12}$C enrichment is concerned, the trend with mass is more straightforward, with 
$\rm M =1~\rm M_{\odot}$ stars ending up with surface mass fractions of the order of 0.005, 
$2~\rm M_{\odot}$ stars reaching $X(^{12}C) \sim 0.02$, and $1.25-1.5~\rm M_{\odot}$
stars exhibiting an intermediate behaviour.

The overall duration of the TP-AGB phase for this class of stars
(see Table \ref{tabgen}) is of the order of 1 Myr. There is not a clear trend between the
initial mass of the star and the AGB time scales. This is because the effect of the larger luminosities attained by higher mass stars, which would make the evolutionary
time scales shorter, is compensated by the higher mass of the envelope,
which requires a higher number of TP before the external mantle is entirely lost.

\subsection{Intermediate mass AGBs}
\label{interm}
In this group we include the stars which experience both TDU and HBB, thus
the variation of the surface chemical composition is determined by the balance
between these two mechanisms. The (initial) mass range spanned by these stars
is $2~\rm M_{\odot} < M < 5~\rm M_{\odot}$\footnote{This results holds for the
$Z=3\times 10^{-5}$ model stars; in the $Z=3\times 10^{-7}$ case it is
$2~\rm M_{\odot} < M < 4~\rm M_{\odot}$}. The lower limit is related to the fact that
a minimum core mass, of the order of $0.8~\rm M_{\odot}$, is required to activate
HBB \citep{ventura13}; on the other hand stars with initial mass higher than 
the upper limit given above, which will be discussed in the following
sub-section, experience negligible TDU, thus their surface chemistry is
almost entirely influenced by HBB. 

Fig.~\ref{f40m} reports the evolution of a $4~\rm M_{\odot}$ model star of
metallicity $Z=3\times 10^{-5}$, with the top panels showing the most relevant physical quantities, i.e. luminosity, temperature at the base of the envelope, stellar radius and mass-loss rate, whereas in the bottom panels we show the change in the surface chemical composition.

In the top, left panel of Fig.~\ref{f40m} we can see the typical imprinting
of HBB, which favours a fast increase in the luminosity \citep{blocker91}, which in this 
particular case rises up to $\sim 7\times 10^4~L_{\odot}$ and in the temperature at the base of the envelope, which rapidly grows to $\sim 10^8$ K. The luminosity of the star, in analogy with higher metallicity models undergoing HBB, decreases during the final AGB phases, because the gradual loss of the envelope weakens the strength of HBB, until turning it off completely \citep[see e.g.][]{ventura13}. This behaviour causes an overall
contraction of the external regions, whose extension decreases down to $\sim 400~R_{\odot}$, after reaching a peak value of $\sim 600~R_{\odot}$,
and a decrease in the mass-loss rate towards the end of the AGB phase.

From the variation of the surface $^{12}$C, shown in the bottom, left
panel of Fig.~\ref{f40m}, we deduce that TDU begins first, after 10 TP,
whereas the effects of HBB appear 4-5 TP later. Initially the
surface of the stars becomes more and more enriched in carbon, whereas during
later phases the $^{12}$C dredged-up after each TDU is later converted into 
$^{14}$N by HBB. $^{16}$O is only marginally affected by HBB in this specific
case, thus the effects of TDU are more relevant for this species, and the surface abundance of $^{16}$O gradually increases, eventually reaching values $\sim 20$ times higher than the initial mass fraction. These stars never
become carbon stars and the final C$/$O ratios are below unity (see Table \ref{tabgen}). As discussed e.g. in Iwamoto et al. (2009, hereinafter I09), the most relevant outcome of the TDU+HBB nucleosynthesis, visibile in the bottom, left panel of Fig.~\ref{f40m}, is the remarkable increase in the surface $^{14}$N, with the final value being three orders of magnitude higher than the initial quantity. Note that the freshly synthesized nitrogen is of primary origin, because as stated previously it is mostly produced by proton capture reactions by $^{12}$C nuclei, produced in the $3\alpha$-burning shell.

The HBB temperatures reached are sufficiently large to the
activation of the Mg-Al nucleosynthesis, as confirmed by the gradual depletion of the surface $^{24}$Mg, shown in the bottom, right panel of Fig.~\ref{f40m}. This nuclear activity reflects into a significant increase in the surface mass fraction of the two heavier magnesium isotopes and in the
synthesis of $^{27}$Al and $^{28}$Si. Production of $^{23}$Na via proton capture by the $^{22}$Ne nuclei originally present in the surface regions of the star and by those transported to the surface by TDU, also takes place (see bottom, right panel of Fig.~\ref{f40m}). Some $^{28}$Si production
occurs during the AGB lifetime, with an overall increase of the
order of a factor $4$.

To understand the chemistry of the gas ejected by this class of stars
during the AGB lifetime, it is crucial to combine the results shown in
the bottom panels of Fig.~\ref{f40m}, reporting the time variation of the
surface chemical composition, with the evolution of the mass-loss rate,
reported in the top, right panel of the same figure. $\dot M$ is
sensitive to the period of the star, thus on the stellar radius. 
The latter quantity is heavily affected by the increase in the surface metallicity and the parallel rise in the luminosity, which make the star to assume a more and more expanded configuration: the stellar radius increases steadily by a factor $\sim 3$ from the beginning of the AGB phase until the luminosity peak is reached. As shown in the top, right panel of Fig.~\ref{f40m}, it is during the phases slightly before the luminosity peak that the mass of the envelope begins to be lost in a significant way, a process which continues until the consumption of the whole external mantle, with mass-loss rates in the $4\times 10^{-5} - 10^{-4}~\rm M_{\odot}/$yr range: all the envelope is lost during the final $\sim 10$ inter-pulse phases, after the surface chemistry was severely modified by the combined effects of TDU and HBB.

\subsection{Massive AGB stars}
\label{massivem}
We discuss the most massive among the stars that experience the AGB phase,
whose surface chemistry reflects the effects of HBB, with no contribution from TDU.
We refer to the stars of initial mass above $\sim 4~\rm M_{\odot}$, including the objects which develop a core composed by oxygen and neon,
discussed in section \ref{preagb}.

\begin{figure*}
\begin{minipage}{0.48\textwidth}
\resizebox{1.\hsize}{!}{\includegraphics{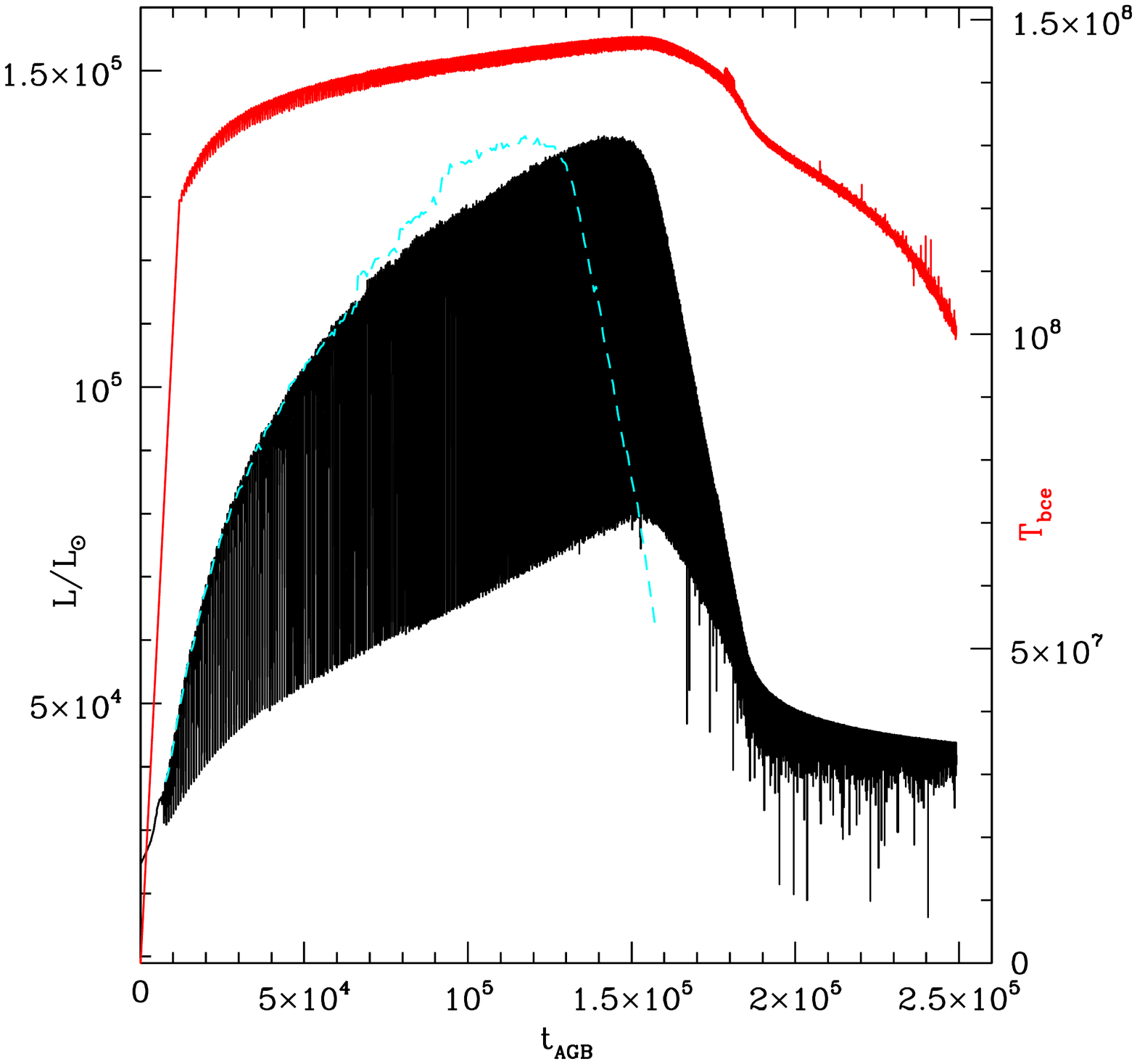}}
\end{minipage}
\begin{minipage}{0.48\textwidth}
\resizebox{1.\hsize}{!}{\includegraphics{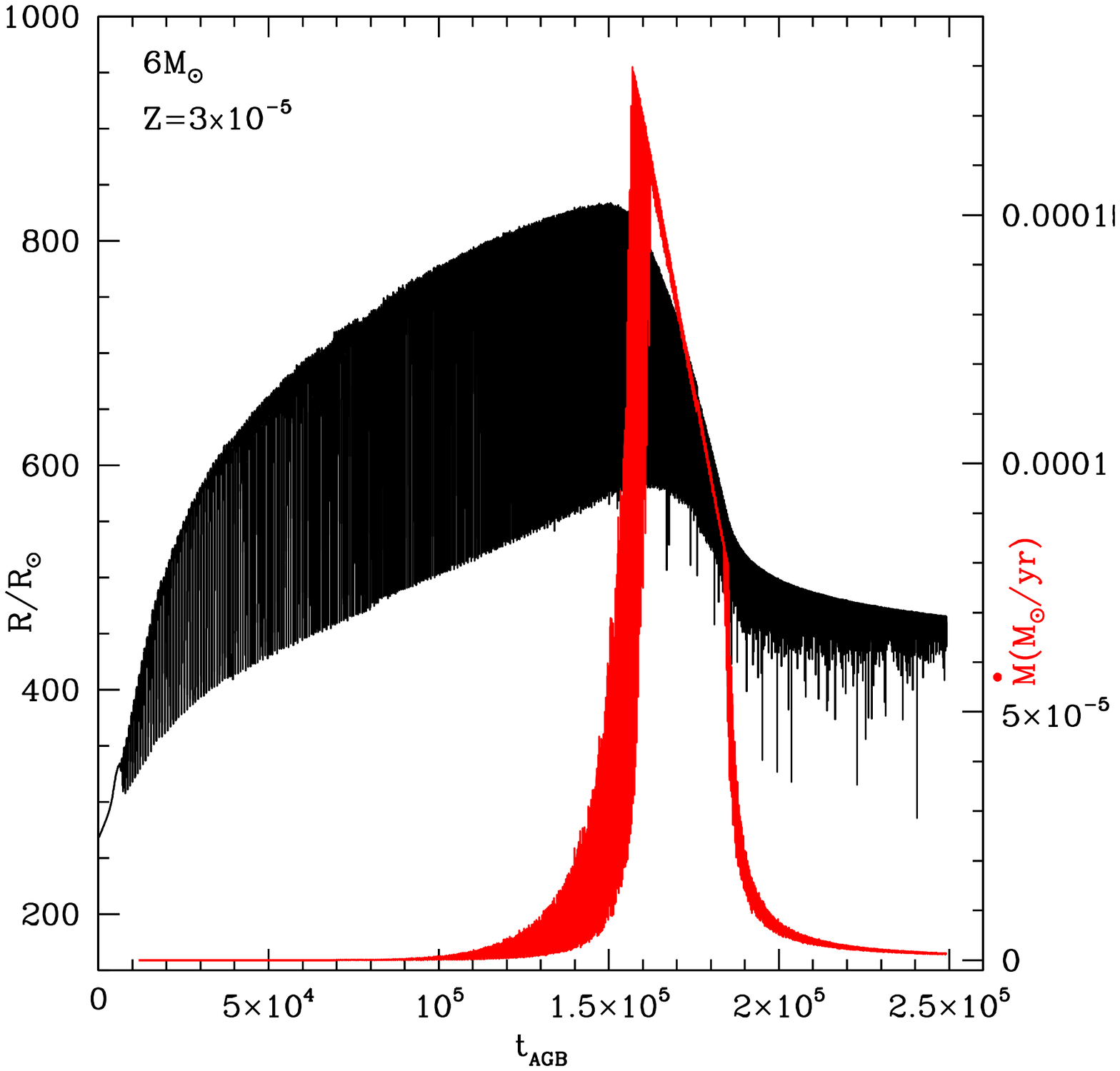}}
\end{minipage}
\vskip-70pt
\begin{minipage}{0.48\textwidth}
\resizebox{1.\hsize}{!}{\includegraphics{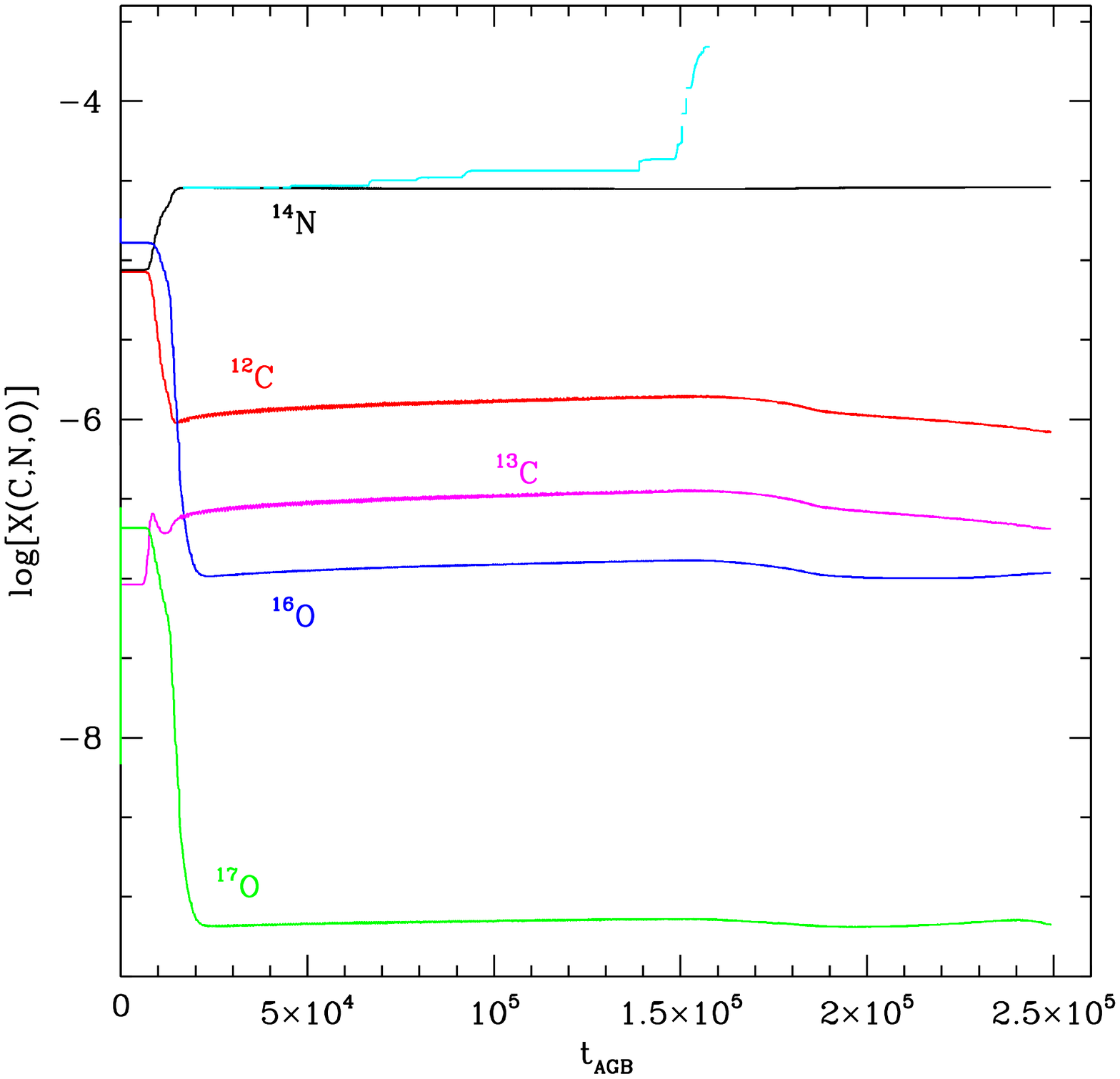}}
\end{minipage}
\begin{minipage}{0.48\textwidth}
\resizebox{1.\hsize}{!}{\includegraphics{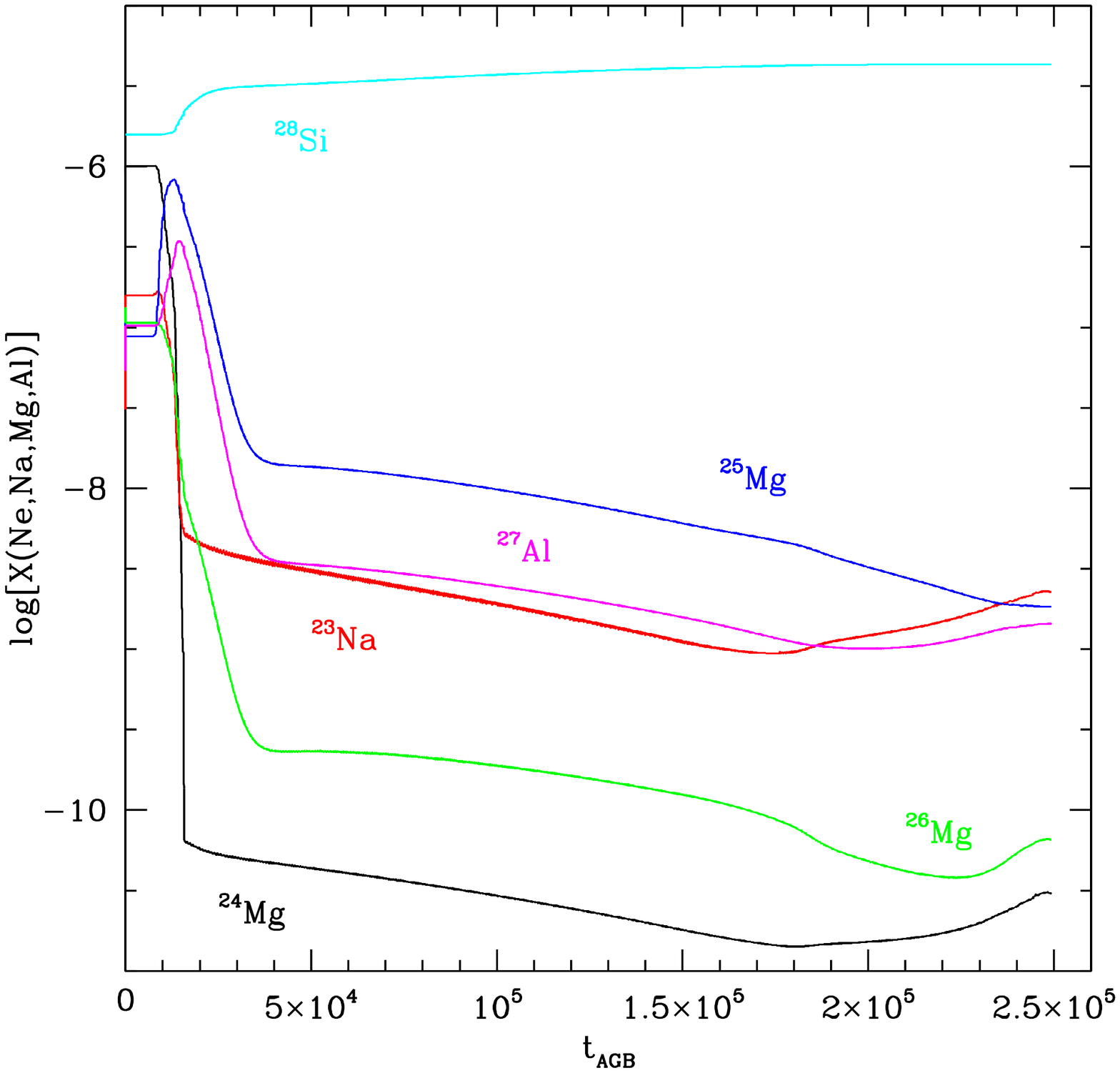}}
\end{minipage}
\vskip-50pt
\caption{Evolution of the $6~\rm M_{\odot}$ model star of metallicity 
$Z=3\times 10^{-5}$ as a function of the time, counted from the beginning of the
TP-AGB phase. The top panels reports the same quantities as in the corresponding
panels of Fig.~\ref{f40m}, with the difference that the lines corresponding to
the temperature at the bottom of the convective envelope and of the mass-loss
rate refer to the inter-pulse phases, for readability. The bottom panels,
showing the variation of the surface chemistry, use the same color-coding as
in Fig.~\ref{f40m}. The cyan dashed lines in the left panels refer to the luminosity (top) and the surface $^{14}$N of a model star calculated by doubling the OS.}
\label{f60m}
\end{figure*}

Similarly to the intermediate mass stars discussed earlier in this section,
the evolution of massive AGBs is characterized by the significant increase in the overall luminosity, which starts with the ignition of HBB, and in the
temperature at the base of the convective envelope. In the case reported in Fig.~\ref{f60m}, a $6~\rm M_{\odot}$ model star with $Z=3\times 10^{-5}$, the luminosity reaches $L \sim 1.5\times 10^5~L_{\odot}$, whereas the temperature at the bottom of the envelope rises up to almost $150$ MK. Also similar to the lower mass counterparts is the general behaviour of luminosity, which decreases after reaching a peak value, owing to the gradual loss of the stellar envelope.

The most important difference with respect to the intermediate mass model stars is that the overall surface metallicity is approximately constant and the relative distribution among the different species reflects the equilibria of the proton-capture nucleosynthesis experienced at the bottom of the convective envelope. This is of considerable importance for the evolution of the star, because the low metallicity prevents a significant expansion of the external layers, keeping the radii of these stars comparable to those of the intermediate mass counterparts presented in the previous section, despite the higher luminosities (this can be seen in the comparison between the radius variations reported in the top, right panels of Fig.~\ref{f40m} and \ref{f60m}). The behaviour of the radius is reflected into the mass-loss
rates experienced, which is extremely low until the peak luminosity is reached, with the consequence that the stars experience a large number of TP before the envelope is entirely lost. The time-scales of the TP-AGB phase of massive AGBs are consequently longer than those of slightly lower mass stars
experiencing TDU, which is the reason for the local maximum in the values reported in the 5th column of Table \ref{tabgen}, found at $5~\rm M_{\odot}$ ($4~\rm M_{\odot}$) for $Z=3\times 10^{-5}$ ($Z=3\times 10^{-7}$).

On the chemical side, the composition of the gas ejected by massive AGBs (see bottom panels of Fig.~\ref{f60m}) reflects the surface chemistry of the star during the final phases, characterized by extremely small quantities of $^{16}$O and $^{24}$Mg, exposed to strong depletion by proton capture reactions at the base of the outer convective region: the final $^{16}$O and $^{24}$Mg are 2 and 4 orders of magnitude smaller than the initial abundances, respectively. $^{12}$C is also extensively destroyed by HBB, though the depletion factor, $\sim 10$, is lower with respect to $^{16}$O, which is the typical situation occurring in presence of a very hot CNO cycling. The idea that massive AGBs produce oxygen- and magnesium-poor ejecta was earlier explored by \citet{ventura13}, who studied AGB models of metallicity $Z=3\times 10^{-4}$, and by \citet{siess10}, who found strong signatures of $^{16}$O and $^{24}$Mg depletion in the ejecta of $\sim 8~\rm M_{\odot}$ stars of metallicity $Z=10^{-4}$. In this case the situation is more extreme, because the metallicities investigated are smaller, thus the HBB temperatures hotter.

These stars will evolve as carbon stars for almost the entire AGB phase, since the activation of HBB. Such a nuclear activity at the bottom of the envelope is reflected into an enhancement of the surface $^{14}$N, although the relative increase is significantly smaller than in the intermediate mass stars, 
because in this case, with no contribution from TDU, the $^{12}$C and $^{16}$O nuclei used to synthesize $^{14}$N are entirely of secondary origin.

Turning to the other species, the temperatures of the internal regions of the convective envelope are so hot to favour a very advanced nucleosynthesis, such that the mass fractions of the heavier magnesium isotopes, sodium and aluminium are smaller than the initial quantities, whereas some $^{28}$Si production takes place.

\subsection{An overall view of the AGB evolution and the change in the
surface chemistry}
The results discussed so far, presented in Fig.~\ref{f15m}, \ref{f40m} and
\ref{f60m}, outlined that the evolution of metal-poor AGBs is driven by
various factors and is sensitive to the delicate interplay between the
various mechanisms able to alter the surface chemistry. Indeed the onset of
fast mass-loss is driven by the metals enrichment of the surface regions,
which leads to a rapid loss of the envelope which ends the AGB phase \citep{marigo02, constantino14}.
In the following we discuss the role played by the mass of the star and
the metallicity on the main aspects of the AGB phase and on the chemical
composition of the gas ejected into the interstellar medium.

\begin{figure*}
\begin{minipage}{0.48\textwidth}
\resizebox{1.\hsize}{!}{\includegraphics{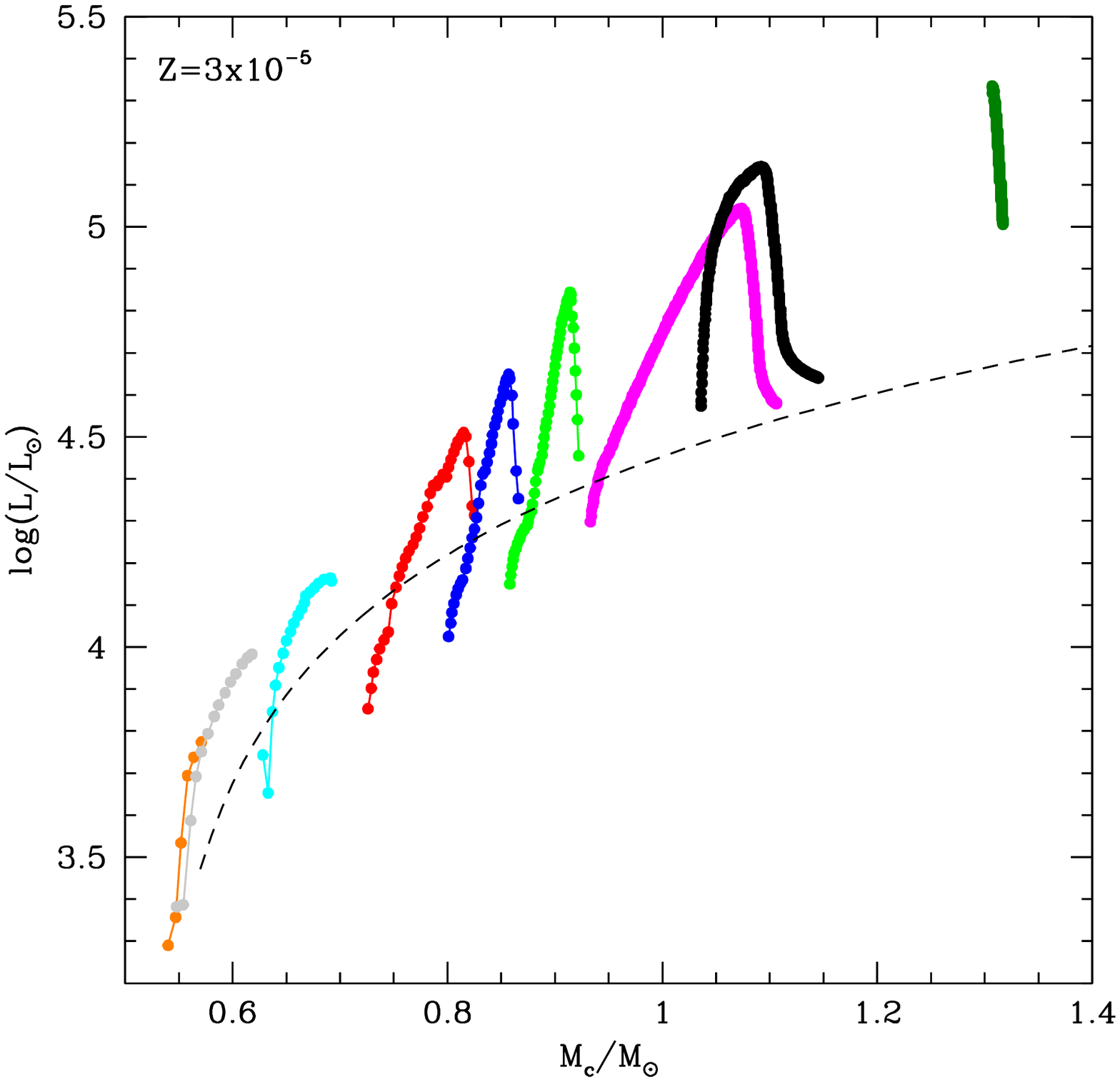}}
\end{minipage}
\begin{minipage}{0.48\textwidth}
\resizebox{1.\hsize}{!}{\includegraphics{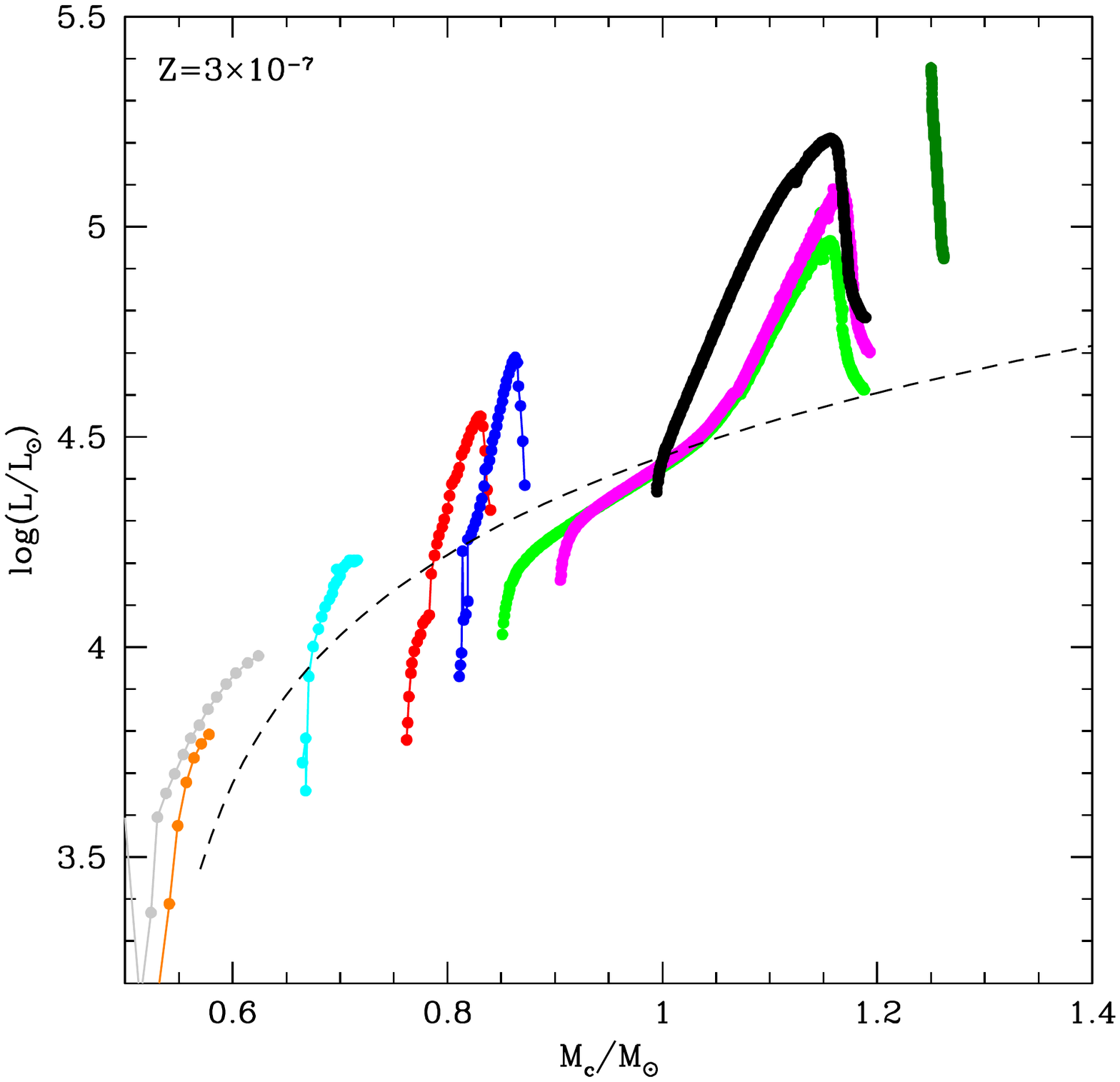}}
\end{minipage}
\vskip-70pt
\begin{minipage}{0.48\textwidth}
\resizebox{1.\hsize}{!}{\includegraphics{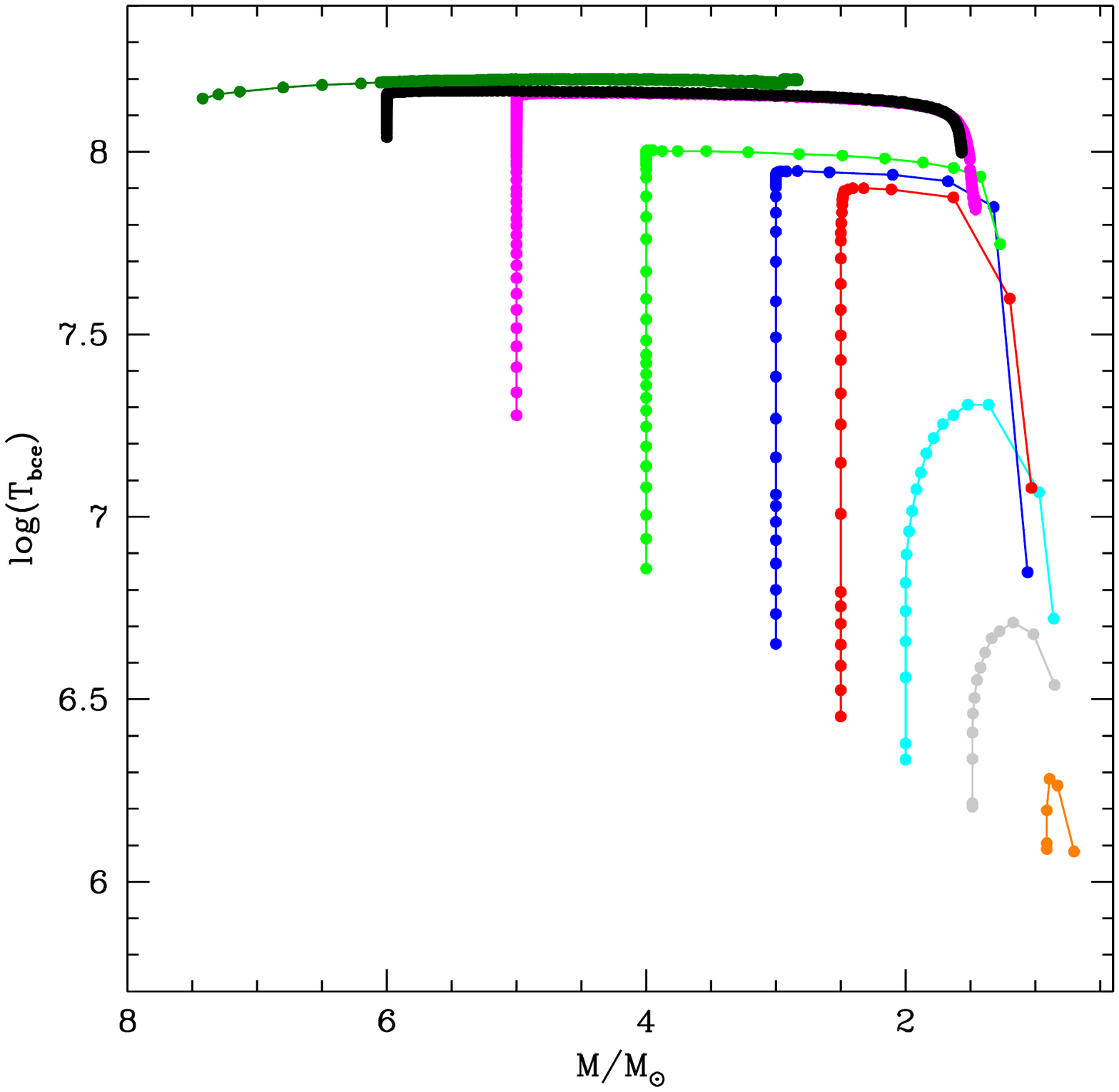}}
\end{minipage}
\begin{minipage}{0.48\textwidth}
\resizebox{1.\hsize}{!}{\includegraphics{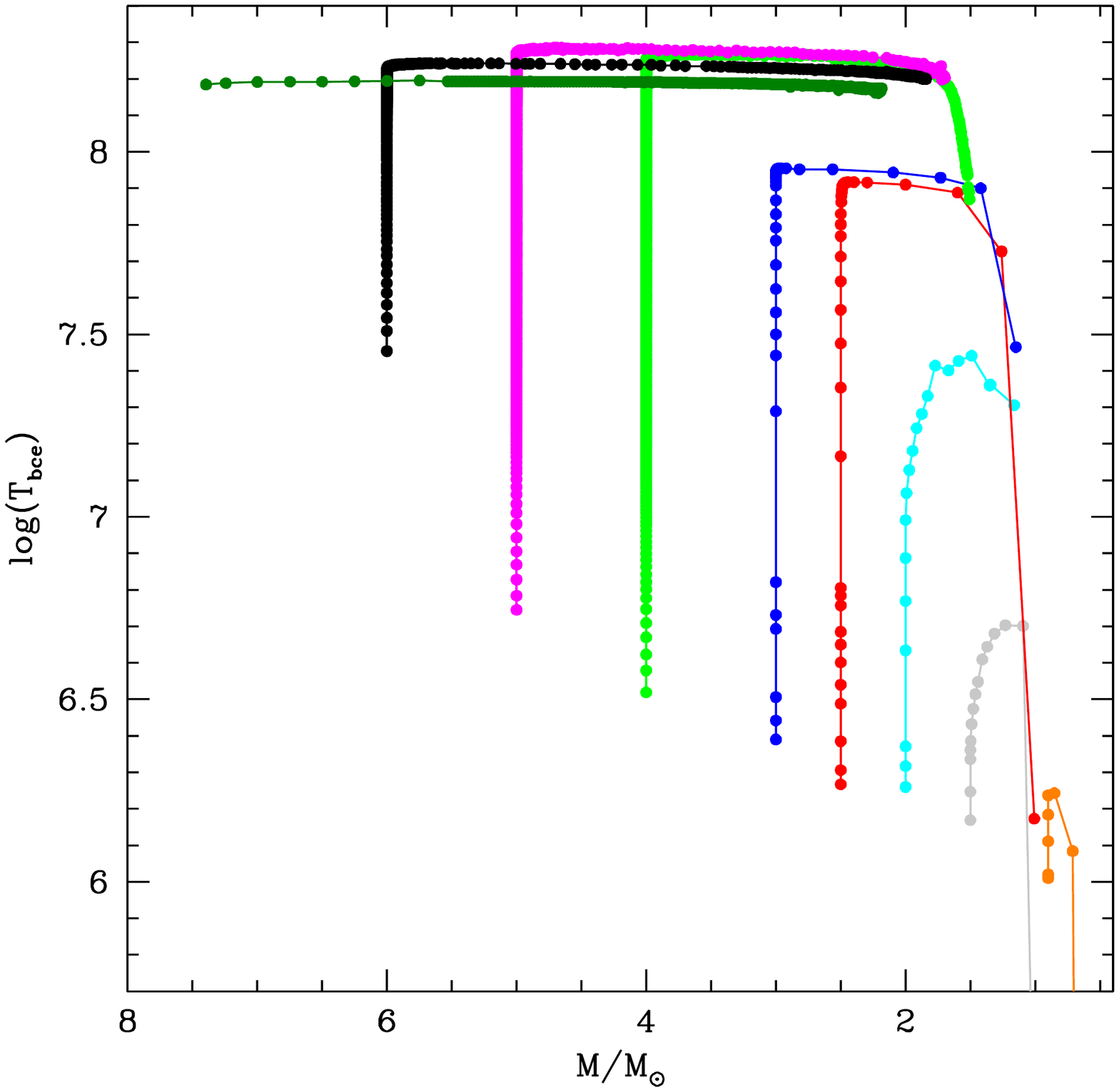}}
\end{minipage}
\vskip-50pt
\caption{Top panels show the variation of the luminosity of the model stars of metallicity $Z=3\times 10^{-5}$ (left) and $Z=3\times 10^{-7}$ (right) as a funnction of the core mass. The different lines refer to the stars of intial mass $1~\rm M_{\odot}$ (orange), $1.5~\rm M_{\odot}$ (grey), $2~\rm M_{\odot}$ (cyan) $2.5~\rm M_{\odot}$ (red), $3~\rm M_{\odot}$ (blue), $4~\rm M_{\odot}$ (light green), $5~\rm M_{\odot}$ (magenta), $6~\rm M_{\odot}$ (black), $7.5~\rm M_{\odot}$ (dark green). The dashed lines indicate the core mass - luminosity relationship by \citet{paczynski}. The bottom panels report the evolution of the temperature at the bottom of the convective envelope of the same model stars shown in the top panels, as a function of the current mass of the star.} 
\label{ftbce}
\end{figure*}

Fig.~\ref{ftbce} shows the variation of the luminosity and of the temperature
at the bottom of the convective envelope during the AGB phase of stars of different initial mass and metallicity $Z=3\times 10^{-5}$ (left panels) and $Z=3\times 10^{-7}$ (right). The luminosity (reported in the top panels) and $T_{\rm bce}$ (bottom panels) are shown as a function of the core mass and of the current mass of the star, respectively.

The luminosities span the $2\times 10^3 - 2\times 10^5~L_{\odot}$ range, the
peak luminosities being larger the higher the initial mass of the star.
This reflects the well known result of AGB evolution modelling, that more massive
stars evolve at larger luminosities, because of the higher core mass.
%We see in Fig.~\ref{ftbce} significant deviations from the \citet{paczynski}'s
%law (indicated with dashed lines in the figure), particularly in the models which %experience HBB; this is consistent with previous investigations on this argument %and with the pioneering exploration by \citet{blocker91}. 

The results shown in the bottom panels of Fig.~\ref{ftbce} indicate that
the temperature at the base of the surface convective regions, similarly
to the luminosity, is extremely sensitive to the initial mass of the star.
For both metallicities it is evident the gap between $\rm M >2~\rm M_{\odot}$
stars, which experience HBB, reaching temperatures above 50 MK, and the
stars of lower mass, which evolve at $T_{\rm bce}$ below 30 MK. 

The vertical trends visible in the first part of each track, particularly in the model stars exposed to HBB, correspond to the series of thermal pulses experienced during the initial and middle part of the AGB evolution, when the core mass and $T_{\rm bce}$ increase, while the mass keeps practically constant, owing to the negligible mass-loss rates experienced. In the $Z=3\times 10^{-7}$ case we note that the $T_{\rm bce}$ vs initial mass trend presents a turning point around $5~\rm M_{\odot}$, with the stars of higher mass evolving at cooler $T_{\rm bce}$. This is because the luminosities of the most
massive stars are so large that strong mass-loss occurs since the early AGB phases, which prevents significant growth in the core mass and inhibits the convective envelope from reaching very large temperatures in the innermost layers.

\begin{figure*}
\begin{minipage}{0.48\textwidth}
\resizebox{1.\hsize}{!}{\includegraphics{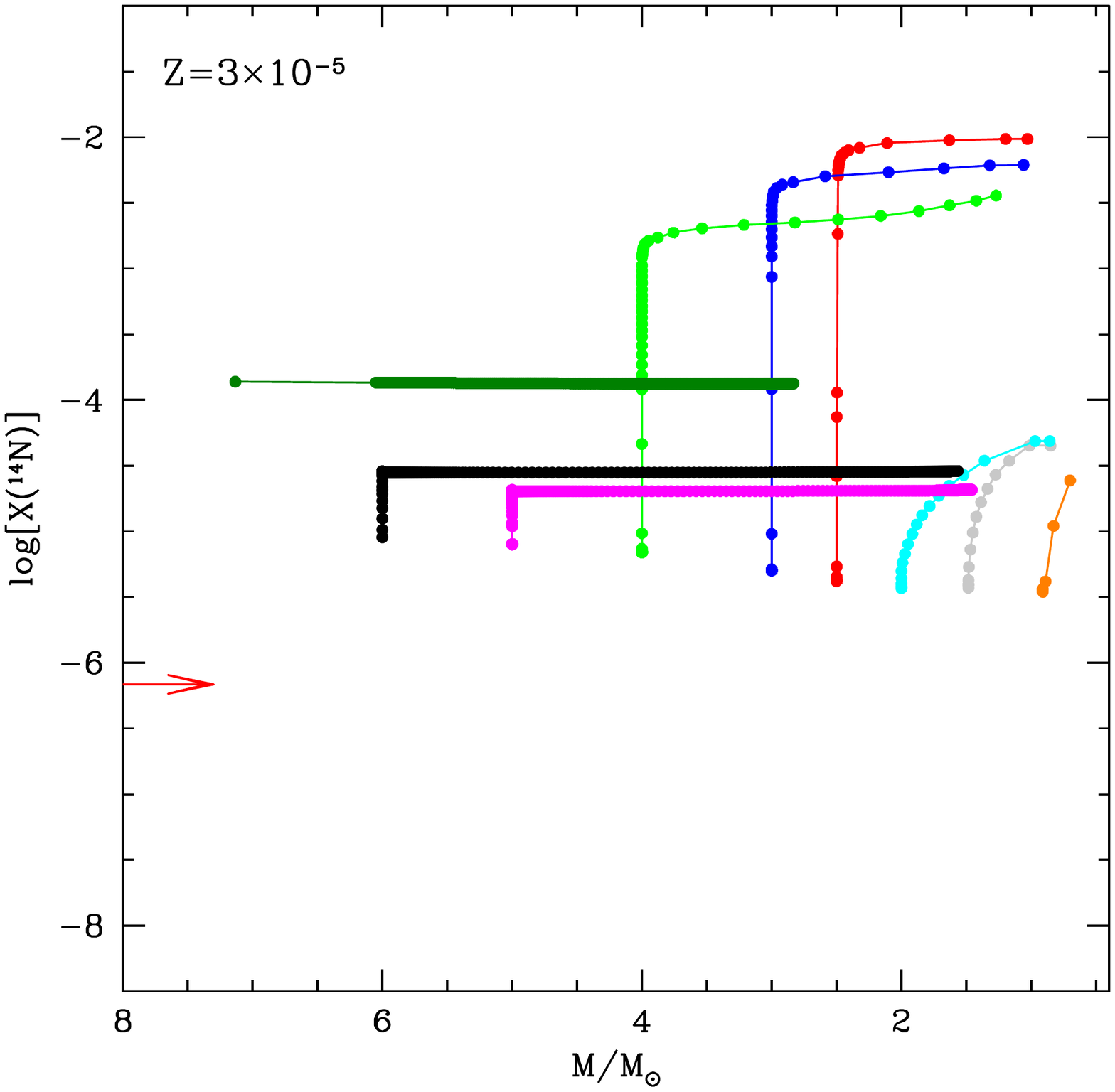}}
\end{minipage}
\begin{minipage}{0.48\textwidth}
\resizebox{1.\hsize}{!}{\includegraphics{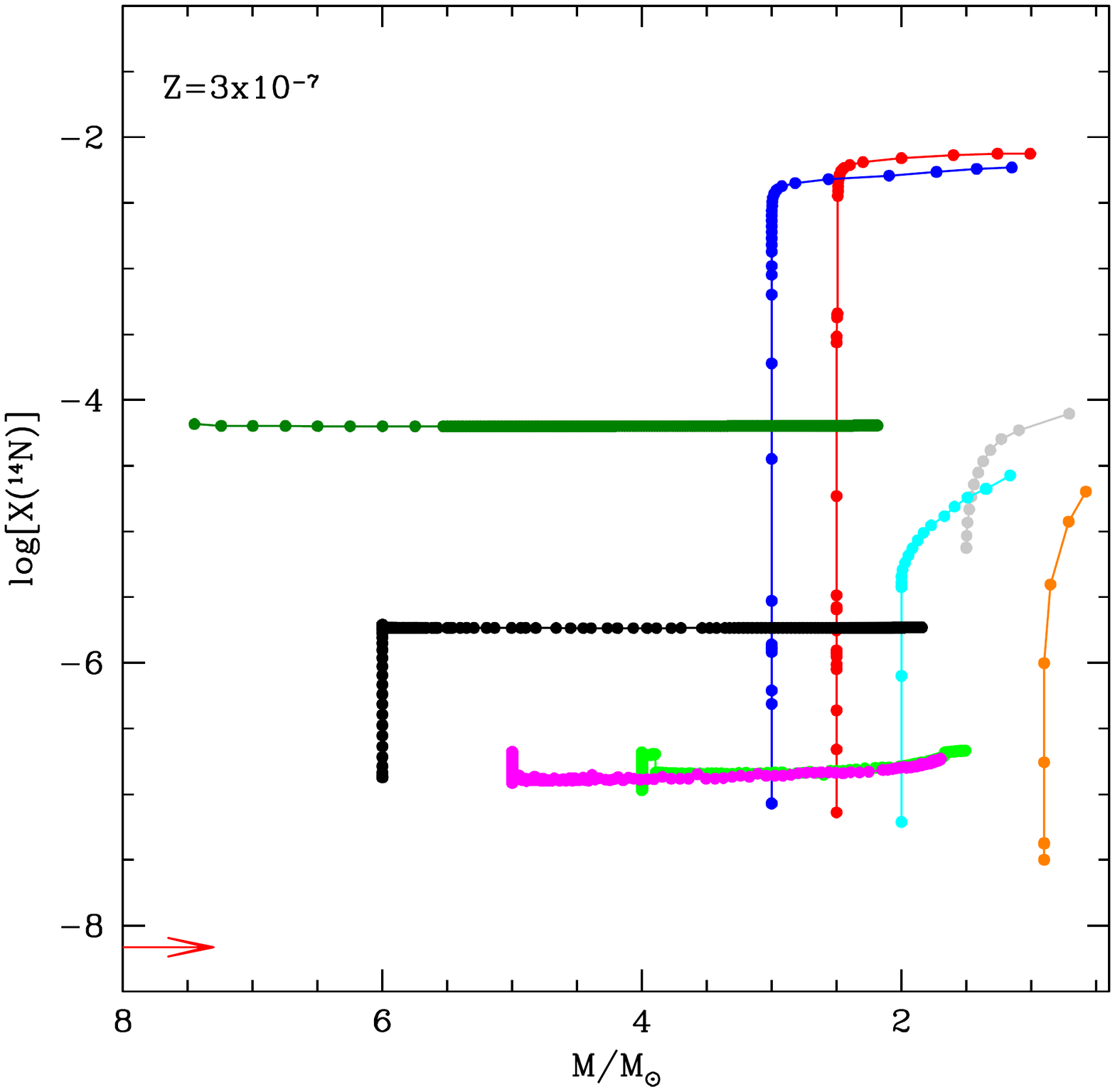}}
\end{minipage}
\vskip-70pt
\begin{minipage}{0.48\textwidth}
\resizebox{1.\hsize}{!}{\includegraphics{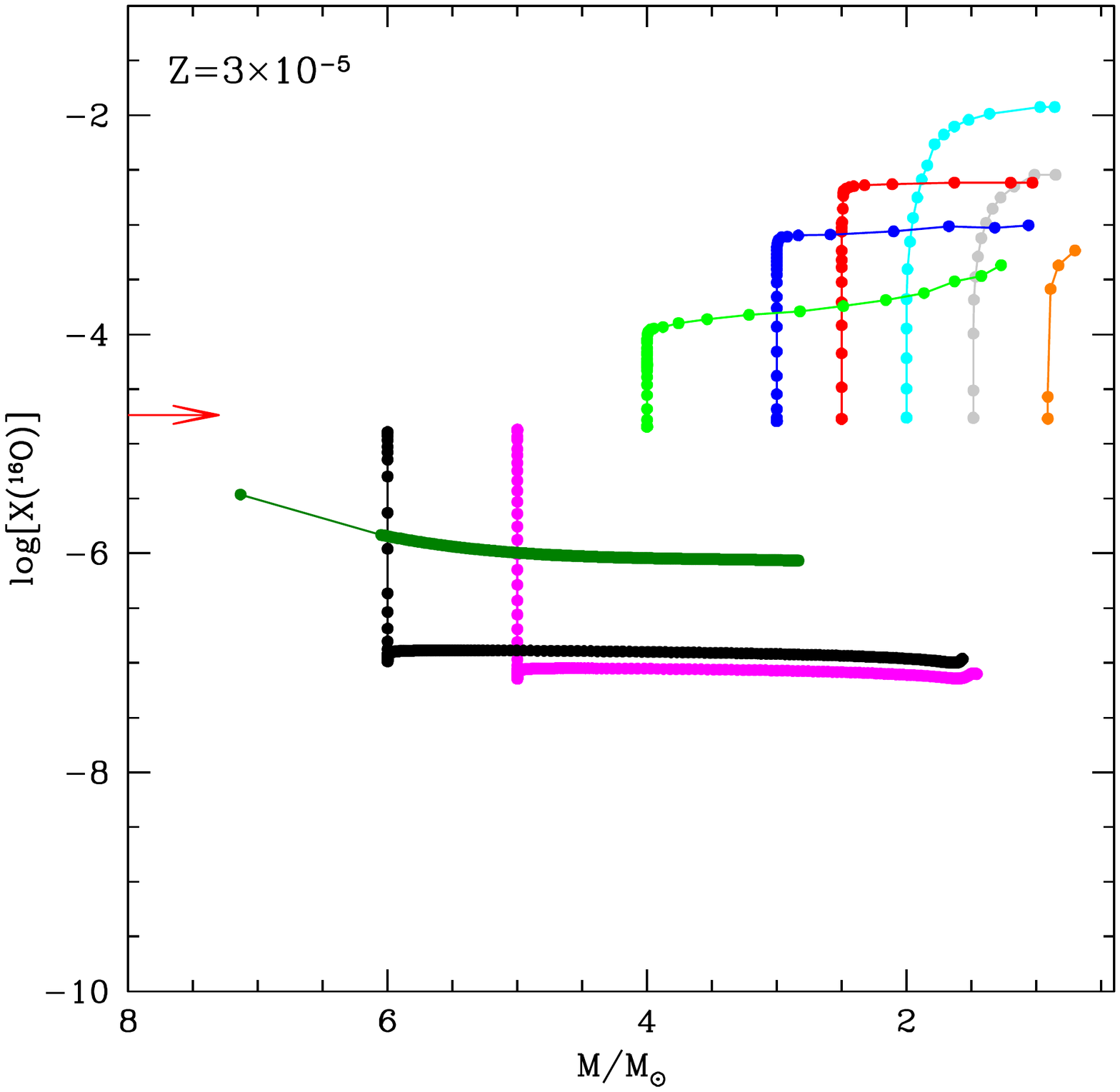}}
\end{minipage}
\begin{minipage}{0.48\textwidth}
\resizebox{1.\hsize}{!}{\includegraphics{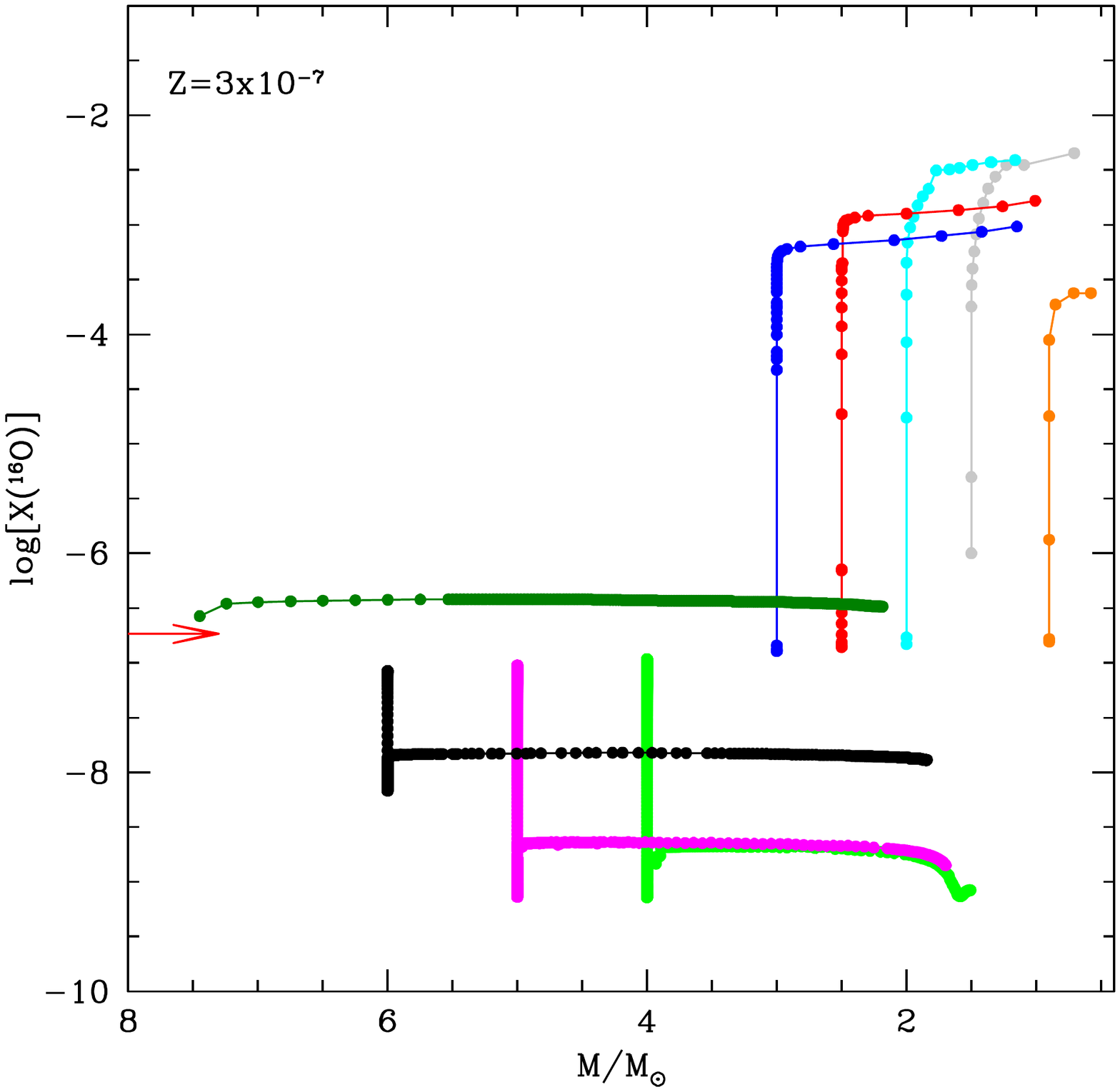}}
\end{minipage}
\vskip-50pt
\caption{Evolution of the surface mass fraction of nitrogen (top panels)
and oxygen (bottom) as a function of the (current) stellar mass of the
$Z=3\times 10^{-5}$ (left) and $Z=3\times 10^{-7}$ (right) model stars
discussed in the present work. The colour coding is the same as in
Fig.~\ref{ftbce}. Red arrows on the left side of each panel indicate the
initial mass fractions.} 
\label{fno}
\end{figure*}

\begin{figure*}
\begin{minipage}{0.48\textwidth}
\resizebox{1.\hsize}{!}{\includegraphics{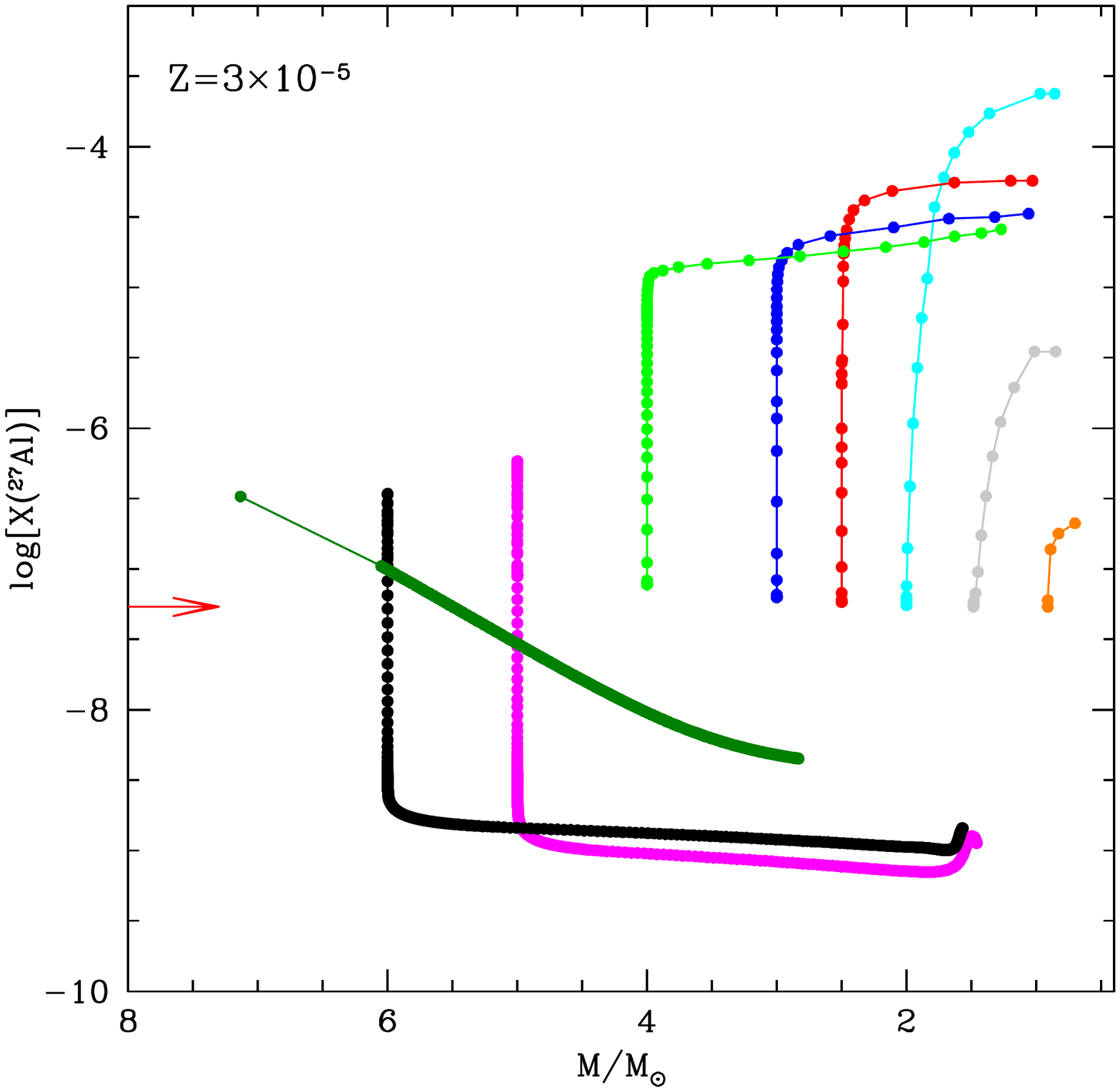}}
\end{minipage}
\begin{minipage}{0.48\textwidth}
\resizebox{1.\hsize}{!}{\includegraphics{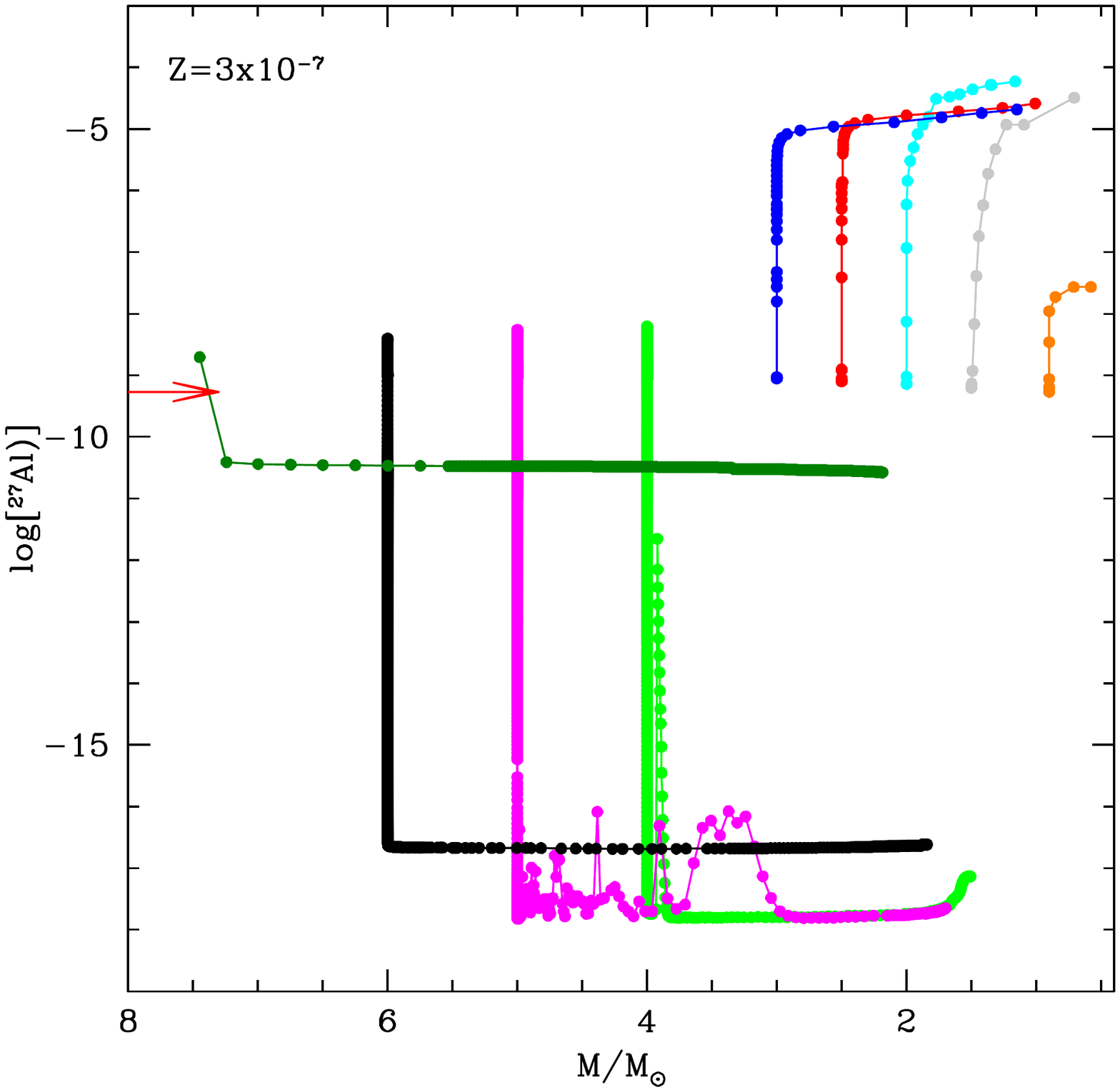}}
\end{minipage}
\vskip-70pt
\begin{minipage}{0.48\textwidth}
\resizebox{1.\hsize}{!}{\includegraphics{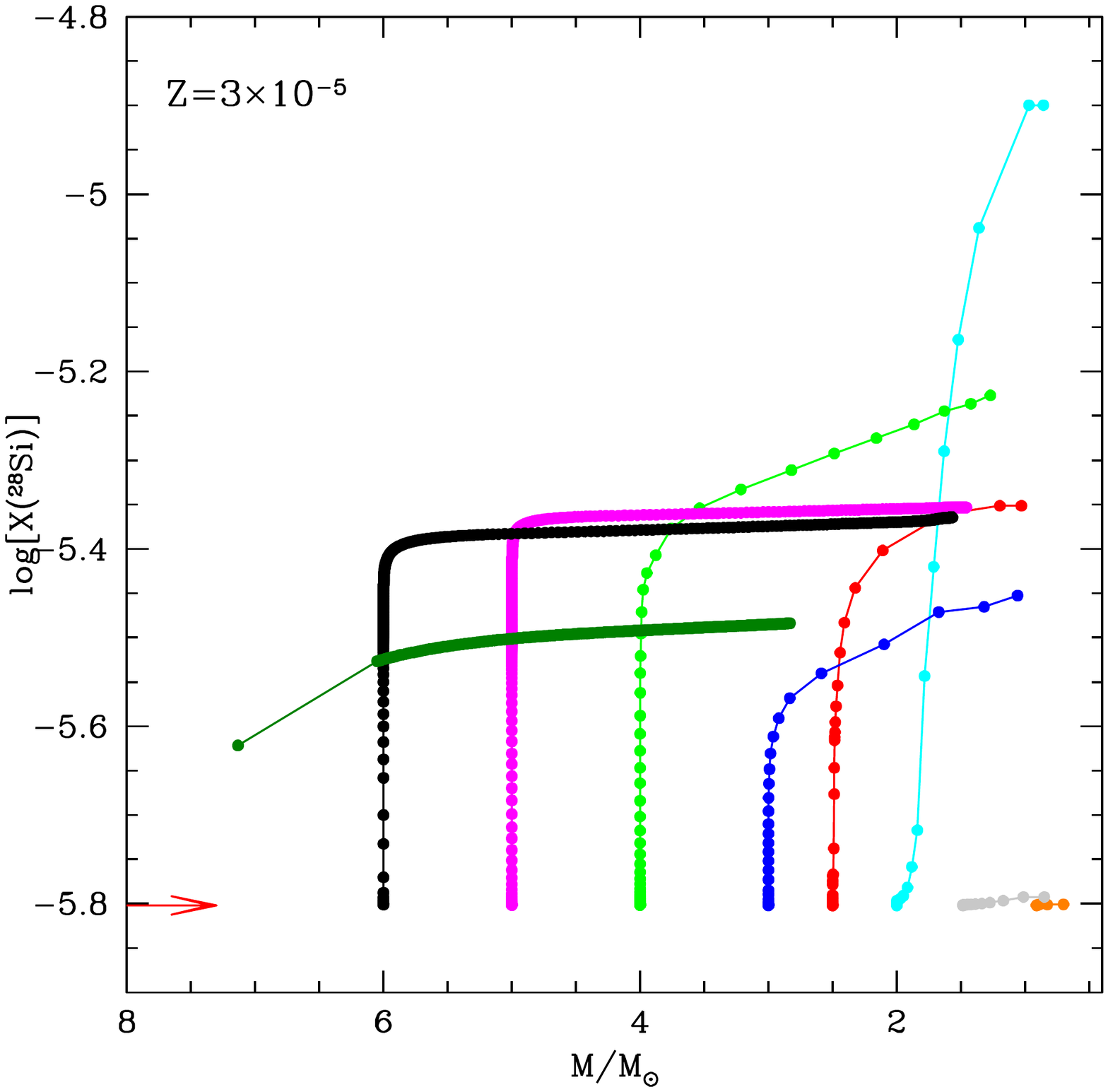}}
\end{minipage}
\begin{minipage}{0.48\textwidth}
\resizebox{1.\hsize}{!}{\includegraphics{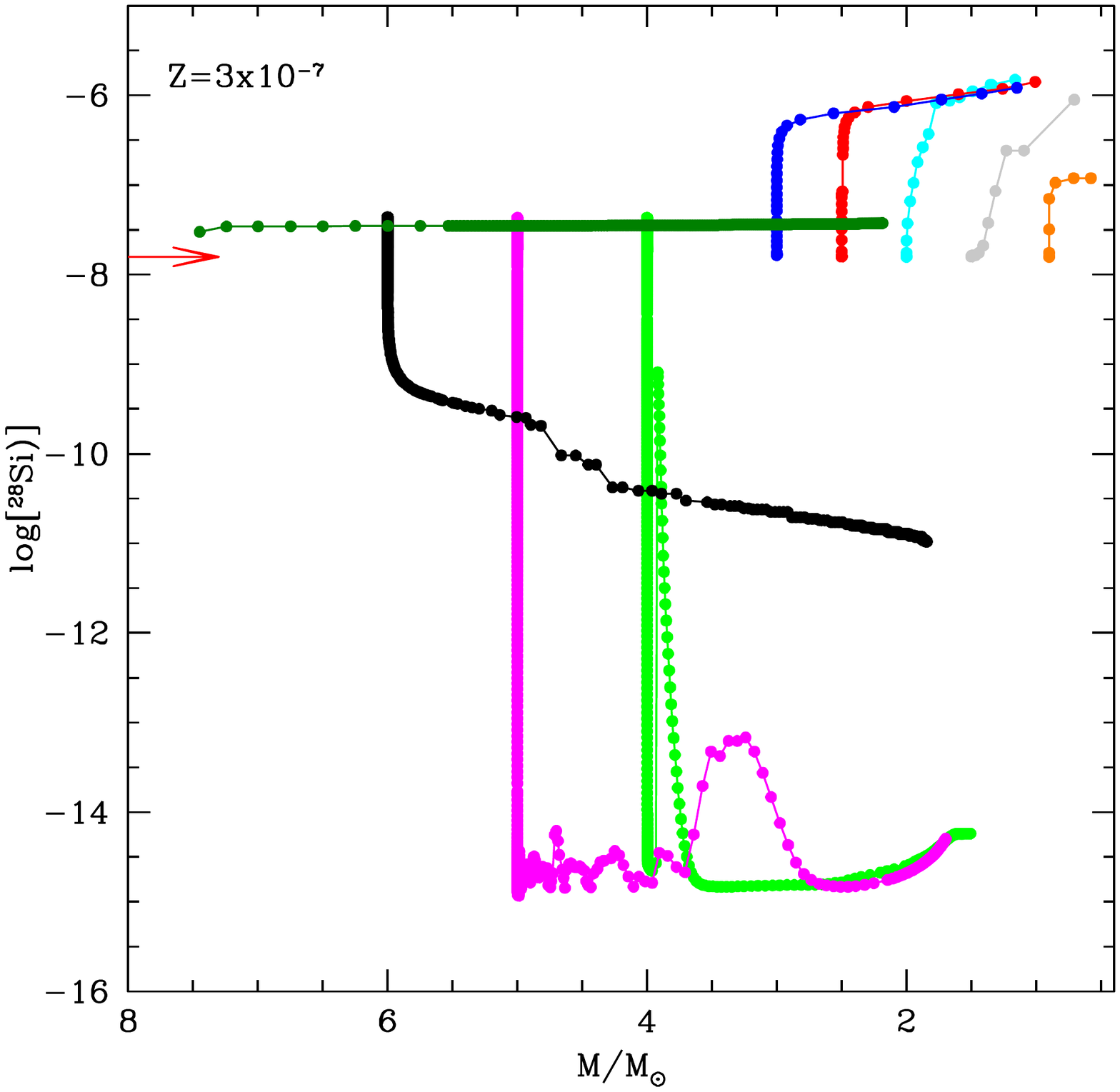}}
\end{minipage}
\vskip-50pt
\caption{Evolution of the surface mass fraction of $^{27}$Al (top panels)
and $^{28}$Si (bottom) as a function of the (current) stellar mass of the
$Z=3\times 10^{-5}$ (left) and $Z=3\times 10^{-7}$ (right) model stars
discussed in the present work. The colour coding is the same as in
Fig.~\ref{ftbce}. Red arrows on the left side of each pane indicate the
initial mass fractions.} 
\label{falsi}
\end{figure*}

To discuss the change in the surface chemical composition of metal-poor AGB stars,
which is relevant to understand the contamination of the interstellar
medium expected from these sources, we report in Fig.~\ref{fno} and \ref{falsi}
the behaviour of $^{14}$N, $^{16}$O, $^{27}$Al and $^{28}$Si, to focus
on the species involved in the CNO cycling and in the Mg-Al nucleosynthesis,
respectively. We use again the current mass of the star as abscissa, to have
an idea of the average chemical composition of the gas ejected.

The change in the surface $^{16}$O exhibits a dichotomous behaviour: the stars
experiencing TDU produce oxygen during the AGB lifetime, whereas those
exposed to HBB only destroy it, releasing gas which is almost oxygen-free.
The largest production of $^{16}$O, of the order of a factor $\sim 10^3$ 
($10^5$) in the $Z=3\times 10^{-5}$ ($Z=3\times 10^{-7}$) case, is found in the most massive models not experiencing HBB, given to the repeated TDU episodes, which gradually increase the surface $^{16}$O. $^{14}$N is produced in all cases, with the largest enhancement, between 3 and 5 orders of magnitude, according to the metallicity, taking place in $2~\rm M_{\odot} < M < 5~\rm M_{\odot}$ stars, where the $^{12}$C dredged-up to the surface is converted into $^{14}$N by HBB operating during the following inter-pulse phase.

The difference between the surface chemistry of models experiencing TDU and 
those suffering HBB only is also clear in the evolution of the surface
$^{27}$Al, shown in the top panels of Fig.~\ref{falsi}. In the former model
stars $^{27}$Al is synthesized owing to the ignition of the Mg-Al nucleosynthesis at the base of the envelope, which starts with the depletion of the surface $^{24}$Mg, and the dredge-up of $^{25}$Mg and $^{26}$Mg produced by $\alpha-$capture nucleosynthesis during the TP (see bottom, right panel of Fig.~\ref{f40m}). Conversely, massive AGBs experience a very advanced proton-capture nucleosynthesis in the innermost regions of the convective envelope, in such a way that 
the $^{27}$Al content of matter lost by the star is on the average lower than
in the gas from which the star formed. 

The behaviour of $^{28}$Si is extremely sensitive to mass and metallicity.
On general grounds, the stars producing most $^{28}$Si are those
experiencing both TDU and HBB. In the massive AGBs domain production of
$^{28}$Si takes place via the activation of the Mg-Al-Si nuclosynthesis at
the base of the envelope. Note that in $Z=3\times 10^{-7}$ case the
temperatures of the HBB nucleosynthesis are so hot that the gas ejecta
are silicon-poor.

\subsection{The role of metallicity}
The comparison among the results obtained for $Z=3\times 10^{-5}$ and $Z=3\times 10^{-7}$ and those published in previous investigations allows understanding how the evolutionary properties of AGB stars are sensitive to
the metallicity. Such an analysis was presented for the metal-poor domain
by \citet{flavia19}, whereas for higher metallicities we address the reader
to the studies by \citet{ventura13} and \citet{ventura18}.

\begin{figure*}
\begin{minipage}{0.99\textwidth}
\resizebox{1.\hsize}{!}{\includegraphics{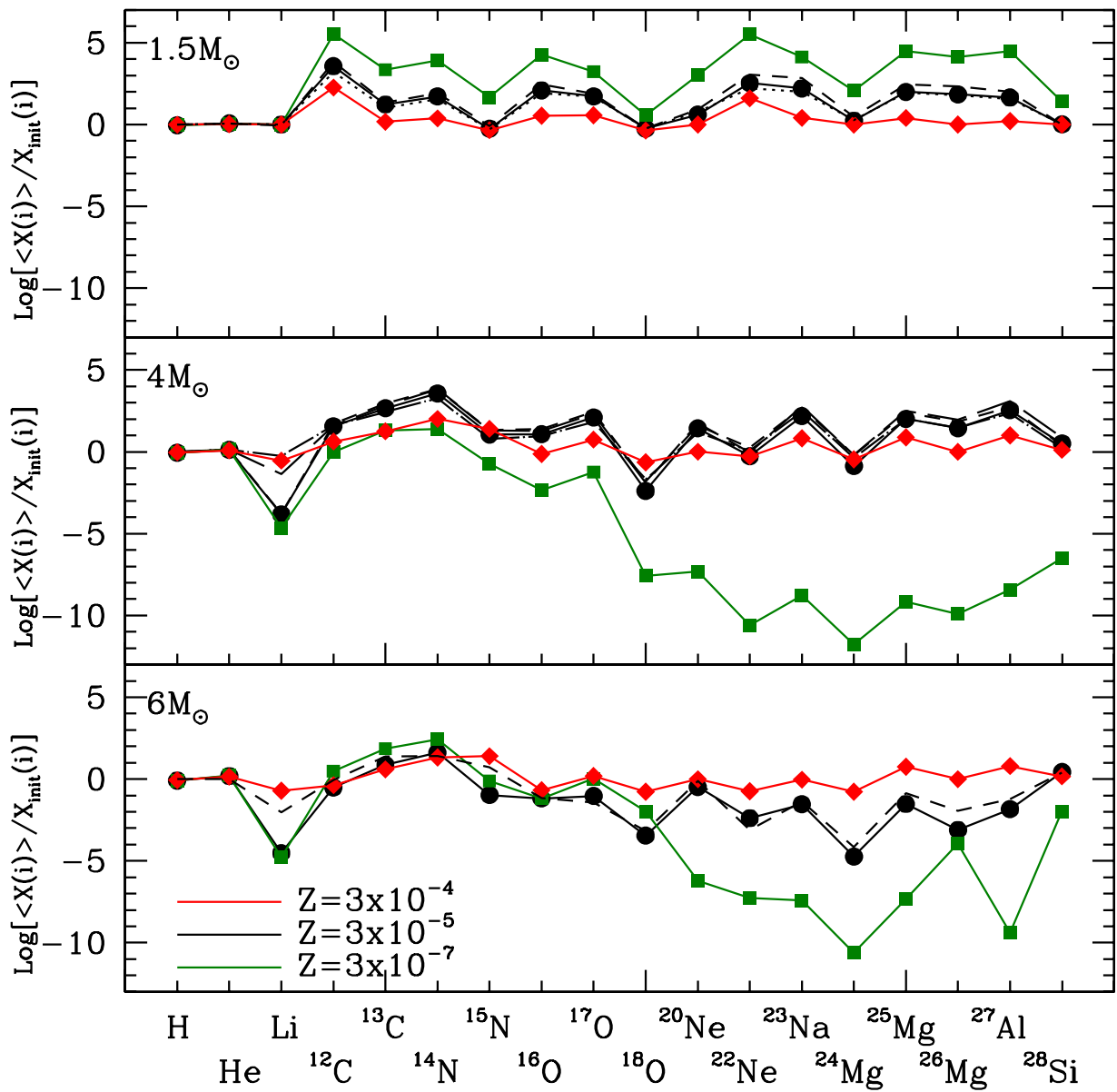}}
\end{minipage}
\vskip-120pt
\caption{The ratio between the average abundances of different chemical
species in the ejecta of AGB stars of initial mass $1.5~M_{\odot}$ (top panel), $4~M_{\odot}$ (middle panel) and $6~M_{\odot}$ (bottom panel) and the initial mass fractions. The different lines refer to the $Z=3\times 10^{-5}$ (black) and $Z=3\times 10^{-7}$ (green) model stars investigated in the present work, and to the $Z=3\times 10^{-4}$ (red) models published in \citet{flavia19}. Dotted (dashed) lines in the panels refer to the
results of $Z=3\times 10^{-5}$ models obtained by assuming an overshoot parameter a factor 2 lower (higher) than in the standard case. Long-dashed and dotted-dashed lines
in the middle panel indicate the results of the evolution of a $4~M_{\odot}$ model star with $Z=3\times 10^{-5}$, obtained by modelling mass-loss according to \citet{blocker95}, by assuming a value of the free parameter of $\eta_R = 0.005$ and $\eta_R = 0.02$, respectively.}
\label{fchem}
\end{figure*}

To ease the comparison among the variation of the surface chemical composition
of stars of different metallicity we show in Fig.~\ref{fchem} the 
chemical composition of the ejecta, averaged over the whole AGB phase,
of stars of different mass, for the two metallicities investigated here
and for the $Z=3\times 10^{-4}$ model stars discussed in \citet{flavia19}.
Overall these results refer to a metallicity range extending over 3 orders
of magnitude. The masses shown in the figure were chosen in order to represent the three groups 
discussed in sections \ref{lowmass}, \ref{interm} and \ref{massivem}.

The threshold mass required to activate HBB in the present metallicity domain is $\sim 2~M_{\odot}$, the same as for $Z=1-3\times 10^{-4}$ \citep{flavia19}; this is consistent with the
results obtained by CL08, who find that $2~M_{\odot}$ model stars experience HBB
for $[$Fe$/$H$] \leq -4$. This threshold mass is significantly smaller than in higher metallicity model stars: recent studies showed that only $\rm M \geq 3.5~\rm M_{\odot}$ stars reach HBB conditions in the solar \citep{ventura18}, super-solar \citep{ventura20} and slightly sub-solar \citep{ventura14a} case, whereas for the $Z=10^{-3}$ chemistry this mass threshold is $\sim 3~\rm M_{\odot}$ \citep{ventura13}. These differences are connected with the higher steepness of the core mass vs initial mass relationship in lower metallicity stars, because the threshold in the core mass required to activate HBB is $\sim 0.8~\rm M_{\odot}$, almost independently of the metallicity \citep{ventura13}.

\citet{flavia19} discussed the sensitivity of the strenght of HBB to the
metallicity, outlining that both $T_{\rm bce}$ and luminosity increase when
$Z$ decreases. The $T_{\rm bce}$ vs mass relationship was shown to get steeper
and steeper as Z decreases, with the most massive AGBs reaching temperatures
at the base of the envelope above 140 MK during the AGB phase. The present
results confirm this trend, with the largest temperatures being $T_{\rm bce} \sim 170$ MK for $Z=3\times 10^{-5}$ and $T_{\rm bce} \sim 195$ MK $Z=3\times 10^{-7}$.
Given the tight relationship between $T_{\rm bce}$ and luminosity of the stars
experiencing HBB, the peak luminosities are also correlated with metallicity,
the largest values being $2.4\times 10^5~L_{\odot}$ and $2.15\times 10^5~L_{\odot}$ for the $7.5~M_{\odot}$ model star of metallicity 
$Z=3\times 10^{-7}$ and $Z=3\times 10^{-5}$, respectively (see table 
\ref{tabgen}).

The degree of the nucleosynthesis taking place at the bottom of the convective envelope, and consequently the variation of the surface chemical composition, 
are sensitive to $T_{\rm bce}$. On this regard, the most clear effect of
metallicity can be seen in the bottom panel of Fig.~\ref{fchem}, showing the
average chemical composition of the gas expelled from stars with initial mass
$6~M_{\odot}$, whose surface chemistry is determined almost exclusively by HBB.
While the ejecta of $Z=3\times 10^{-4}$ stars are enriched in some of the species involved in the advanced p-capture nucleosynthesis, such as $^{23}$Na, $^{27}$Al and $^{28}$Si, at lower metallicities all the species from neon to silicon are
destroyed by HBB; this is evident in the $Z=3\times 10^{-7}$ case and particularly in the behaviour of $^{24}$Mg, which is consumed almost in its
entirely in the surface regions of the star. $^{12}$C and $^{16}$O are also
exposed to proton capture when HBB is activated; however, the results regarding
these species for the $6~M_{\odot}$ stars are less sensitive to the metallicity than the heavier elements, because of the effects of the dredge-out, experienced
by the $6~M_{\odot}$, $Z=3\times 10^{-7}$ star (see discussion in 
section \ref{preagb}), which increases the overall
CNO and consequently the equilbrium abundances of all the CNO species. This is
the reason why the lines corresponding to the different metallicities in the
bottom panel of Fig.~\ref{fchem} are extremely close in correspondence of the
CNO elements.

The results shown in the middle panel of Fig.~\ref{fchem} outline notable
differences between the $4~M_{\odot}$ stars of metallicity $Z=3\times 10^{-7}$ and
the counterparts of same mass and higher Z. As in the case of $6~M_{\odot}$ stars
discussed earlier, the ejecta of the most metal-poor model stars exhibit a
must stronger signature of HBB. This is partly due to the hotter $T_{\rm bce}$'s;
however, an additional reason for this difference is that the chemistry of the
most metal-poor, $4~M_{\odot}$ stars is influenced by HBB only, whereas in the
$Z=3\times 10^{-4}$ and $Z=3\times 10^{-5}$ cases the action of TDU increases
the mass fractions of the various chemical species considered.

The evolution of the stars not experiencing HBB, discussed in section \ref{lowmass}, is driven by the gradual enrichment in the surface $^{12}$C,
which eventually leads to the formation of carbon stars. \citet{flavia19} found that $1-2~M_{\odot}$ stars evolved as carbon stars for a fraction of the 
AGB lifetime in the $50-75\%$ range for the metallicities $Z=1-3\times 10^{-4}$.
On this regard, the present results for the $Z=3\times 10^{-5}$ chemistry are similar to \citet{flavia19}, whereas for the lowest metallicity $Z=3\times 10^{-7}$ we find slightly higher percentages, owing to the smaller initial fraction of $^{16}$O, which makes the achievement of the C-star stage easier.

In terms of carbon enrichment we find, consistently with \citet{flavia19}, that the results are not particularly sensitive to the metallicity, because
the $^{12}$C dredged-up to the surface is of primary origin, thus independent 
of $Z$. The different lines shown in the top panel of Fig.~\ref{fchem}, 
reporting the production factor of the various species for the three metallicities
analysed in the present discussion, share the same trend, the values being
higher the lower the metallicity, which is connected with the lower initial mass
fractions of all the species but hydrogen and helium in lower-Z stars.

\subsection{The impact of convection and mass-loss modelling}
It is generally recognized that the results from AGB modelling are sensitive
to the description of some physical phenomena still unknown from first
principles, the two most relevant being convection and mass loss \citep{karakas14b}. 

As far as the determination of the convective temperature gradient is concerned,
the FST modelling adopted here was shown to lead to stronger HBB conditions than in the results
based on the mixing length theory (MLT), in the model stars achieving temperatures at the base of the envelope exceeding $\sim 30$ MK \citep{ventura05a}, which, in the metallicity domain explored here, reflects into initial masses above $2~M_{\odot}$. 
The nucleosynthesis at the bottom of the convective envelope experienced by
model stars calculated by adopting MLT modelling for turbulent convection
is less advanced than in their FST counterparts, which also reflects into
lower luminosities and slower evolution \citep{ventura05a}. These arguments
will be reconsidered in section \ref{comp}, where the present findings will
be compared with MLT-based computations form other research groups. For what
attains low-mass stars discussed in section \ref{lowmass}, no significant 
differences are expected between the FST and the MLT results.

Still in the convection modelling context, a further source of uncertainty is associated to the treatment of convective borders, which rules the extension of extra-mixing, the depth of TDU, thus the amount of $^{12}$C and of material processed by helium-burning nucleosynthesis convected to the surface after each TDU event. In the present computations, as discussed 
in section \ref{agbinput}, we adopted an OS $l_{OV} = 0.002H_p$, calibrated against the observations of carbon stars in the Magellanic Clouds \citep{ventura14b}. 

To test how changes in this quantity reflect into the results obtained in the low-mass stars domain, we run evolutionary sequences of a $1.5~M_{\odot}$ model star, based on $l_{OV} = 0.001H_p$ and $l_{OV} = 0.004H_p$, which represent a factor 2 change with respect to the standard assumption. These results are indicated in Fig.~\ref{f15m} with dotted and dashed lines, which refer to the $l_{OV} = 0.001H_p$ and $l_{OV} = 0.004H_p$ cases, respectively. For readability we only show the results regarding the luminosity and the surface (C$-$O) and $^{22}$Ne. A change in the assumed $l_{OV}$ reflects into the extent of each TDU event, thus on the amount of material transported to the surface regions after each TP. We find that doubling (halving) the OS results in a $\sim 40 \%$ increase (decrease) in the surface (C$-$O) and a factor 2 increase (decrease) in the $^{22}$Ne accumulated in the surface regions. Regarding the other chemical species, $\sim 30\%$ changes in the final abundances are found. The choice regarding $l_{OV}$ also affect the general evolution of the star, because a deeper TDU decreases the core mass of the star, which therefore
evolves at lower luminosities and consequently on slower time scales: this
is shown in the top, left panel of Fig.~\ref{f15m}, where we note a clear
difference between the various results, with the luminosity increasing
faster in the $l_{OV} = 0.001H_p$ case. These differences do not significantly affect the overall duration of the AGB phase, which are similar within $10\%$ in the three cases, because, as discussed in section \ref{lowmass}, most of the mass is lost in the very final TPs.

Turning to the stars experiencing HBB, described in sections \ref{interm} and \ref{massivem}, the only additional case which we explore in the analysis of the effects of the treatment of convective borders is the $l_{OV} = 0.004H_p$ one; we do not expect to find significant differences between the $l_{OV} = 0.001H_p$ and $l_{OV} = 0.002H_p$ cases, as the TDU experienced was found to be extremely weak in the latter case, as discussed in section \ref{interm}. The finding with the $l_{OV} = 0.004H_p$ assumption are indicated with dashed, cyan lines in the left panels of Fig.~\ref{f40m} and in Fig.~\ref{f60m}. 

Both in the $4~M_{\odot}$ and $6~M_{\odot}$ cases doubling the OS associated 
to the penetration of the convective bubbles into the
radiatively stable regions favours a deeper TDU with respect to the standard
case, with the transportation of larger quantities of $^{12}$C to the surface.
The variation of the surface $^{12}$C does not show up significant difference
between the two cases, because of the action of HBB, which destroys the  
surface $^{12}$C via proton fusion; on the other hand the production of 
$^{14}$N is enhanced by a factor $\sim 5$, as can be seen in the  bottom, left
panels of Figg.~\ref{f40m} and \ref{f60m}. The choice of a larger extra-mixing
from the borders of the convective zones also affect the general physical
evolution of the stars, because the increase in the surface metallicity favours the expansion of the star and the rise in the mass-loss rate: the models
calculated with a higher $l_{OV}$ evolve faster, with the overall duration
of the AGB phase being $\sim 10\%$ shorter. This is seen in the run of 
the luminosity, reported in the top, left panels of Fig.~\ref{f40m} and 
Fig.~\ref{f60m}

The description of mass-loss is a major issue in AGB modelling, considering the
poor knowledge of this physical mechanism and the relevant impact on the
AGB evolution \citep{ventura05b}. Most of the calibrations presented so far
are based on stars of metallicity significantly higher than those explored 
here, which suggests to consider the implications relative to possible
sensitivity of mass-loss on the chemical composition \citep{pauldrach89}. 
In the low-mass domain we believe that the description of mass-loss adopted 
here is at the same level of accuracy of model stars of higher metallicity,
because these stars experience a significant increase in the surface
metallicity, related to the effects of TDU, which enrich the external regions 
with primary carbon, produced in the $3\alpha$ burning shell; the surface
metal enrichment is practically independent of metallicity, which has only
an indirect influence, however considered in the treatment by \citet{wachter08}
adopted here, connected with the hotter effective temperatures of metal-poor
stars when compared to the higher-Z counterparts. These stars produce large
quantities of carbonaceous dust (see discussion in section \ref{dust}) for most of their life, thus their mass-loss is driven by the radiation pressure effects on
the dust grains, which is considered in the determination of the rates given
by \citet{wachter08}, used in the present analysis. 

The case of the stars experiencing HBB is more tricky, as we find that
no significant increase in the overall metallicity takes place, as discussed in sections \ref{interm} and \ref{massivem}. We adopt the description by
\citet{vw93}, consisting into a direct relation between $\dot M$ and pulsation period, found via an analysis of CO microwave observations of AGB stars. In this
description the metallicity of the star enters via the period, which is 
generally shorter the lower is Z. Alternative description of mass-loss of
bright AGB stars is given by \citet{blocker95}, based on dynamical models
of the winds of M-type stars. To explore the sensitivity of the results on the
mass loss description we considered two $4~M_{\odot}$ models based on the
\citet{blocker95} treatment of mass-loss, in which the free parameter 
entering the \citet{blocker95}'s formula was taken as $\eta_R=0.005$ and
$\eta_R=0.02$\footnote{The latter quantity is the standard assumption in
the AGB calculations of solar and sub-solar metallicity, calibrated on the
basis of the luminosity distribution of lithium-rich stars in the Magellanic
Clouds by \citet{ventura00}}. These results are shown in Fig.~\ref{f40m}
as green, dotted-dashed ($\eta_R=0.02$) and long-dashed ($\eta_R=0.005$) lines.

The results obtained with the \citet{blocker95} description are found to bracket
those obtained with the \citet{vw93} treatment. The $\eta_R=0.02$ model
experiences a mass-loss rate significantly larger than the \citet{vw93} counterpart, an effect of the steep sensitivity of $\dot M$ on the luminosity 
of the star included in the \citet{blocker95} formula. As a consequence the peak
luminosity is $\sim 30\% $ smaller and the TP-AGB phase is $\sim 30\% $ shorter.
The variation of the surface chemistry is also affected by the faster
consumption of the envelope, because the number of TDU events experienced
is smaller. This is not going to affect the species exposed to proton captures
during HBB, such as $^{12}$C, but has serious consequences on the species
synthesized when HBB is activated, such as $^{14}$N, whose final surface mass
fraction is $\sim 30\%$ lower than in the \citet{vw93} case. In stars of higher
mass we do not expect significant dissimilarities on the chemical side, because
their chemistry is almost entirely determined by HBB, with little contribution
from TDU. On the other side, when $\eta_R=0.005$ is assumed in the 
\citet{blocker95} recipe, the mass-loss rates are generally smaller than in
the \citet{vw93} case, thus the evolutionary time-scales are longer, the
difference being within $\sim 30\%$. The HBB experienced is slightly stronger, 
but the main effect on the surface chemistry is once more due to the
higher production of primary nitrogen, related to the higher number of
TDUs experienced: the final $^{14}$N is $\sim 50\%$ higher than in the 
\citet{vw93} case.

We believe important to underline here that while these numerical explorations 
can provide a broad estimate of how the present results are sensitive
to the mass-loss description, use of the \citet{blocker95} description can
be hardly applied to massive AGBs in the metallicity domain explored in the
present investigation: the \citet{blocker95} prescription is based on 
dynamical models of the winds of massive AGB stars which consider the
effects of radiation pressure on the dust particles, but in the present
context very poor dust formations is expected to take place in the circumstellar
envelope of M-type stars, owing to the scarcity of gaseous silicon.

\section{Gas yields}
\label{yields}
We discuss the yields of the different chemical species of the model stars
discussed in the present work.

We follow the classical definition of the stellar yield, i.e., the yield of a given element $i$ is the net amount of newly- produced element that is ejected in the interstellar medium by a star during its life:
$$
Y_i=\int{[X_i-X_i^{\rm init}]}\dot M \rm{dt}
$$

The integral is calculated over the entire stellar lifetime, $X_i^{\rm init}$ 
is the mass fraction of species $i$ at the beginning of the evolution, 
and $\dot M$ is the mass-loss rate. If the element is destroyed in the stellar interior, then the yield is negative.\footnote{The yields for the main chemical species of all the model stars presented in this work are available at  www.oa-roma.inaf.it/arca/} 

\begin{figure*}
\begin{minipage}{0.48\textwidth}
\resizebox{1.\hsize}{!}{\includegraphics{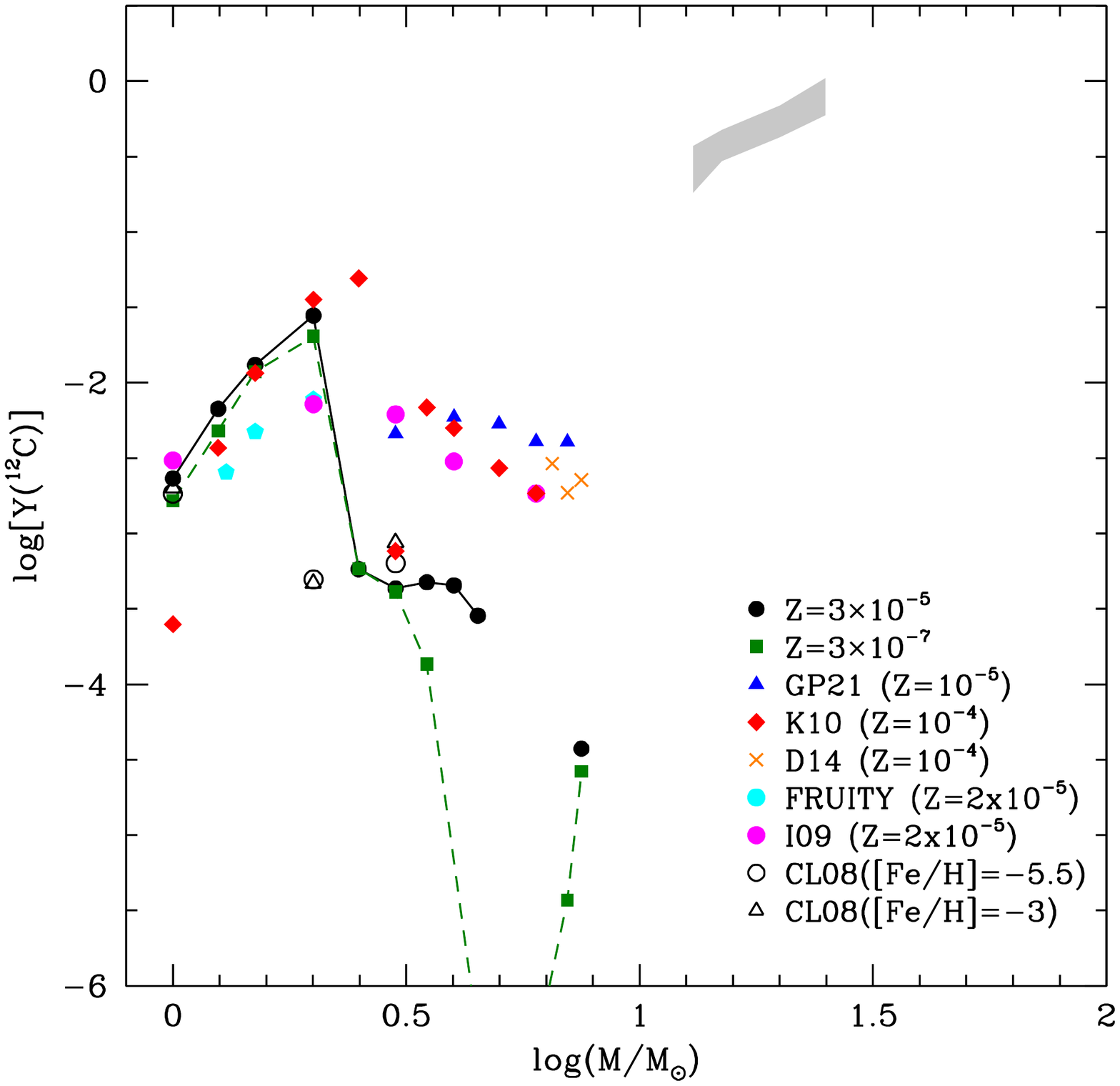}}
\end{minipage}
\begin{minipage}{0.48\textwidth}
\resizebox{1.\hsize}{!}{\includegraphics{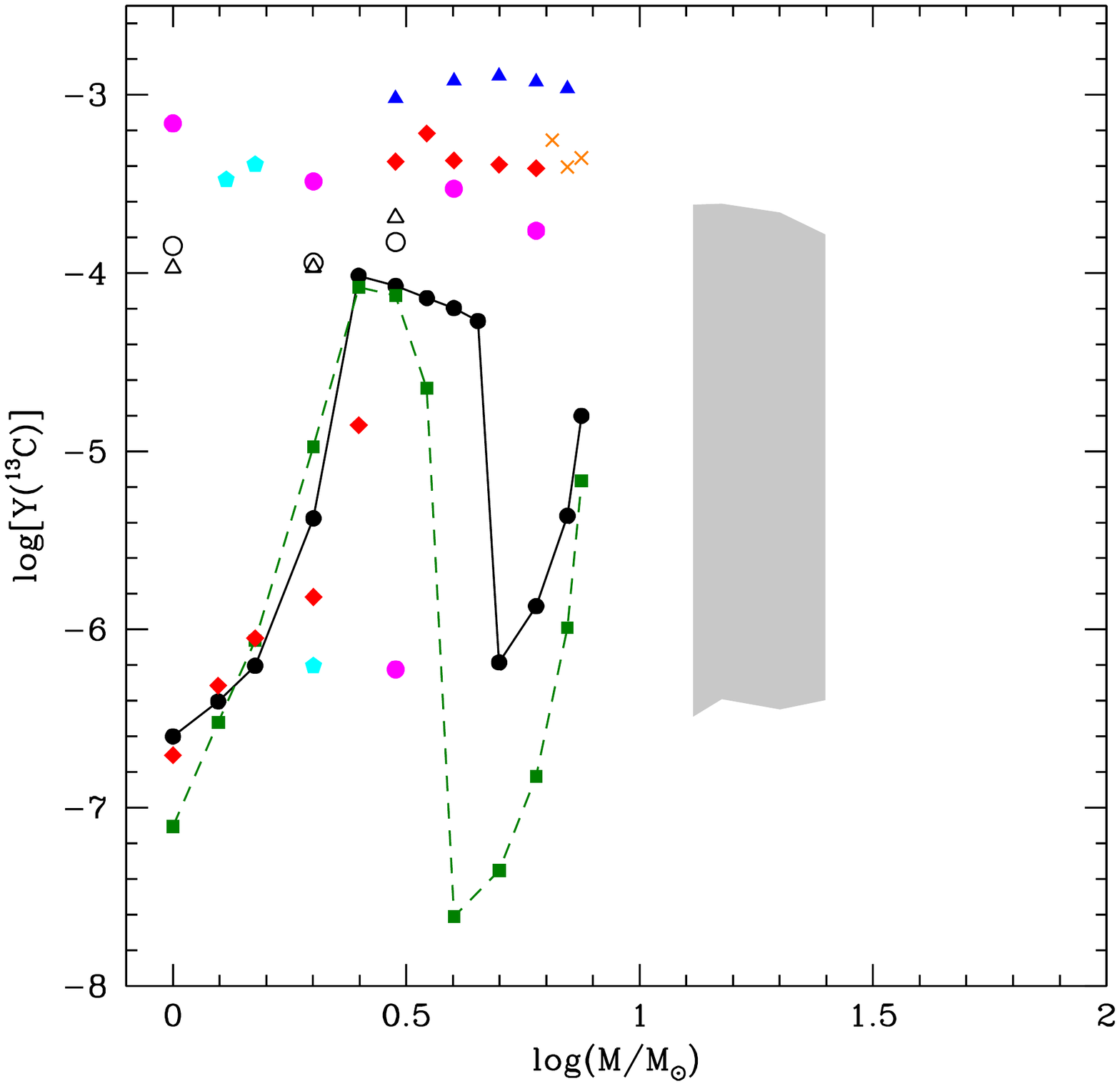}}
\end{minipage}
\vskip-70pt
\begin{minipage}{0.48\textwidth}
\resizebox{1.\hsize}{!}{\includegraphics{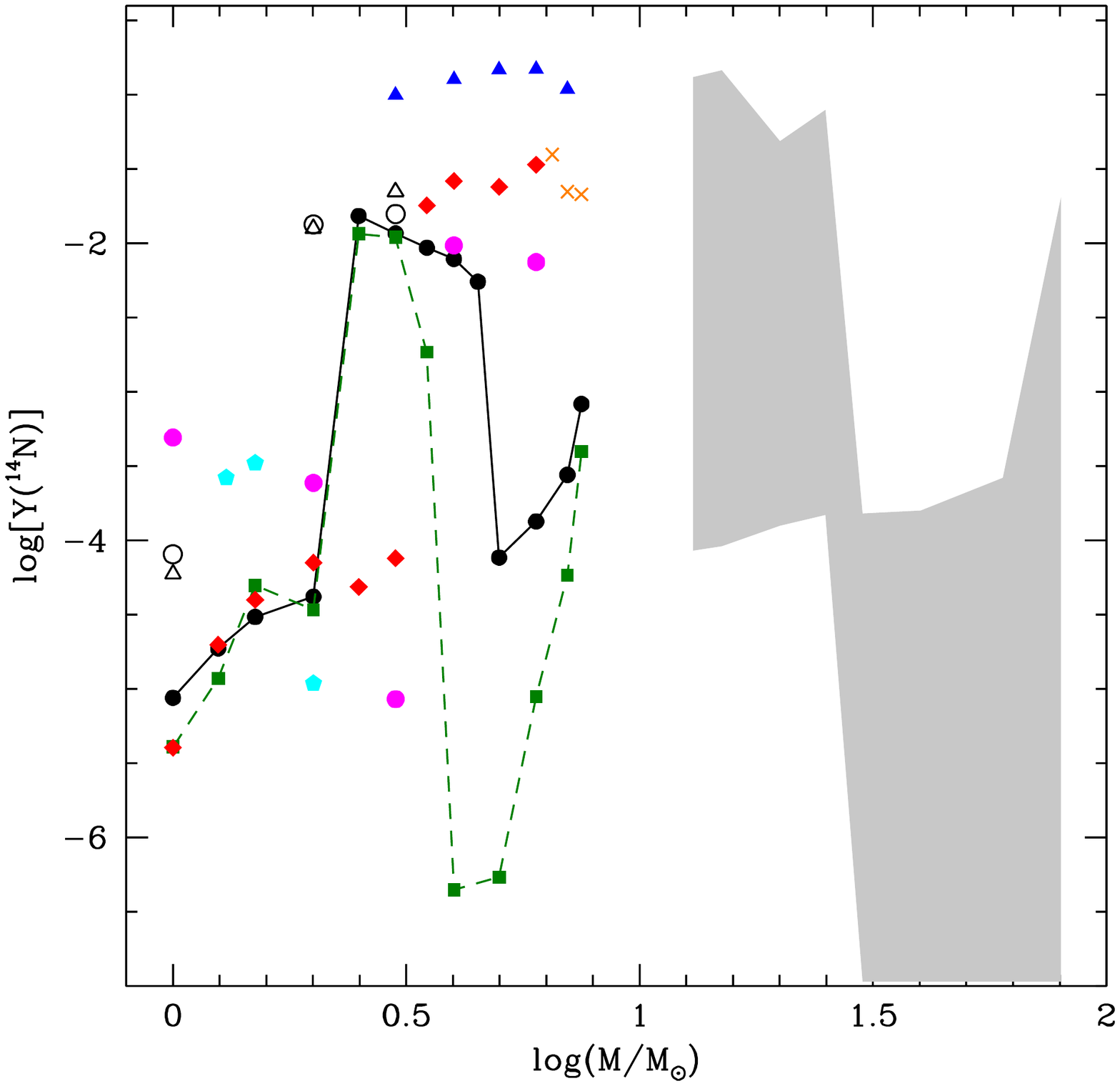}}
\end{minipage}
\begin{minipage}{0.48\textwidth}
\resizebox{1.\hsize}{!}{\includegraphics{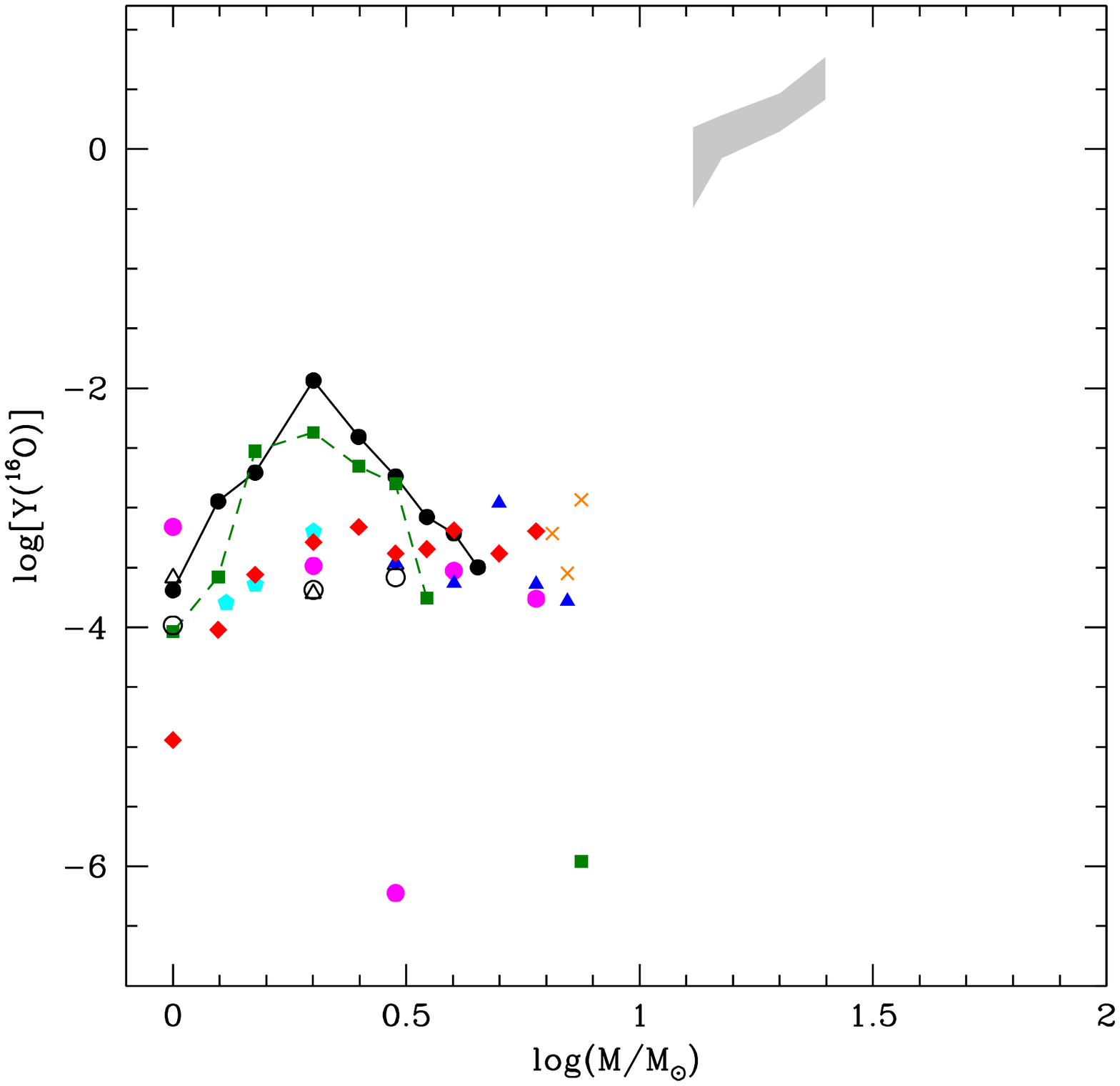}}
\end{minipage}
\vskip-50pt
\caption{Yields (solar masses, see definition in section \ref{yields}) of the chemical species involved in CNO cycling are shown as a function of the initial mass of the star, for the metallicities $Z=3\times 10^{-5}$ (black points) and $Z=3\times 10^{-7}$ (green squares). The corresponding yields from GP21, \citet{karakas10} and \citet{doherty14} are indicated with blue triangles, red diamonds and orange crosses, respectively. Shaded areas on the right side of the panels indicate the values covered by the yields from \citet{limongi18}, for different initial rotation velocities. The values of the mass for which the points or the shadowing is missing indicate that the corresponding yields are negative. } 
\label{fycno}
\end{figure*}

Fig.~\ref{fycno} and \ref{fymgal} show the gas yields of the models presented
here as a function of the initial mass of the star. We also show the yields
from massive stars presented by LC18, which were evolved from the same chemical composition of the $Z=3\times 10^{-5}$ models investigated here. This allows a more exhaustive analysis on the relative impact of stars of different mass on the chemical evolution of the host galaxy. LC18 present an exhaustive discussion on the role of rotation on the internal mixing of massive stars and the related effects on the gas yields, by comparing results obtained by varying the assumed equatorial velocity from zero to 300 Km$/$sec. The results reported in Fig.~\ref{fycno} and \ref{fymgal} are given in terms of the range of values of the yields of a star of a given mass corresponding to different rotational velocities.

\subsection{CNO yields}
The yields reported in Fig.~\ref{fycno} refer to the CNO species. 
In the low-mass domain the $^{12}$C yields (see top, left panel of
the figure) of the AGB stars exhibit a growing trend with the initial mass of the star, ranging from a few $10^{-3}~\rm M_{\odot}$, for $1~\rm M_{\odot}$ stars, to almost $0.1~\rm M_{\odot}$, for $2~\rm M_{\odot}$ stars. This is because the number
of TP experienced, hence of the TDU events which take place during the
AGB phase, is higher the larger the initial mass of the star. We note that the $^{12}$C yields are not particularly sensitive to the metallicity in this mass domain, because the $^{12}$C dredged-up from the helium burning shell is of primary origin and is much larger than the $^{12}$C quantity (and more generally of the overall C+N+O) initially present in the star.

The sudden drop in the $^{12}$C yields for $\rm M >2~\rm M_{\odot}$ is connected with the ignition of HBB, which destroys the surface carbon. For the massive AGB stars the $^{12}$C yields are negative, because the HBB equilibria are shifted to $^{12}$C mass fractions lower than in the gas from which the stars formed. This can be seen, e.g., in the time variation of the surface $^{12}$C of the $6~\rm M_{\odot}$ model star, shown in the bottom, left panel of Fig.~\ref{f60m}. This trend is reversed at $M \geq 7~M_{\odot}$, because as discussed in section \ref{preagb} these stars experience the dredge-out mechanism, which favours a significant increase in the surface $^{12}$C during the phases preceding the series of thermal pulses: the $^{12}$C yields are positive in this mass domain, with numerical values of the order of $10^{-5}-10^{-4}~\rm M_{\odot}$, much smaller than in $\rm M<2~\rm M_{\odot}$ stars. The action of HBB causes the drop in the $^{12}$C yields, thus interrupting the growing trend with mass, which would ideally connect the yields of $\rm M\leq 2~\rm M_{\odot}$ stars with those from $13-25~\rm M_{\odot}$ stars calculated by LC18, which span the $0.2-1~\rm M_{\odot}$ range.

The trends with mass of the yields of $^{13}$C and $^{14}$N are similar.
For both elements the largest yields, of the order of $10^{-4}~\rm M_{\odot}$ and
$\sim 0.02~\rm M_{\odot}$ respectively, are provided by the intermediate mass
stars, that experience both TDU and HBB. For what attains $^{14}$N, these
findings are fully consistent with the results shown in the top panels of
Fig.~\ref{fno}. Even in this case the quoted values are not sensitive to the metallicity, 
as the chain of reactions leading to the synthesis of $^{13}$C and $^{14}$N begins by proton captures on $^{12}$C nuclei, mostly of primary origin at the low metallicities considered in this work. 
In the massive AGBs domain the metallicity is more relevant, because these
stars experience negligible TDU, thus the equilibrium abundances of the various species are mainly sensitive to the initial overall C+N+O (see again top
panels of Fig.~\ref{fno}). The rise in the trend of the $^{13}$C (top, right panel of Fig.~\ref{fycno}) and $^{14}$N (bottom, left panel) yields with mass, which is seen in the massive AGBs domain, is due to the occurrence
of dredge-out, which increases the overall C+N+O in the surface regions,
thus causing higher equilibrium abundances of the CNO species during the
AGB phase.

The $^{14}$N yields of the intermediate mass models presented here are
within the range of values of those of the $13-25~\rm M_{\odot}$ model stars by LC18. The $^{14}$N yields yields of the non-rotating models by LC18 are
$\sim 2$ orders of magnitude lower than those reported here, whereas
the yields corresponding to the rotating models, of the order of
$0.1~\rm M_{\odot}$, are a factor of $\sim 10$ higher. The same holds for
$^{13}$C, with the difference that the gap between the largest yields 
by LC18 and those presented here is reduced to a factor $\sim 2$.

The $^{16}$O yields can also be interpreted on the basis of the efficiency of
the physical processes able to change the surface chemistry, and are
consistent with the results shown in the bottom panels of Fig.~\ref{fno}. Some $^{16}$O production occurs in the low-mass domain, owing to the dredge-up of oxygen nuclei from the $3\alpha$ burning shell. The largest yields, slightly above $10^{-2}~\rm M_{\odot}$, are produced by $\sim 2~\rm M_{\odot}$ stars (see the cyan tracks in the bottom panels of Fig.~\ref{fno}). On the other hand, the $^{16}$O ejecta of massive AGBs are oxygen-poor, with the gas from $Z=3\times 10^{-7}$ stars being almost oxygen-free. As expected the
$^{16}$O yields, at most $\sim 0.01~\rm M_{\odot}$, are much lower than the corresponding yields of the $13-25~\rm M_{\odot}$ model stars by LC18, which
are on the average above $1~\rm M_{\odot}$.

\subsection{Sodium}
We now turn to the chemical species not involved in CNO cycling, starting
with sodium, whose yields are shown in the top, left panel of Fig.~\ref{fymgal}. The surface sodium is extremely sensitive to the 
relative importance of the different mechanisms able to modify the
surface chemical composition. The effects of HBB on the surface sodium
changes according to the temperature at the base of the envelope: when $T_{\rm bce}<70$ MK, sodium is produced via proton capture by $^{22}$Ne nuclei, whereas for higher temperatures sodium is destroyed, as the destruction channel via proton capture prevails \citep{mowlavi99}. 
This is the reason why the $^{23}$Na yields exhibit a positive trend with mass only in the $\rm M <3~\rm M_{\odot}$ mass domain, with the
maximum yield being slightly above $10^{-4}~\rm M_{\odot}$. In higher mass stars the temperatures at the base of the envelope are hotter than the 70 MK threshold given above (see bottom panels of Fig.~\ref{ftbce} and the
values of the largest temperatures reached, reported in col. 7 of table
\ref{tabgen}), thus the trend with mass is reversed and
eventually the sodium yields are negative. In all the model stars
with initial mass $M < 5~M_{\odot}$, which experience several TDU events,
most of the sodium is indeed of primary origin, as a large fraction of
the $^{22}$Ne converted into sodium is produced by a chain of reactions
taking place in the $3\alpha$ burning shell formed at the ignition of each thermal pulse. 
Compared to LC18, the sodium
yields presented here are systematically smaller than those corresponding
to the $12-25~\rm M_{\odot}$ stars: the largest yields of sodium, corresponding to $2.5~\rm M_{\odot}$ stars, are between 2 and $\sim 30$ times smaller than in LC18.

\subsection{The elements involved in the Mg-Al-Si chain}
\begin{figure*}
\begin{minipage}{0.48\textwidth}
\resizebox{1.\hsize}{!}{\includegraphics{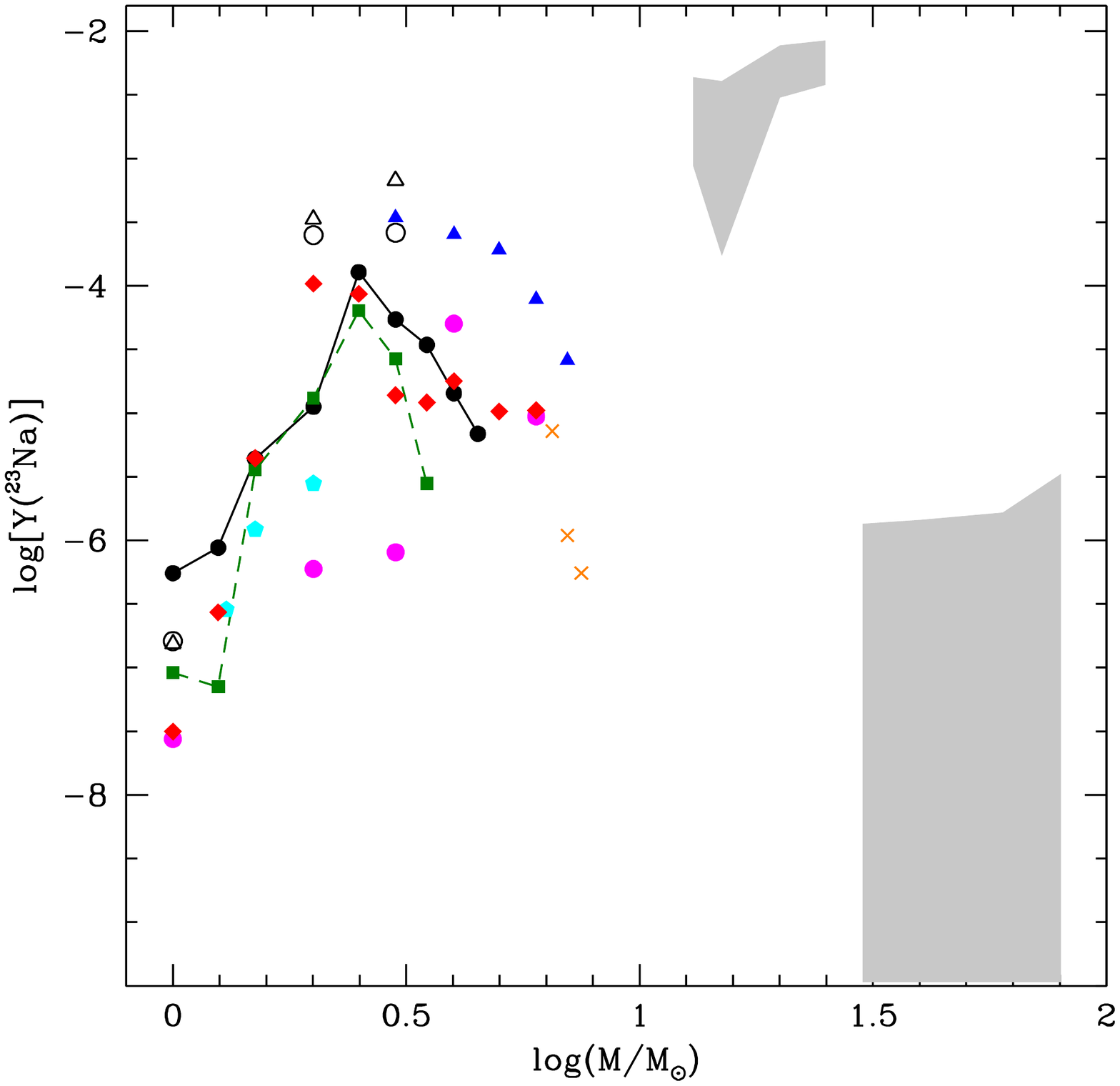}}
\end{minipage}
\begin{minipage}{0.48\textwidth}
\resizebox{1.\hsize}{!}{\includegraphics{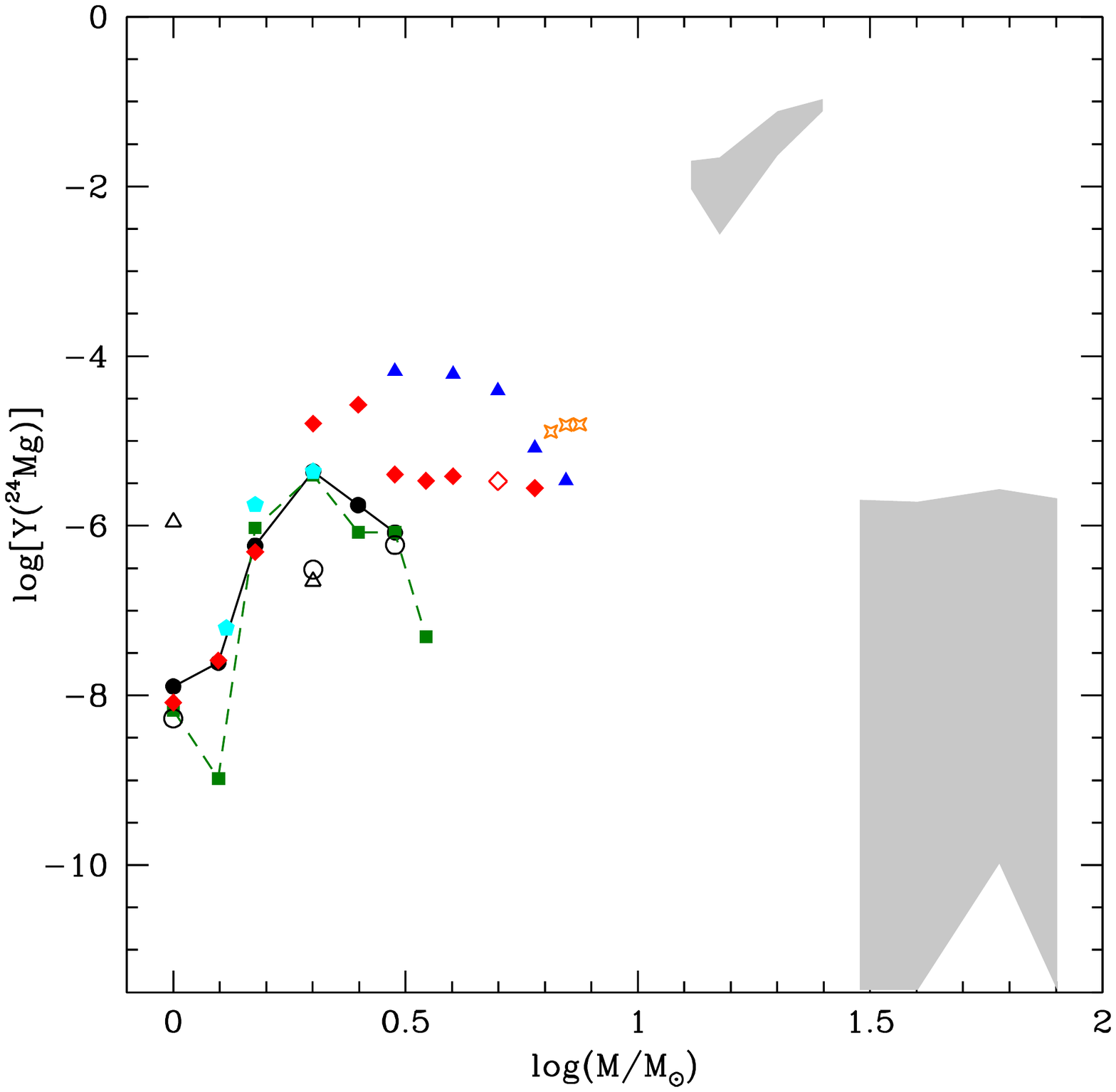}}
\end{minipage}
\vskip-70pt
\begin{minipage}{0.48\textwidth}
\resizebox{1.\hsize}{!}{\includegraphics{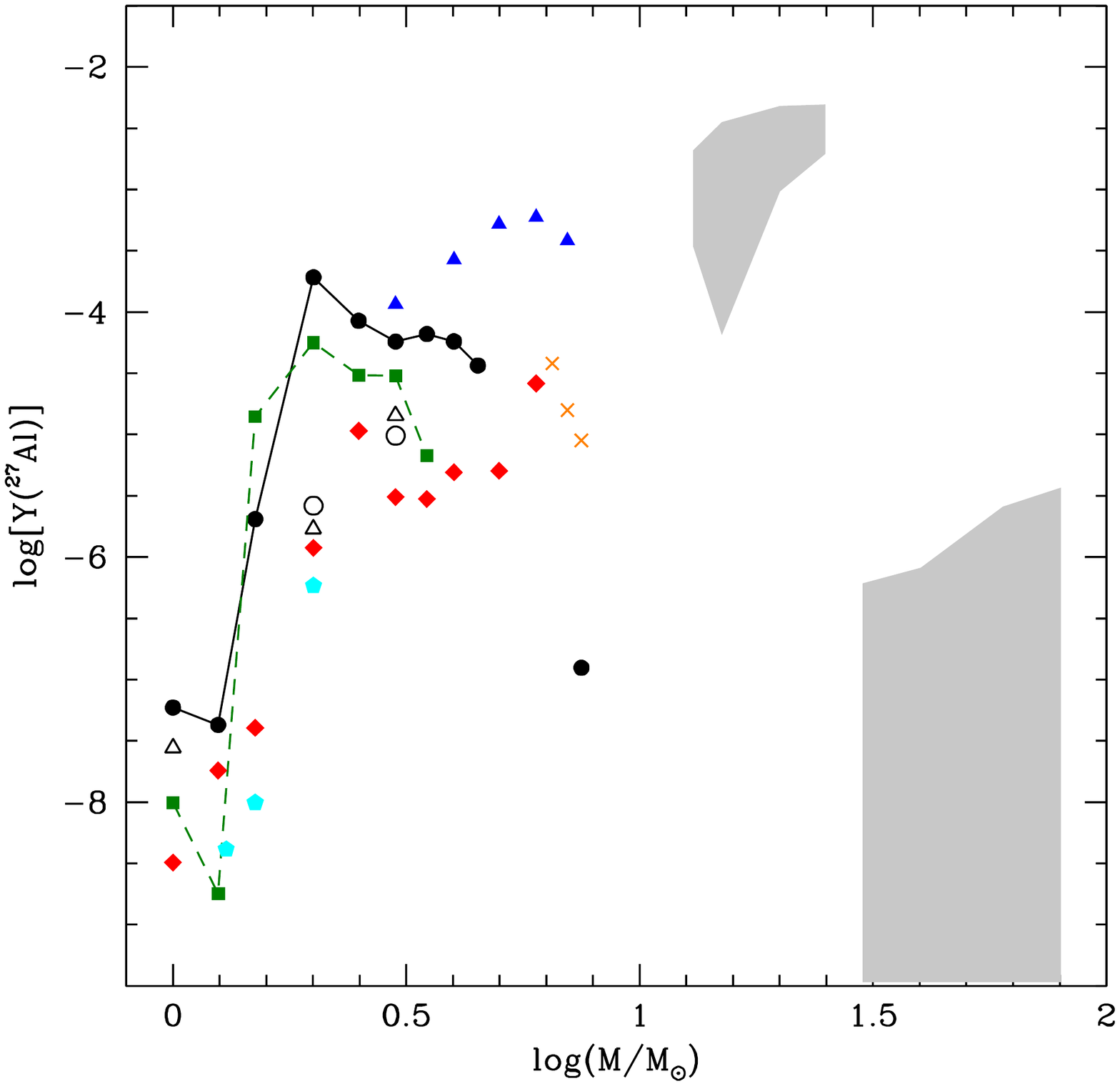}}
\end{minipage}
\begin{minipage}{0.48\textwidth}
\resizebox{1.\hsize}{!}{\includegraphics{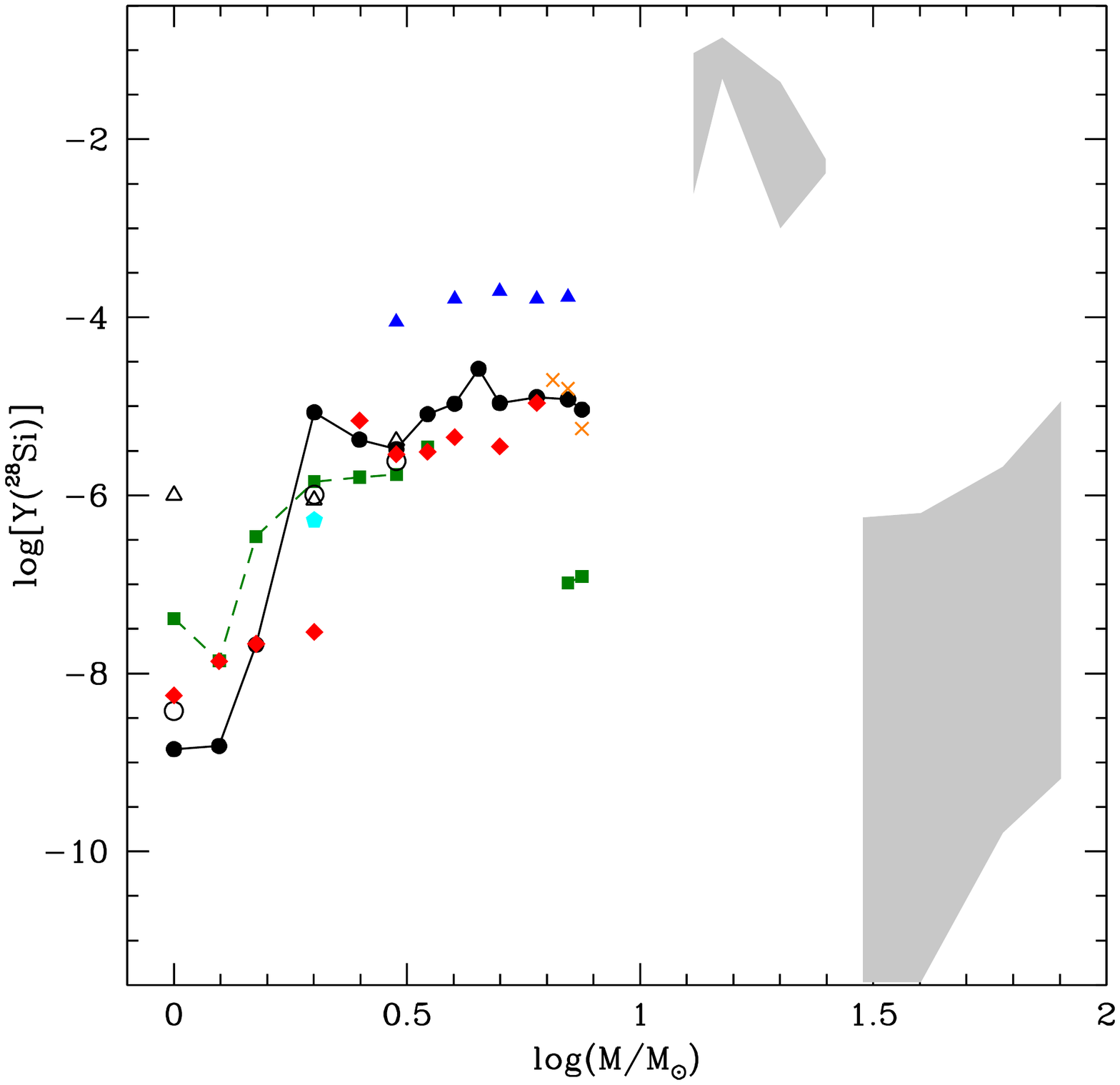}}
\end{minipage}
\vskip-50pt
\caption{Yields of sodium (top, left panel), $^{24}$Mg (top, right),
$^{27}$Al (bottom, left) and $^{28}$Si (bottom, right) as a function of
the initial mass of the star. The symbols and the colour coding adopted
is the same as in Fig.~\ref{fycno}.} 
\label{fymgal}
\end{figure*}

The very advanced degree of the nucleosynthesis experienced by intermediate and massive AGB stars in the extremely low metallicity domain investigated
here is even more evident in the behaviour of the yields of $^{24}$Mg,
$^{27}$Al and $^{28}$Si, shown in the top-right, bottom-left and
bottom-right panels of Fig.~\ref{fymgal}, respectively. For both 
$^{24}$Mg and $^{27}$Al the yields get negative in the $\rm M >3~\rm M_{\odot}$
mass domain, whereas for $^{28}$Si this behaviour is found in the lowest
metallicity models only. 

Regarding $^{24}$Mg and $^{28}$Si, the yields presented here are 
$\sim 3$ orders of magnitude smaller compared to the $12-25~\rm M_{\odot}$
stars by LC18. Conversely, the $^{27}$Al yields of $2-5~\rm M_{\odot}$
stars, of the order of $10^{-4}~\rm M_{\odot}$, fall inside the range
of values spanned by the LC18 models and are $\sim 20$ times smaller
than the largest values of the LC18 work.

\begin{figure}
\resizebox{\hsize}{!}{\includegraphics{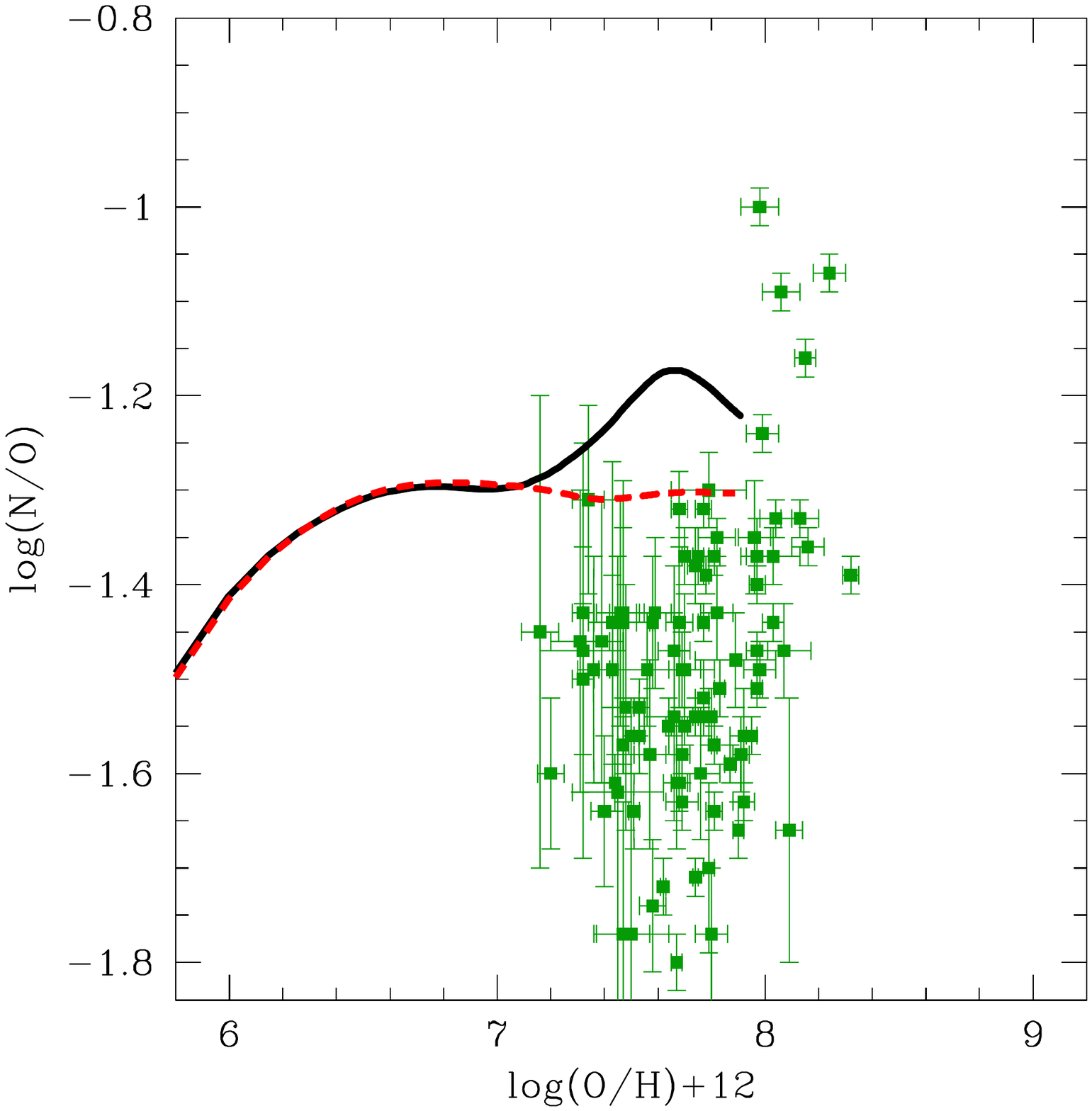}}
\vskip-60pt
\caption{
The evolution of the N/O abundance ratio in a dwarf galaxy with stellar mass $2 \times10^6$~M$_\odot$, modelled according to the method documented in \citet{romano13} and \citet{romano15} (see text for details on the various assumptions). Dashed, red line indicates the results obtained 
by \citet{romano19}, when the yields of very metal-poor low- and intermediate-mass stars were extrapolated from those corresponding to the higher metallicity counterparts ($Z = 3 \times 10^{-4}$). The black lines refers to the results obtained when the present yields are considered. Green squares with error bars are data for low-mass star-forming galaxies from \citet{berg12} and \citet{izotov12}.
}
\label{fyield}
\end{figure}

\subsection{The need for gas yields from extremely metal-poor low and intermediate mass stars}
As mentioned in the Introduction, detailed yields for extremely metal-poor stars like the ones presented in this work are especially useful for studying the chemical evolution of small galaxies. In fact, the metallicity of large galaxies such as the Milky Way increases fast and only a few LIMS are formed with Z$\le 10^{-5}$. Thus, their effects on the global galactic evolution are almost negligible. In small galaxies, on the other hand, the less efficient star formation, together with a galaxy wide IMF likely skewed in favour of low- and intermediate-mass stars, makes the extremely metal-poor low- and intermediate-mass stars have non-negligible effects on the evolution of particular elements. 

For example, in Fig.~\ref{fyield} we show the predicted evolution of the nitrogen-to-oxygen abundance ratio in the interstellar medium of a dwarf galaxy with a stellar mass of $\sim 2 \times 10^{6}$~M$_\odot$. The adopted chemical evolution model is an updated version of the one used in \citet{romano13} and \citet{romano15} to describe the evolution of the Sculptor dwarf spheroidal and B\"ootes~I ultrafaint dwarf, respectively. Here, it is not used to follow the evolution of a specific galaxy; rather, it is only used for illustration purposes. The modeled dwarf galaxy forms stars with a low efficiency ($\nu = 0.25$ Gyr$^{-1}$) according to a Kennicutt-Schmidt star formation law, $\psi(t) = \nu M_{\mathrm{gas}}^k(t)$, with $k = 1$ (see Romano et al. 2015, their section~3.1, for details). The dashed (red) line shows the results in the log(N/O)--log(O/H)+12 plane that are obtained assuming the same nucleosynthesis prescriptions of model MWG-11 of Romano et al. (2019), i.e., the yields of Ventura et al. (2013, 2020) for low- and intermediate-mass stars and those of LC18 for massive stars (that are assumed to rotate fast below [Fe/H]~$= -1$ dex and do not rotate at all above such metallicity threshold). The solid (black) line is the very same model, implementing the yields for AGB stars presented in this work for $Z < 0.0003$ ($Z = 0.0003$ was the lowest metallicity value considered by Ventura et al. 2013). It can be clearly seen that the inclusion of the new yields, specifically referring to the most metal-poor stars, produces sizeable effects on the model predictions, i.e., log(N/O) values $\sim$0.1 dex higher are expected in the interstellar medium at the end of the evolution. For illustration purposes, the green squares in Fig.~\ref{fyield} show the spread in log(N/O) data for dwarf irregular galaxies. It appears that part of the spread could be due to the presence/absence of a sizeable fraction of low- and intermediate-mass stars formed at the lowest metallicities, as a consequence of the different histories of mass assembly and star formation suffered by the different galaxies, that are likely to follow very different evolutionary paths (see, e.g., Tolstoy, Hill \& Tosi 2009).

Finally, it is important to stress that our considerations above are based on the outputs of galactic chemical evolution models that assume I.R.A. (instantaneous recycling approximation). Hydrodynamical simulations that take into account the inhomogeneities of the interstellar medium (see, e.g., Emerick et al. 2020) could come to different conclusions, especially regarding the impact of the yield sets presented in this work on the early evolution of the Galactic halo.

\section{The comparison with previous investigations}
\label{comp}
As discussed in the introduction the very metal-poor metallicity domain
investigated here has been treated very rarely in the literature. 
CL08 presented primordial to extremely metal-poor stars yields up to $3~\rm M_{\odot}$. Though limited in mass, at present this is the most extended exploration, in terms of metallicity range. I09 published AGB models with metallicity $[$Fe$/$H$]=-3$, in the $1-6M_{\odot}$ mass range. More recent results were presented by GP21, who studied $Z=10^{-5}$ model stars with mass in the $3-8~\rm M_{\odot}$ range, and 
in the FRUITY database developed by S. Cristallo, who included $Z=2\times 10^{-5}$
stars with initial mass $1.3,1.5, 2~M_{\odot}$. Turning to slightly higher metallicities, we consider also the works by Karakas (2010, hereinafter K10), who 
describe the evolution of $Z=10^{-4}$, $1-6~\rm M_{\odot}$ stars, and 
by Doherty et al. (2014, hereinfter D14), who investigate the evolutionary properties of $Z=10^{-4}$ stars undergoing the super-AGB evolution. The gas yields from the  works mentioned above are compared with the results obtained in the present investigation in Fig.~\ref{fycno} and \ref{fymgal}. For what concerns the results from CL08 we only show the results corresponding to the metallicities close
to those discussed in the present work, namely $[$Fe$/$H$]=-5.5$ and $[$Fe$/$H$]=-3$.

In the low-mass domain, below $3~\rm M_{\odot}$, we see that overall the present results are extremely similar to K10, except for a slight discrepancy in the $^{16}$O yields, which are higher in the present case. In the $1-1.5~M_{\odot}$ mass range the model stars by I09, CL08 and those in the FRUITY database follow a different behaviour, with the signatures of CNO cycling evident in the
large yields of $^{13}$C (see top, right panel of Fig.~\ref{fycno}) and
$^{14}$N (bottom, left panel of Fig.~\ref{fycno}); this is connected with the
occurrence of a few hydrogen ingestion episodes found in the latter works, which 
triggers a fast increase in the abundances of CNO elements in the early AGB
phases. \citet{ventura18, ventura20} found consistency between the ATON and MONASH results of low-mass stars of solar and super-solar chemical composition, thus suggesting that the treatment of
convective borders leads to similar results in the two cases. In the present analysis the
interpretation is more tricky: these results indicate that the extent of mixing, particularly 
of proton-ingestion episodes, are similar, but this could be partly due
to the fact that the metallicity of K10 models is close to the threshold $Z$ below which
proton-ingestion is expected to take place.

The comparison among AGB models of intermediate and high mass is generally more tricky than in the low-mass case, because the surface chemical enrichment is determined by the detailed interplay  between TDU and HBB. In the $3-7~\rm M_{\odot}$ mass domain we will confront our results preferably with those from GP21, because they are the most
recent, based on a metallicity close to the higher $Z$ stars discussed in this work. We will restrict the comparison to the GP21 models based on the \citet{vw93} prescription for mass-loss, the same adopted here.

By comparing the results reported in Table \ref{tabgen} with those in Table 1 of GP21 we note the following: a) the core masses at the beginning of
the AGB phase are rather similar, with practically no difference for $3~\rm M_{\odot}$ and differences within $0.05~\rm M_{\odot}$ for $\rm M  \sim 6-7~\rm M_{\odot}$; b) significant differences characterize the temperatures at the bottom of the convective envelope, which in the present models are between $20$ Mk (for $3~\rm M_{\odot}$ stars) and $50$ Mk (for $7~\rm M_{\odot}$) hotter than in GP21; c) the TDU is very efficient in the models by GP21, characterized by $\lambda\sim 0.8-1$, whereas in the present case we find that the TDU is generally less efficient in this mass domain and tends to vanish for $\rm M \geq 5~\rm M_{\odot}$; d) as a consequence
of the larger luminosities and $T_{\rm bce}$, the model stars presented
here evolve faster than in GP21, with a factor $\sim 3$ difference in the duration of the TP-AGB phase, as deduced from the comparison between the times reported in col. 5 of table \ref{tabgen} and those in col. 2 of table 1 of GP21.

From point (a) above we deduce that the pre-AGB evolution of the model stars in the two different sets is rather similar, and that the difference in the initial core mass in the most massive stars considered is likely due to a slightly more efficient core overshooting during the main sequence phase.

The differences between the temperatures at the base of the envelope (point b) are remarkable, and are mainly connected to convection modelling, which is based on the FST treatment in the present models, whereas GP21 use the classic mixing length theory (MLT) description; differences in the
assumption regarding extra-mixing from the base of the convective envelope might also play a role
here. The strong impact of convection modelling on the evolution of massive AGB stars was discussed in a seminal paper by \citet{ventura05a} and was extensively explored in more recent investigations, where results from evolution codes differing in the treatment of the convective instability were compared in detail \citep{ventura16, ventura18}. On the other hand the present findings indicate that the discrepancies between models calculated with different convection models become more and more important as the metallicity decreases, with the FST models experiencing a much stronger HBB with respect to the MLT ones; we will see that this has dramatic consequences on the change in the surface chemistry, as the nucleosynthesis experienced by FST models at the bottom of the external envelope is far more advanced than in MLT-based calculations.

The results given in point (c) above indicates a substantial difference in the depth of TDU, which is partly related to the diversity in the treatment of the convective borders and partly to the point (b) above. In the present modelling we adopt an exponential decay of velocities from the neutrality point to the radiatively stable region to simulate overshoot, whereas GP21, K10 and D14 models, whose yields are shown in Fig.~\ref{fycno} and \ref{fymgal}, use the neutrality criterion described in \citet{frost96}. The different approach in managing the inner border of the envelope during the TDU events might per se explain part of the differences found. On the other hand, the stronger HBB found in the present computations is also playing a role in this context, as model stars exposed to efficient HBB experience weaker TP, which decreases the TDU efficiency. Whatever is the main actor in the determination of the TDU efficiency, the higher $\lambda$ models are exposed to a larger surface metal enrichment, with the synthesis of primary elements, originating from the matter dredged-up from the helium burning shell. In the models discussed here such enrichment, of lower extent, is limited to $\rm M \leq 5~\rm M_{\odot}$ stars, being negligible in more massive objects.

In terms of the gas yields, important differences are seen in Fig.~\ref{fycno} and concern all the CNO species. The $^{12}$C yields by GP21 and K10 are a factor $\sim 10$ higher than ours in the $2.5-5~\rm M_{\odot}$ mass range, whereas for higher mass stars the situation is even more dramatic, with the present yields turning negative, while in the other cases they keep similar to the slightly lower mass counterparts, i.e. $\sim 0.003~\rm M_{\odot}$.
Significant differences are found also for $^{14}$N (see top, right panel of Fig.~\ref{fycno}), with the GP21 yields of $2.5-5~\rm M_{\odot}$ stars being almost a factor 10 above ours, the difference increasing to 2 orders of magnitude in the massive stars regime. The models by K10 and D14 exhibit smaller differences with ours with respect to GP21, their $^{14}$N yields being approximately a factor 2 smaller than GP21.

The interpretation of the $^{16}$O yields is more cumbersome. In the intermediate mass case the yields presented here are similar to those by K10,
whereas those from GP21 are slightly higher. In the case of massive
stars the context is similar to $^{12}$C, with our yields turning negative, whereas the yields from GP21, K10 and D14 keep positive.

The differences among the yields from the various groups just discussed can be explained based on the evolutionary properties of the models presented here and those from GP21, K10 and D14, discussed earlier in this section.
The analysis of the CNO yields allows understanding the role played by the different physical mechanisms on the variation of the surface chemical composition. In this regard, the results relating to $^{14}$N are emblematic.
The yields from GP21 and, to a lesser extent, K10 and D14 too, are generally higher, a signature of the higher TDU efficiency; this conclusion holds independently of the strength of HBB, because the activation of the CNO nucleosynthesis naturally produces an enrichment in $^{14}$N.

\begin{figure*}
\begin{minipage}{0.48\textwidth}
\resizebox{1.\hsize}{!}{\includegraphics{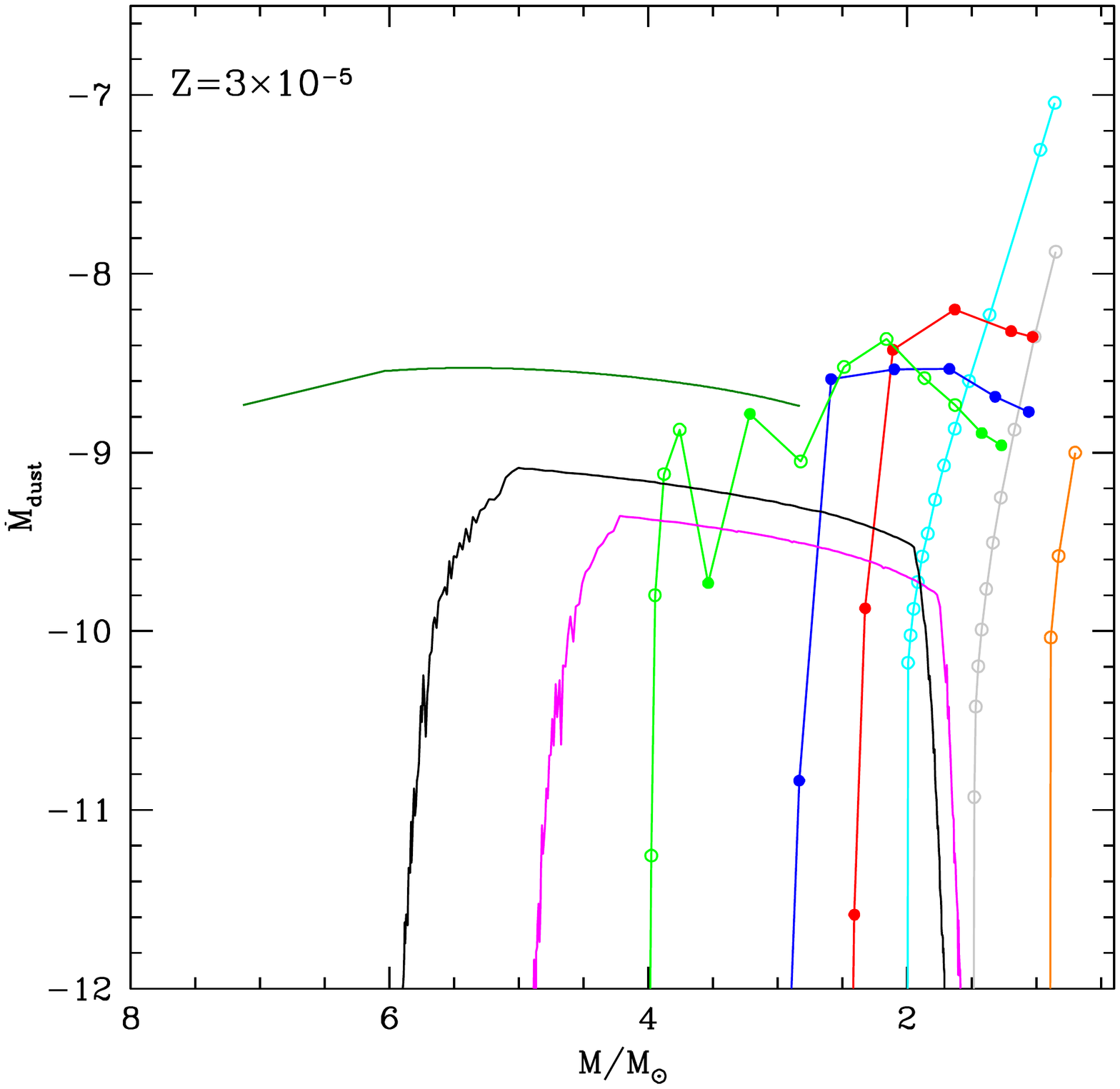}}
\end{minipage}
\begin{minipage}{0.48\textwidth}
\resizebox{1.\hsize}{!}{\includegraphics{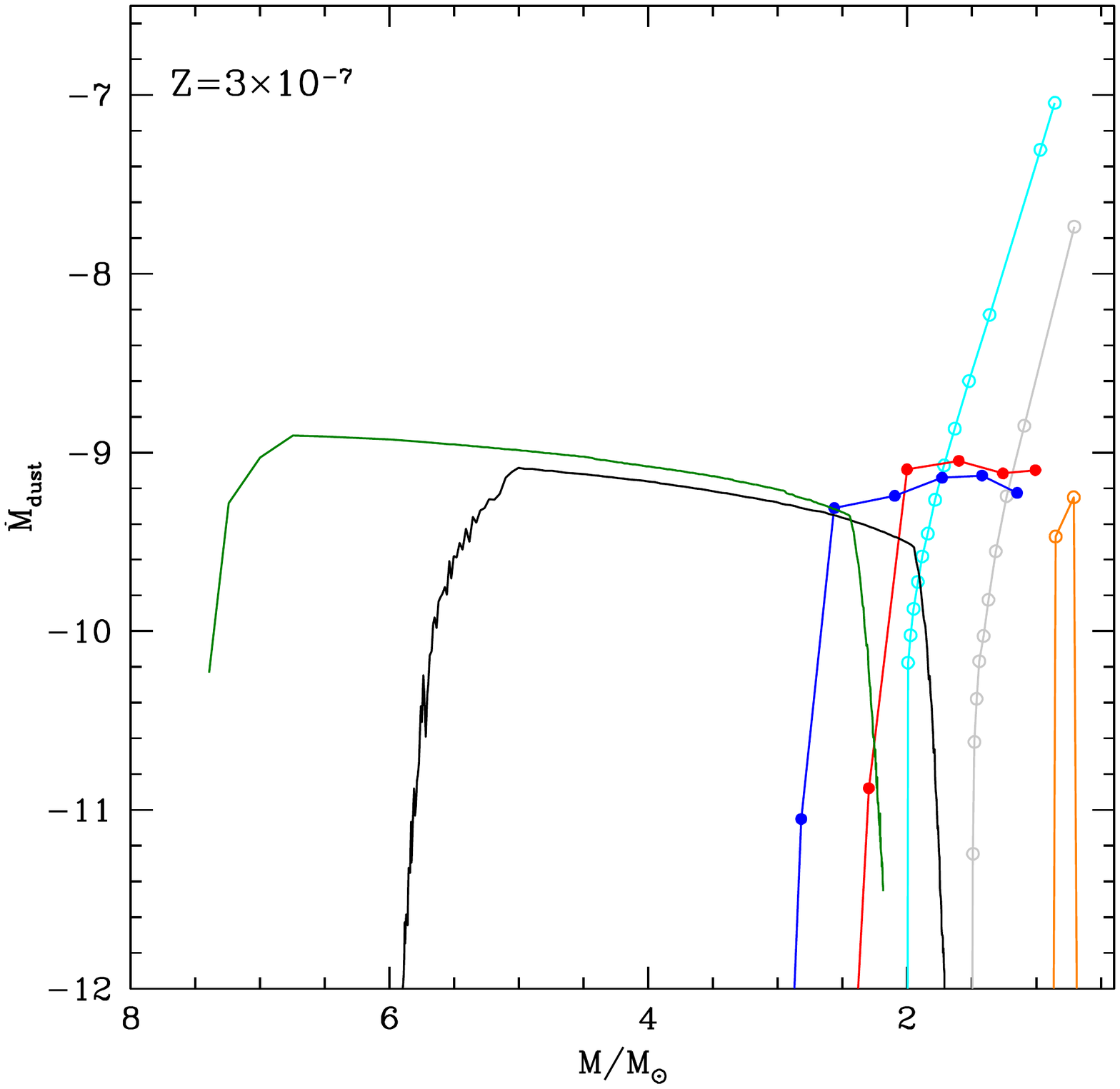}}
\end{minipage}
\vskip-60pt
\caption{The variation of the dust production rate as a function of the
(current) mass of the star for some of the model stars discussed in
Fig.~\ref{ftbce}, for the metallicity $Z=3\times 10^{-5}$ (left panel)
and $Z=3\times 10^{-7}$ (right panel). The colour coding is the same as
in Fig.~\ref{ftbce}. Full points correspond to silicate dust, whereas
open points indicate carbonaceous dust.}
\label{fdust}
\end{figure*}

The analysis of the yields of $^{12}$C and $^{16}$O confirm the much stronger HBB experienced by the present models with respect to GP21, K10 and D14. This is particularly evident in massive AGBs, where the destruction of $^{12}$C and $^{16}$O is so strong that the present yields are negative, unlike those from the other groups. Note that the present models of massive AGB stars represent the only case where it is possible to produce ejecta which are substantially oxygen-free.

Fig.~\ref{fycno} also shows the CNO yields of
$4~M_{\odot}$ and $6~M_{\odot}$ model stars by I09. In this
case the comparison and the understanding of the differences among the results is
more cumbersome than in the case of GP21, because of the several differences in
the physical ingredients adopted, which involve convection, mass-loss modelling,
low-T molecular opacities and the mixing scheme. The CN yields from I09
are intermediate between the present results and those by
GP21, which allows us to infer that the HBB experienced by the present models
is stronger than in I09 (this is expected, as the latter models adopt the MLT, whereas we use the FST) and that the extent of extra-mixing from the 
formal borders of the convective regions is larger in the I09 models.

Turning to heavier species, the activation of the Ne-Na and
Mg-Al-Si chains confirm this understanding. The activation of the Ne-Na nucleosynthesis takes place in all the model stars with initial mass
$\rm M \geq 3~\rm M_{\odot}$, as witnessed by the positive sodium yields of the intermediate mass AGBs, shown in the top, left panel of Fig.~\ref{fymgal}, which span the $10^{-5}-10^{-4}~\rm M_{\odot}$ range, in our case as also in K10. The GP21 sodium yields are substantially higher, a further evidence of a deeper TDU, which brings to the surface primary $^{22}$Ne, converted into sodium during the following inter-pulse phase. In the massive AGB domain the differences are more remarkable, with the present sodium yields turning into negative values, because the HBB experienced is so strong that the equilibria of the Ne-Na nucleosynthesis are shifted to lower sodium abundances \citep{ventura09}. Note that the poor TDU experienced by the massive AGBs presented here also plays a role in this context, as there is no transportation of $^{22}$Ne, the sodium reservoir, to the surface regions.

In metal-poor AGB stars, descending from $\rm M \geq 3~\rm M_{\odot}$ progenitors,  $^{24}$Mg is destroyed by HBB in the surface regions; on the other hand $^{24}$Mg is transported to the external layers via TDU. Therefore in intermediate mass and massive AGBs the evolution of the surface $^{24}$Mg, similarly to $^{12}$C, is determined by the balance between these two mechanisms. In light of this, and on the basis of the differences in the efficiencies of HBB and TDU discussed earlier in this section, it is not surprising that the models presented here are those characterized by the lowest $^{24}$Mg yields, which are negative in the $\rm M \geq 3~\rm M_{\odot}$ mass range, a further confirmation that HBB plays the dominant role. Conversely,
the $^{24}$Mg yields by GP21, K10 and D14 are positive for the masses considered here, although with significant differences among them: 
the GP21 $^{24}$Mg yields are on the average a factor 10 higher than K10,
a further indication that the models by GP21 are those in which the relative weight of TDU on the modification of the surface chemistry is the largest.

We end up the analysis of the differences in the gas yields of the various species with $^{27}$Al (see bottom, left panel of Fig.~\ref{fymgal}).
For $\rm M <5~\rm M_{\odot}$ stars the present $Z=3\times 10^{-5}$ $^{27}$Al
yields are higher than K10, owing to the stronger HBB conditions, which
favour a very advanced Mg-Al nucleosynthesis. The GP21 yields are once
more the highest, owing to deeper TDU events, which dredge-up to the 
surface $^{25}$Mg and $^{26}$Mg nuclei synthsized in the helium burning
shell, then converted into $^{27}$Al by HBB. In the case of massive AGBs the
situation is similar to $^{23}$Na and $^{24}$Mg: the present yields are
negative, because they experience such an advanced HBB nucleosynthesis 
that $^{27}$Al is destroyed at the base of the envelope,
whereas the yields from GP21, K10 and D14 are positive, with the
GP21 $^{27}$Al yields being approximately a factor 10 higher than in
K10 and D14.

The analysis presented in this section can be indirectly used to assess the degree of uncertainty of the results obtained from AGB modelling. The dissimilarities between
results published by different groups, specifically between those from
our group and those obtained by the MONASH code for stellar evolution, were extensively discussed in the case of solar metallicity models by \citet{ventura18} and for super-solar chemistries by \citet{ventura20}. These works have highlighted
some differences, particularly in the strength of HBB experienced by $\rm M >3~\rm M_{\odot}$ stars, but overall the discrepancies in the yields were within $\sim 20\%$,
with the sole exception of the stars in the (narrow) mass range close
to the threshold to activate HBB. Conversely, Dell'Agli et al. (2019, hereinafter D19), in the comparison between $Z=1,3\times 10^{-4}$ models and those by K10,
outlined significant dissimilarities, especially with regard to $^{14}$N, which were ascribed to a greater contribution from primary material in the K10 models, in turn associated with deeper TDU events. In the present work we find that the differences are even more pronounced than in D19, with the yields of some species found by different groups, such as $^{12}$C, $^{16}$O, $^{24}$Mg and $^{27}$Al, being also of opposite sign. 

The uncertainties characterizing the macro-physics adopted, particularly
the convective instability, have a much more relevant effect at extremely
low metallicities. This is partly due to the intrinsic property of metal-poor AGB stars, that experience stronger HBB when compared to the higher metallicity counterparts: this evidently widens the differences between results based on different treatments of the convective instability. But there is also another characteristic of low metallicity stars, which makes their structure and evolution particularly sensitive to the efficiency of the mechanisms able to alter their surface chemical composition. These stars form from gas which is almost metal-free, thus the percentage change in the surface mass fraction of a given species, associated e.g. to the effects of TDU, is much larger than in the more metal-rich counterparts. While in metal-rich stars dissimilarities in the amount of matter
dredged-out to the surface will cause small differences in the increase of the surface metallicity, in the extremely metal-poor domain the differences in the metal enrichment and therefore in the evolutionary properties, can be considerable.

\section{A binary origin for CEMP stars}
\label{cemp}

\begin{figure*}
\begin{minipage}{0.48\textwidth}
\resizebox{1.\hsize}{!}{\includegraphics{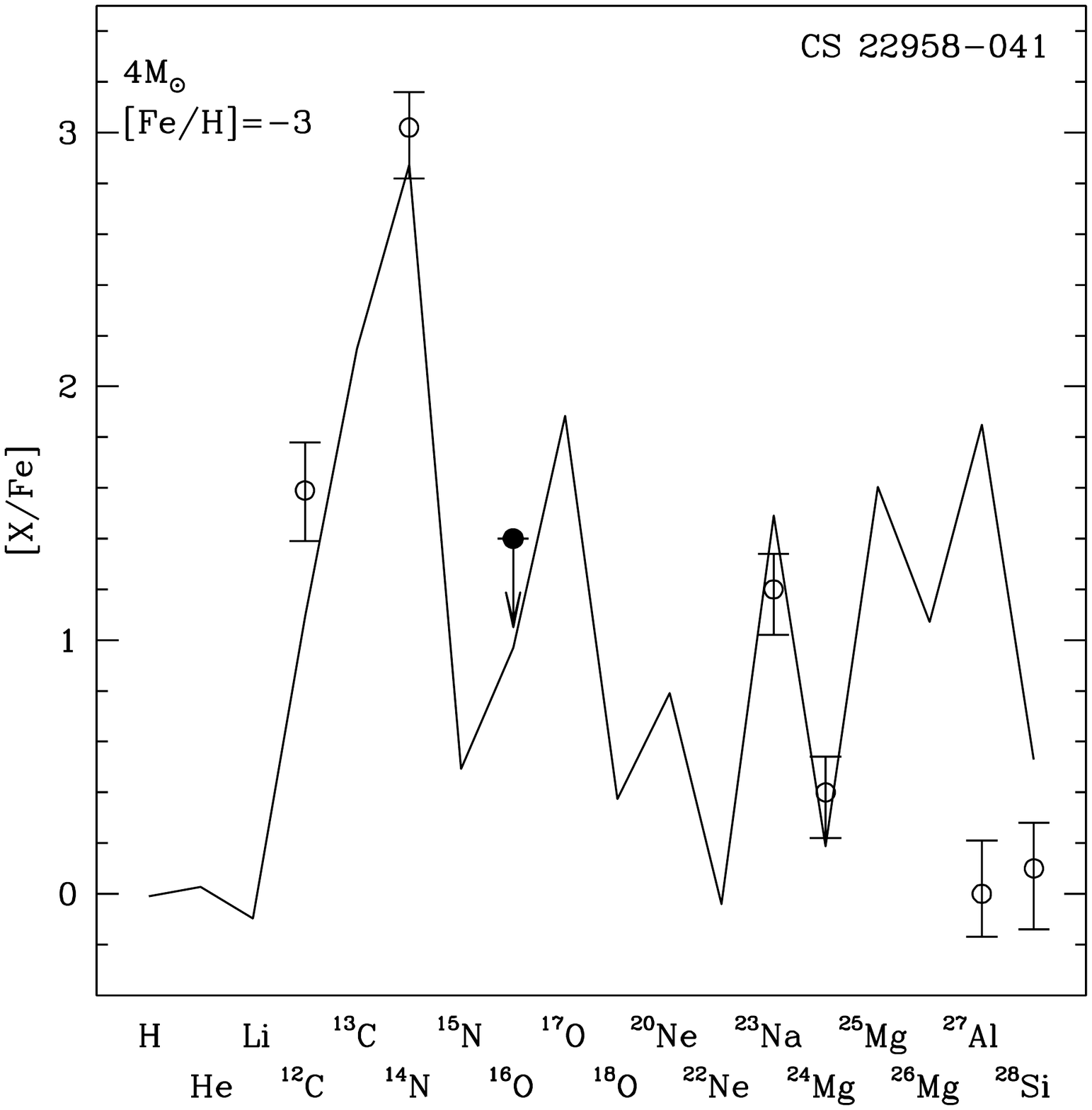}}
\end{minipage}
\begin{minipage}{0.48\textwidth}
\resizebox{1.\hsize}{!}{\includegraphics{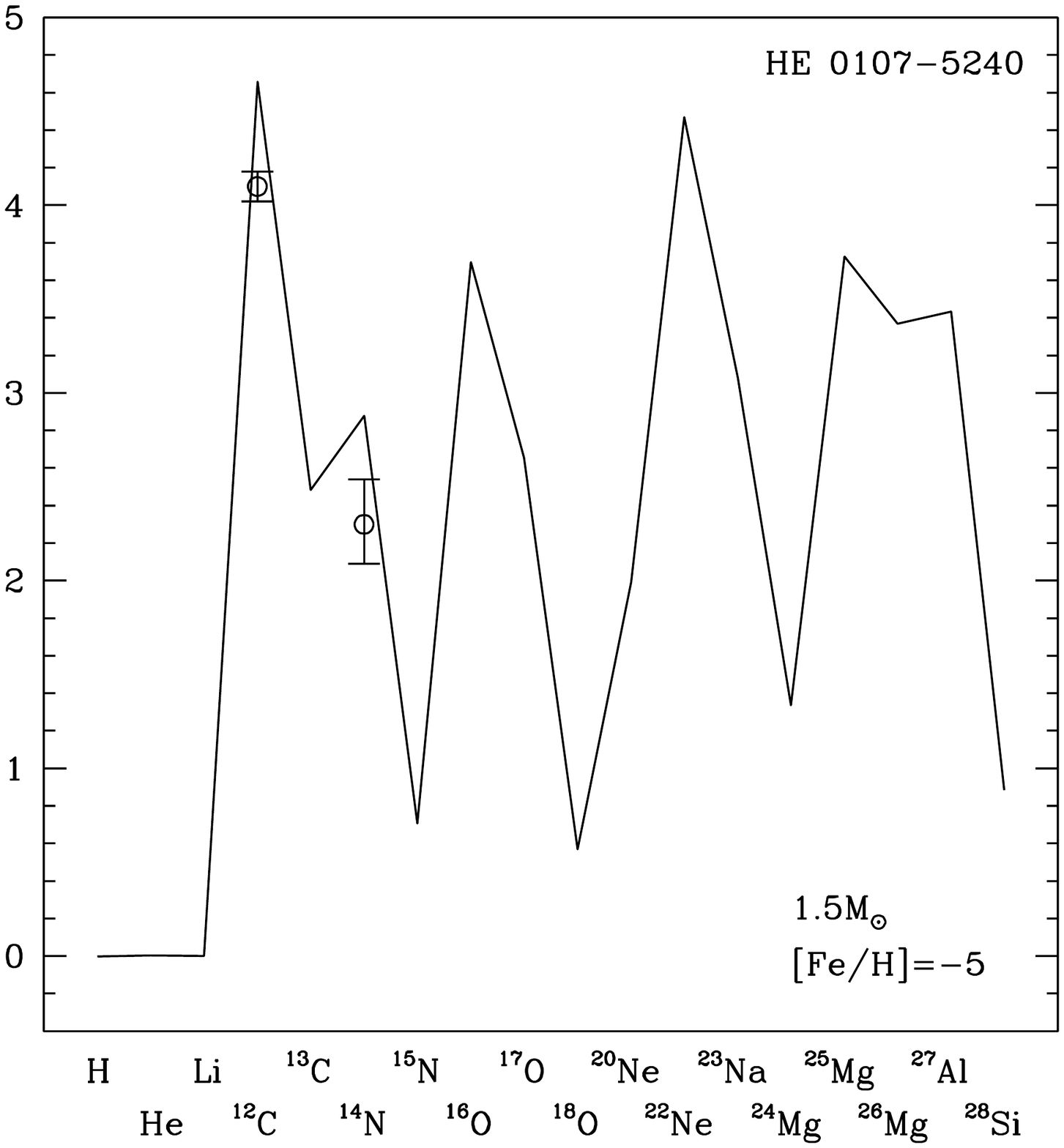}}
\end{minipage}
\vskip-3cm
\begin{minipage}{0.48\textwidth}
\resizebox{1.\hsize}{!}{\includegraphics{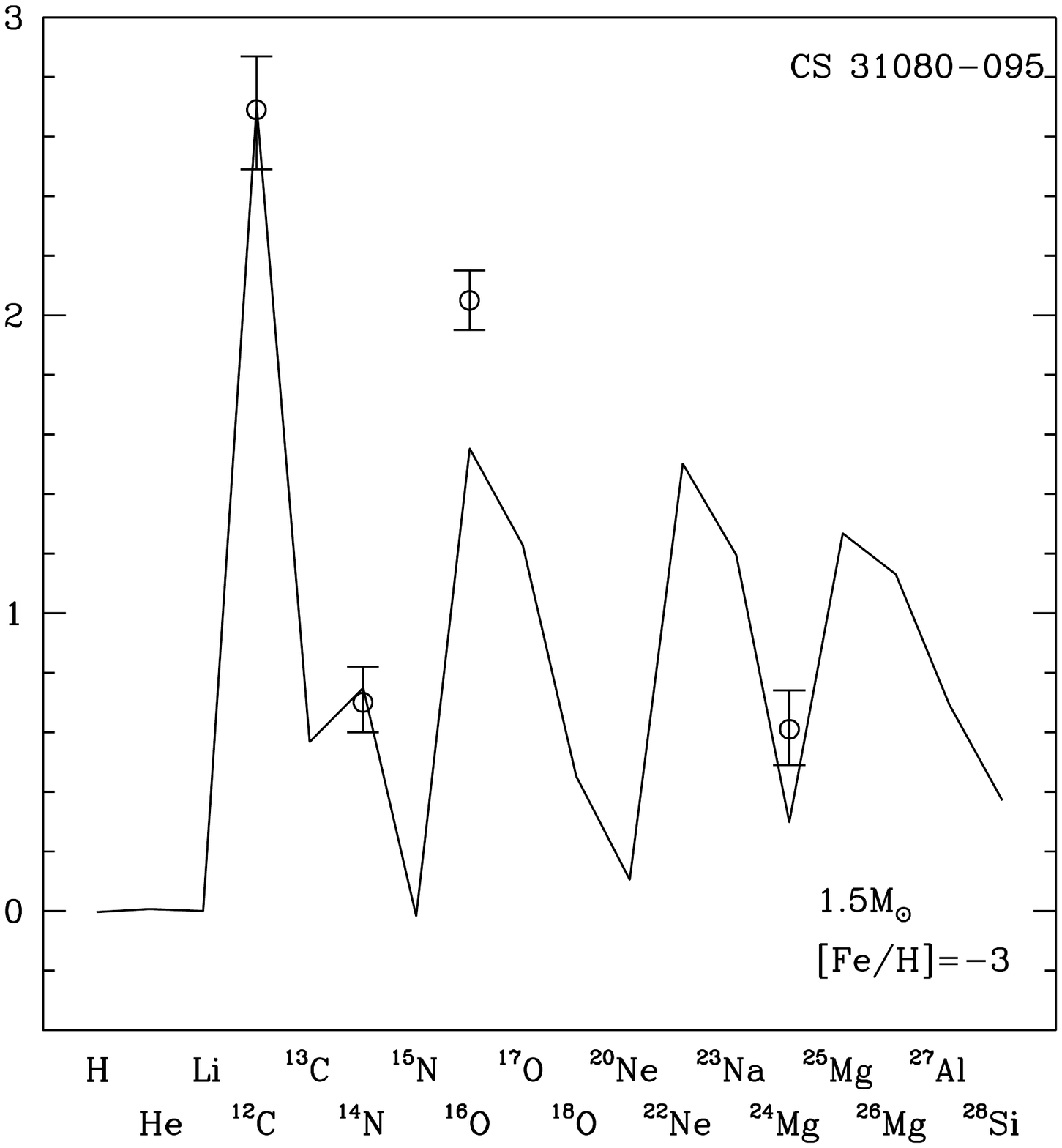}}
\end{minipage}
\begin{minipage}{0.48\textwidth}
\resizebox{1.\hsize}{!}{\includegraphics{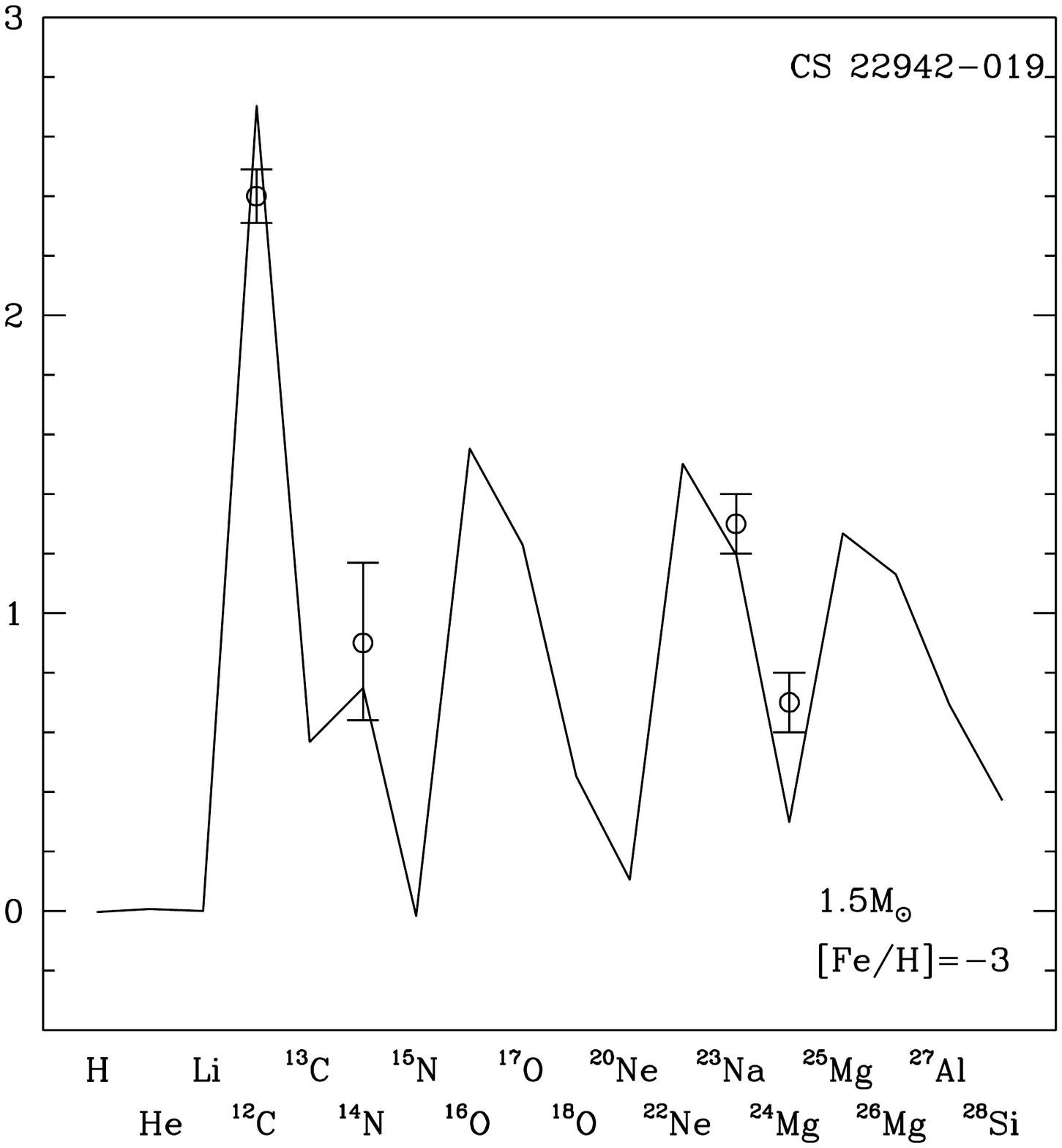}}
\end{minipage}
\vskip-30pt
\caption{
Solid curves: abundance patterns predicted at the surface of unevolved low-mass stars after mixing $1\%$ of the matter ejected by our AGB models to their $0.2~M_{\odot}$ envelopes (see GP21). The AGB models are identified by their initial mass and metallicity, as reported in the bottom right/top left corner of each panel. The circles show the abundances measured in extremely metal-poor stars selected from the SAGA database (see top right label in each plot).
}
\label{fcemp}
\end{figure*}

CEMP stars are low-metallicity ($[$Fe$/$H$] < -2$) dwarfs and giants with substantial carbon enrichment ($[$C$/$Fe$] > 1$; Beers \& Christlieb 2005). A significant fraction ($\sim 80\%$) of 
these objects, known as CEMP-s stars,  show over abundance of elements produced from slow neutron-capture process (s-process). Multi-epoch observations of CEMP-s stars reported that the
majority of them are members of binary systems \citep{lucatello05}, which favours the idea of mass transfer from a companion AGB star which is more massive compared to the primary, and therefore would have evolved to a cool white dwarf \citep{abate15}. The CNO abundances observed and the 
s-process element distribution can be interpreted by assuming pollution from a low-mass, low-metallicity AGB star (see e.g. Cristallo et al. 2009, 2016, and the recent work by Susmitha et 
al. 2021, and references therein).

\citet{suda04} proposed a binary scenario for the formation and evolution of
CEMP stars, according to which CEMP stars belong to binary systems with
a long enough orbital period that accrete part of the material expelled
from a primary companion evolving though the TP-AGB phase, without experiencing
any common envelope evolution.

The debate on whether the accreted material from the donor can be mixed through most of the secondary star or rather is shared only with the outermost layers of the latter is still open \citep{stancliffe07, aoki08}. The analysis of the different factors affecting the depth of the innermost layers reached by accreted material in a low-mass mass star is beyond the scope of the present work.
In the present context we are simply interested to understand whether the
results presented here are consistent with the chemistry of CEMP stars of
similar metallicity. The approach followed is similar to GP21:
we consider dilution of matter with the average chemistry of the AGB ejecta with gas characterized by the original chemical composition, and compare 
with the derived abundances of a few selected CEMP stars belonging to the SAGA
database \citep{suda08}.

Fig.~\ref{fcemp} shows the above comparison for the selected CEMP stars,
which have approximately the same metallicity of the model stars presented  here. We overall find a satisfactory agreement, with three out of the 4 selected stars showing up the imprinting of a dominant contribution from TDU to the AGB chemistry, whereas in one case, namely CS 22958-041, it is clear in the large nitrogen enhancement the combination of HBB and TDU.

\begin{figure}
\resizebox{\hsize}{!}{\includegraphics{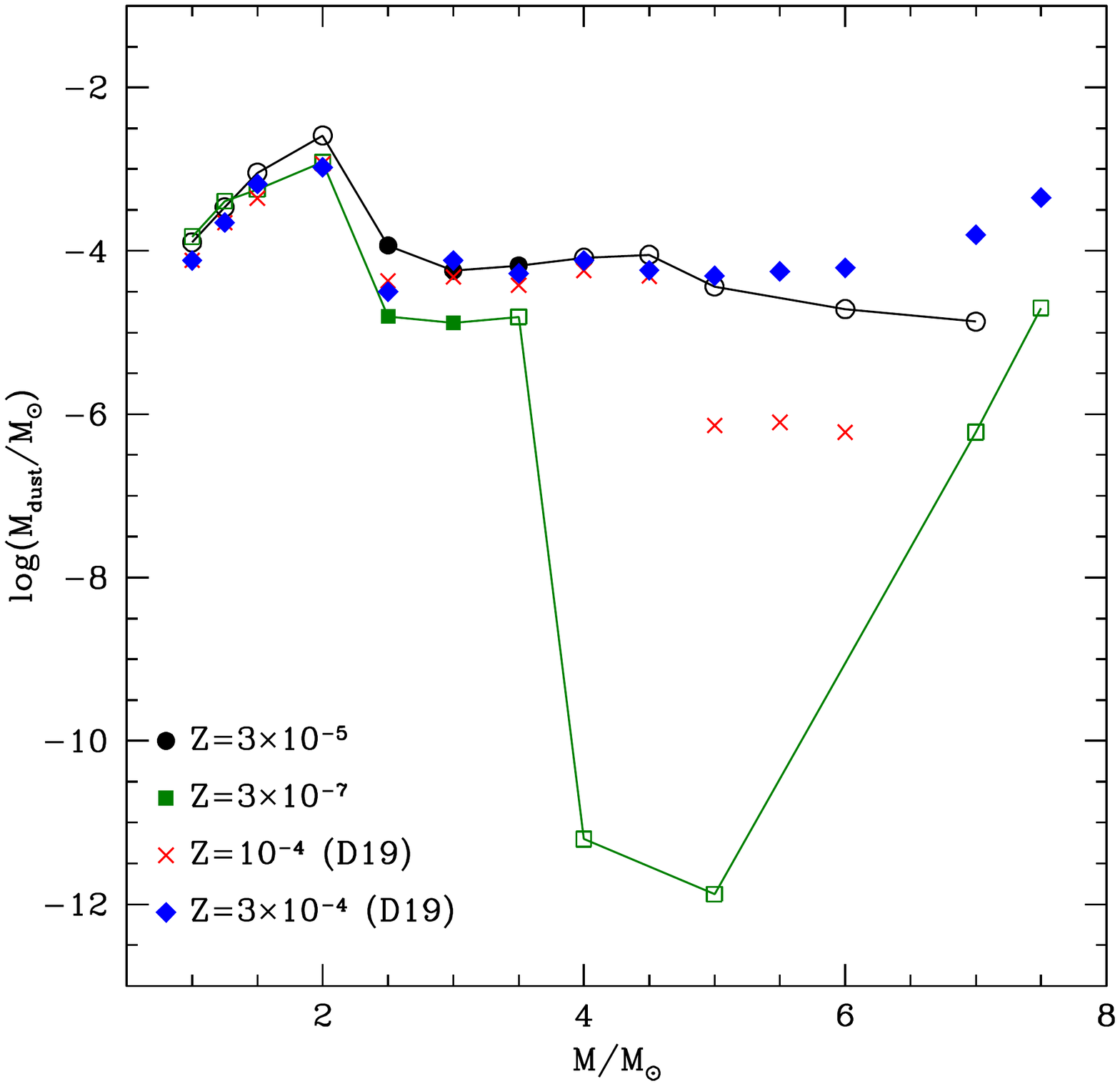}}
\vskip-60pt
\caption{Dust yields from the model stars of metallicity $Z=3\times 10^{-5}$
and $Z=3\times 10^{-5}$. Results from D19 are also shown.
}
\label{fdtot}
\end{figure}

\section{Dust production}
\label{dust}
AGB stars are potentially important dust manufacturers, given the thermodynamic conditions of their wind, with are sufficiently cool and dense to allow significant condensation of gaseous molecules into solid grains \citep{fg06}.
The most relevant dust species formed are silicates and carbonaceous dust.
The former species are formed in oxygen-rich environments, whereas the
latter kind of dust, composed by a majority of solid carbon grains with
some contribution from SiC, forms in the wind of carbon stars, and is
sensitive to the carbon excess (C-O), namely the excess of carbon molecules with
respect to oxygen in the surface layers. The formation 
of dust in metal-poor AGB was recently addressed by D19, who explored the metallicities $Z=1,3\times 10^{-4}$.
In this work we extend the analysis by D19 to more metal-poor environments.

Fig.~\ref{fdust} shows the variation of the dust production rate (DPR),
defined as the rate, in solar masses per year, at which dust is released into the interstellar medium by the star. The different lines refer to the same
model stars presented in Fig.~\ref{ftbce}, \ref{fno} and \ref{falsi},
with the $Z=3\times 10^{-5}$ case in the left panel and the $Z=3\times 10^{-7}$ one on the right\footnote{The evolutionary sequences of all the
model stars discussed in the present investigation, in terms of the
dust production rate from the main dust species and other quantities
connected to dust production, are available at www.oa-roma.inaf.it/arca/,
together with the table of the dust yields.}. 

\subsection{Evolution of the dust production rate during the AGB phase}
AGB stars with masses below $2~ M_{\odot}$, after becoming C-stars, produce carbonaceous dust, mainly
under the form of solid carbon; the contribution from SiC is negligible,
considering the low quantity of silicon in the surface layers.
In these stars the DPR increases during the whole AGB phase, as
clear in the trend defined by the orange, grey and cyan lines in Fig.~\ref{fdust}. The behaviour of the DPR is connected with the gradual accumulation of $^{12}$C in the surface regions of the star owing to the TDU episodes: this enhances dust formation, because of the higher number
of carbon molecules available to condensation and for the consequent
expansion and cooling of the external regions of the star, which partly
inhibits vaporisation. The largest DPR, $\sim 10^{-7}~\rm M_{\odot}/$yr, are reached by $2~\rm M_{\odot}$ stars, as they experience the largest number of
TDU events, thus accumulating higher quantities of $^{12}$C at the surface.

$2.5-4~\rm M_{\odot}$ stars evolve as M-type stars for most of their life
(see bottom, left panel of Fig.~\ref{f40m}). Among the wide metallicity range explored so far, this is the first case where the main dust component formed during the O-rich phase is alumina dust, while silicates form in almost negligible quantities. The reasons for this peculiar behaviour are: a) alumina dust is a stable compound, forming closer to the surface of the star
than silicates, in high density regions of the wind; b) the surface abundance of Al, the key-element for alumina dust, is largely enhanced in this extremely metal-poor regimes (see Fig.~\ref{falsi}); c) the carbon nuclei carried to the
surface regions by TDU episodes lock the majority of oxygen nuclei in CO molecules, leaving only a small fraction of oxygen free for dust condensation. In these conditions the condensation of alumina dust in the inner region of the wind absorbs a large fraction of the oxygen available ($\delta_{Al_2O_3}\sim 0.014$ is the condensation efficiency of oxygen bound in $Al_2O_3$). The DPR reached by these stars, generally below $10^{-8}~\rm M_{\odot}/$yr, are significantly lower than in $1.5-2~\rm M_{\odot}$ stars, because of the much smaller amounts of aluminum and silicon with respect to carbon in the surface layers. Metallicity plays a role here, with the $Z=3\times 10^{-5}$ DPRs being almost a factor of 10 larger than their $Z=3\times 10^{-7}$ counterparts; this is due to the quantities of aluminium  available to form dust, which scale with $Z$.

Massive AGBs (initial mass above $5~\rm M_{\odot}$) evolve as carbon stars for
most of the AGB lifetime, owing to the strong effects of HBB, which destroy
the surface $^{16}$O; an example of this behaviour is shown in 
Fig.~\ref{f60m}, particularly in the bottom, left panel, which reports
the evolution of the CNO species. The C$/$O$>1$ condition leads to the
formation of carbonaceous dust, which accounts for almost $100\%$ of the
overall dust formation by these stars. At $Z=3\times 10^{-5}$ this is mainly under the form of SiC, which is the most stable compound formed in C-rich winds. Even in this case, the carbon atoms free for condensation are almost entirely consumed in the SiC grains, to the detriment of solid carbon which forms in an outer region. In the lower metallicity case, the paucity of Si and its destruction through p-capture nucleosynthesis reduce the amount of Si available for dust formation. The DPR reached, in all cases
below a few $10^{-9}~\rm M_{\odot}/$yr, are significantly smaller than in the
lower mass counterparts, mainly because of the low carbon excesses reached,
favoured by the strong HBB activity, which depletes the surface $^{12}$C
since the early AGB phases (see Fig.~\ref{f60m}). 

\subsection{Dust yields from extremely metal-poor AGB stars}
A summary of the results obtained, in terms of the dust yields produced by stars of different mass (reported on the abscissa), is shown in Fig.~\ref{fdtot}. In the same figure the results published in D19, which refer to metallicities $Z=1,3\times 10^{-4}$, are also shown.

Most of the dust, under the form of solid carbon, is produced by low-mass AGBs. The largest dust production is provided by $2~\rm M_{\odot}$ stars,
which pour a quantity of dust slightly below $0.01~\rm M_{\odot}$ into the
interstellar medium. For $\rm M \leq 2~\rm M_{\odot}$ stars the results turn out
to be independent of metallicity, as confirmed by the comparison among
the metallicities investigated here and those studied in D19. The remarkable similarity in the dust yields of stars of different metallicity is due to the fact that the surface $^{12}$C, the indirect source of the dust formed in
the winds of these stars, is almost entirely of primary origin, being
dredged-up from the He-burning shell which forms at the ignition of each TP.

Stars of higher mass produce smaller quantities of dust, with an overall
dust mass below $10^{-4}~\rm M_{\odot}$. In the comparison among stars of different metallicity, it is clear in Fig.~\ref{fdtot} that the lowest
$Z$ model stars studied here produce quantities of dust significantly smaller than their higher metallicity counterparts. This is due to the lower amount of silicon and/or aluminum in
the matter from which the stars formed and the strong HBB experienced, which
heavily destroys the surface $^{12}$C and $^{16}$O, thus severely inhibiting
the formation of carbonaceous dust and silicates. Furthermore, with respect to the results found in D19 where the main dust components formed are silicates, in the metallicities investigated in this work there is a predominance of alumina dust.

In the context of massive AGBs we see a fundamental difference between the dust produced by the model stars presented here and those by D19. In the latter case the dust is mainly under the form of alumina dust and silicates and the trend with mass is slightly increasing towards the highest masses
investigated, as clear in the trend defined by the blue diamonds in 
Fig.~\ref{fdtot}, which correspond to the $Z=3\times 10^{-4}~\rm M_{\odot}$
stars published in D19. This behaviour is determined by the fact that
the larger the mass of the star, the higher the mass-loss rate experienced,
the denser the wind. On the other hand in the massive AGB model stars presented here the dust is under the form of carbonaceous dust, owing to the afore mentioned strong effects of HBB in extremely metal-poor AGB stars.

The present results, in agreement with D19, indicate that in extremely metal-poor environments the dust contribution from AGB stars becomes relevant only after $\sim 500$ Myr, considering the poor dust production by stars that evolve faster than this time limit. For what regards the contribution from
AGB stars to the cosmic dust yield, these results confirm the conclusions
reached in D19, that at redshifts above $5-6$ the dominant dust factories
are high-mass stars, exploding as supernovae.

\section{Conclusions}
We study the evolution of extremely metal-poor, $M < 8~M_{\odot}$ stars, across the asymptotic giant branch, until the almost complete loss of the external mantle. The metallicities investigated, $Z=3\times 10^{-5}$ and $Z=3\times 10^{-7}$, correspond to $[$Fe$/$H$]=-3$ and $[$Fe$/$H$]=-5$, respectively.

We find that in stars of initial mass $M\leq 2~M_{\odot}$ the surface chemical composition is altered by the effects of TDU. In higher mass stars the temperature at the bottom of the convective envelope reach (or exceed) $\sim 40$ MK, thus activating HBB: in these cases the relative importance of TDU and HBB is the key factor in the change of the surface chemistry. The present results show that in the stars of highest mass, with $M\geq 5~M_{\odot}$, the depth of TDU is so small that the surface chemistry reflects the pure effects of HBB.

In the metallicity domain investigated here the description of the convective instability is crucial for the evolution of the stars. Considering the very small initial metallicity, the depth of TDU, which is extremely sensitive to the treatment of convective borders, proves important to the determination of the surface metal enrichment, which, in turn, affects the evolution of the stellar radius and thus of the mass-loss rate. Numerical simulations exploring the possible effects of changing the assumed OS confirm that the main aspects of the evolution across the AGB phase and the gas pollution from these stars are extremely sensitive to the underlying assumptions regarding the description of the regions close to the convective borders. Convection modelling also affects the strength of HBB, with notable consequences on the largest luminosity attained by the star, hence on the mass-loss rate. The description of mass-loss, highly uncertain in the metallicity domain discussed
here, is also crucial for the determination of the evolutionary history of these stars.

The different impact of TDU and HBB in stars of different mass reflects on the gas yields. Low-mass AGBs, experiencing TDU only, produce gas largely enriched in $^{12}$C and, to a smaller extent, in the other chemical species. This gas is of primary origin
almost in its entirely, as the enrichment in the mass fractions of the various chemicals
is due to chains of nuclear reactions started by helium-burning in the pulse driven
convective shell. When compared to results from other investigations focused on low-mass stars 
of metallicity similar to those explored in the present work, the present yields of
$^{13}$C and $^{14}$N are significantly smaller, owing to the weaker proton-ingestion episodes
experienced. The stars descending from $2~M_{\odot} < M < 5~M_{\odot}$ progenitors
produce gas, once more mostly of primary origin, altered by both TDU and HBB. The most
important feature in this context is the extraordinary production of $^{13}$C and $^{14}$N,
both elements being produced by proton capture on $^{12}$C nuclei, synthesized during the
thermal pulse and convected to the surface via TDU. The most massive AGB stars experience
HBB only, with poor contribution from TDU. The yields of different chemical species
reflect a mostly secondary contribution, with the exception of $M\geq 7M_{\odot}$ stars,
which experience a deep convective mixing after the core helium burning phase, with
transportation of primary carbon to the surface regions. In all the stars of initial
mass above $\sim 5~M_{\odot}$ the nucleosynthesis reflects the equilibria of very hot 
p-capture processing. The very hot HBB temperatures produce remarkable effects on the
surface chemistry: the nucleosynthesis experienced at the base of the envelope is such
that even the yields of chemical species which are found to be largely produced in popI 
and popII AGBs, such as sodium and aluminium, are negative in the present cases.

Dust production by extremely metal-poor AGBs is limited to $M\leq 2~M_{\odot}$ stars, as in higher mass objects the very small abundances of silicon prevent significant formation of silicates; note that in these stars formation of carbonaceous dust is
also negligible, owing to the effects of HBB, which destroys the surface carbon. In the low-mass domain solid carbon dust is formed, in quantities spanning the $10^{-4}-5\times 10^{-3}~M_{\odot}$ range, the larger quantities of dust being  formed by the stars of higher mass, which accumulate more carbon in the surface regions. These findings support the idea of a dominant contribution
of massive stars to dust production in the Universe in the early epochs following the Big Bang.

\label{end}

\begin{acknowledgements}
P.V. and D.R. benefited from the International Space Science Institute (ISSI, Bern, CH, and ISSI-BJ, Beijing, CN) thanks to the funding of the team "Chemical abundances in the ISM: the litmus test of stellar IMF variations in galaxies across cosmic time". M.C. acknowledges support from "progetto INAF Mainstream" (PI: S. Cassisi). M.Lugaro acknowledges the support of the Hungarian National Research, Development and Innovation Office (NKFI), grant KH\_18 130405. E.M. acknowledges support from
the {\it Angelo Della Riccia} foundation and is indebted to the Instituto de Astrofisica de Canarias for the kind hospitality and the scientific support. M.T. acknowledges support from MIUR through the FARE project R164RM93XW SEMPLICE (PI: Milone) and from MIUR under PRIN program 2017Z2HSMF (PI: Bedin)
\end{acknowledgements}


\end{document}